\documentclass[manuscript]{aastex62}

\submitjournal{ApJ}

\hypersetup{breaklinks=true}
\begin{document}

\title{The Transition from Diffuse Molecular Gas to Molecular Cloud 
Material in Taurus}

\correspondingauthor{S. R. Federman}
\email{steven.federman@utoledo.edu, johnathan.s.rice@gmail.com, 
ritchey.astro@gmail.com, hkim@gemini.edu, 
lacy@astro.as.utexas.edu, paul.f.goldsmith@jpl.nasa.gov, 
nflagey@stsci.edu, gmace@astro.as.utexas.edu, 
dll@astro.as.utexas.edu}

\author[0000-0002-8433-9663]{S. R. Federman}
\affiliation{University of Toledo, Department of Physics and Astronomy, 
Toledo, OH 43606, USA}

\author[0000-0003-4675-314X]{Johnathan S. Rice}
\affiliation{University of Toledo, Department of Physics and Astronomy, 
Toledo, OH 43606, USA}

\author[0000-0002-3659-4192]{A. M. Ritchey}
\affiliation{Eureka Scientific Inc., Oakland, CA 96402, USA}

\author[0000-0003-4770-688X]{Hwihyun Kim}
\affiliation{Gemini Observatory/NSF’s NOIRLab, Casilla 603, La Serena, Chile}

\author[0000-0001-6783-2328]{John H. Lacy}
\affiliation{Department of Astronomy, 
University of Texas at Austin, Austin, TX 78712, USA}

\author[0000-0002-6622-8396]{Paul F. Goldsmith}
\affiliation{Jet Propulsion Laboratory, 
California Institute of Technology, Pasadena, CA 91109, USA}

\author[0000-0002-8763-1555]{Nicolas Flagey}
\affiliation{Canada France Hawaii Telescope Corporation, 
Kamuela, HI 96743, USA}
\affiliation{Space Telescope Science Institute, 3700 San Martin Drive, Baltimore, 
MD 21218, USA}

\author[0000-0001-7875-6391]{Gregory N. Mace}
\affiliation{W. J. McDonald Observatory and Department of Astronomy, 
University of Texas at Austin, Austin, TX 78712, USA}

\author[0000-0003-1814-3379]{David L. Lambert}
\affiliation{W. J. McDonald Observatory and Department of Astronomy, 
University of Texas at Austin, Austin, TX 78712, USA}
\nocollaboration

\begin{abstract}
We study four lines of sight that probe the transition from diffuse molecular 
gas to molecular cloud material in Taurus.  Measurements of atomic and 
molecular absorption are used to infer the distribution of species and the 
physical conditions toward stars behind the Taurus Molecular 
Cloud (TMC).  New high-resolution spectra at visible and near infrared 
wavelengths of interstellar Ca~{\small II}, Ca~{\small I}, K~{\small I}, 
CH, CH$^+$, C$_2$, CN, and CO toward HD~28975 and HD~29647 are combined with 
data at visible wavelengths and published CO results from ultraviolet 
measurements for HD~27778 and HD~30122.  Gas densities and temperatures are 
inferred from C$_2$, CN, and CO excitation and CN chemistry.  
Our results for HD~29647 are noteworthy because the CO column 
density is 10$^{18}$~cm$^{-2}$ while C$_2$ and CO excitation 
reveals a temperature of 10~K and density about 1000~cm$^{-3}$, 
more like conditions found in dark molecular clouds.  
Similar results arise from 
our chemical analysis for CN through reactions involving 
observations of CH, C$_2$, and NH.  Enhanced potassium depletion and a 
reduced CH/H$_2$ column density ratio also suggest the presence of a 
dark cloud.  The directions toward HD~27778 and HD~30122 probe 
molecule-rich diffuse clouds, which can be considered CO-dark gas, 
while the sight line toward HD~28975 represents an intermediate case.  Maps 
of dust temperature help refine the description of the material along the 
four sight lines and provide an estimate of the distance between 
HD~29647 and a clump in the TMC.  An Appendix provides 
results for the direction toward HD~26571; this 
star also probes diffuse molecular gas.
\end{abstract}


\section{Introduction}

There are numerous observational studies of diffuse atomic and molecular gas, 
based of a combination of absorption measurements spanning ultraviolet to
radio wavelengths as well as emission lines at cm and mm wavelengths. 
More recent examples include a number of comprehensive studies  
\citep{lis96, lis98, lis02, wel01, pan05, son07, she08, jen09, bur10, ger10, jen11, ind12}. 
The same can be said of dark molecular clouds from observations at 
infrared through radio wavelengths  
\citep[e.g.,][]{mye83a, mye83b, gai84, pat95, pat98, miz01, pin08, ros08, pin10, lac17}.  
However, there are few published studies of dark clouds with a focus on absorption 
at visible wavelengths for direct comparison with the efforts on diffuse atomic and 
molecular clouds.  Here we define dark clouds as having CO column densities, 
$N$(CO), in excess of 10$^{17}$ cm$^{-2}$ 
so that a substantial fraction of elemental carbon 
is in gas-phase CO.  Such column densities are significantly greater than those in 
sight lines used as benchmarks for diffuse molecular gas, such as toward $\zeta$ 
Oph, where the CO column density is $\sim$10$^{15}$ cm$^{-2}$ \citep{lam94}.  This 
regime of dark clouds with $N$(CO) of 10$^{17}$ cm$^{-2}$ contains the final 
stages of the chemical transitions from atomic to molecular gas for carbon 
(C$^+$ to CO), nitrogen (N to NH$_3$), and oxygen (O to H$_2$O) and where CO, CO$_2$, 
and H$_2$O ices begin to coat interstellar dust grains.  In this paper, we present a 
detailed study of this regime for portions of the Taurus Molecular Cloud (TMC) 
probed by the background stars HD~28975 and HD~29647.  In particular, data on 
Ca~{\small II}, K~{\small I}, CN, CH$^+$, CH, C$_2$, and CO are 
analyzed and compared with available results as well as newly acquired data 
for the nearby sight lines toward HD~27778 and HD~30122.

Previous results for two directions, those toward HD~29647 and HD~200775, 
are used as guides on how to proceed.  It is important to note that these 
sight lines are dominated by a single velocity component.  \citet{cru85} 
combined measurements of atomic and molecular absorption at visible 
wavelengths with millimeter-wave emission from CO and its isotopologues, CN, HCN, 
and HCO$^+$.  His detailed analysis revealed $N$(CO) of about 
$1.5\times10^{17}$ cm$^{-2}$, H$_2$ densities of 800 cm$^{-3}$, and a 
kinetic temperature of  10 K.  Earlier C$_2$ measurements were described by 
\citet{hob83} and \citet{ lut83}, and subsequently published results 
on CH, C$_2$, and CN appeared in \citet{tho03}.  All the previously 
published data at visible wavelengths for HD~29647 were acquired at 
moderate spectral resolution (about 7 to 15 km s$^{-1}$).  A comprehensive 
analysis of atomic and molecular absorption toward HD~200775 \citep{fed97} at 
high-spectral resolution (2 to 3 km s$^{-1}$) was followed by measurements 
on CO absorption at ultraviolet wavelengths with the {\it International 
Ultraviolet Explorer} \citep{kna01} and the {\it Far Ultraviolet Spectroscopic 
Explorer} (FUSE) \citep{she08}.  These measurements probed the Photon 
Dominated Region (PDR) in NGC~7023 in front of HD~200775.  The results 
indicate $N$(CO) $=$ $1.9\times10^{17}$ cm$^{-2}$, proton densities 
[$n$(H) $+$ 2$n$(H$_2$)] between 300 and 900 cm$^{-3}$, and a 
kinetic temperature of  40 K.  Here, we extend this body of work through 
high-resolution observations (about 2.5 and 6.5 km s$^{-1}$, respectively)
in the visible and near infrared toward 
HD~28975 and HD~29647, combined with similar-quality results for 
the nearby directions toward HD~27778 and HD~30122.

In order to place our study into context, Figure 1 shows a map of 
$^{13}$CO intensity \citep{pin10} indicating the lines of sight to 
HD~27778, HD~28975, HD~29647, and HD~30122.  The more 
reddened stars (HD~28975 and HD~29647) lie behind regions containing 
filaments seen in $^{13}$CO.  \citet{fed94} noted that HD~27778 samples 
gas near L~1506, while HD~28975 and HD~29647 probe material in L~1529 
and Heiles Cloud-2, respectively (see their Fig. 4c).  The direction to HD~30122 
lies beyond the $^{13}$CO emission in Fig. 1.  Also indicated in Fig. 1 are 
the sight lines toward embedded and background sources in Taurus that were 
used by \citet{lac17} in their study of the CO/H$_2$ ratio.  These targets lie 
along paths with more intense and filamentary structures seen in $^{13}$CO 
emission.  The results from \citet{lac17} are discussed further below when 
we consider the small, parsec scale variations in $N$(CO).

\begin{figure}[tbh]
\includegraphics[scale=0.65]{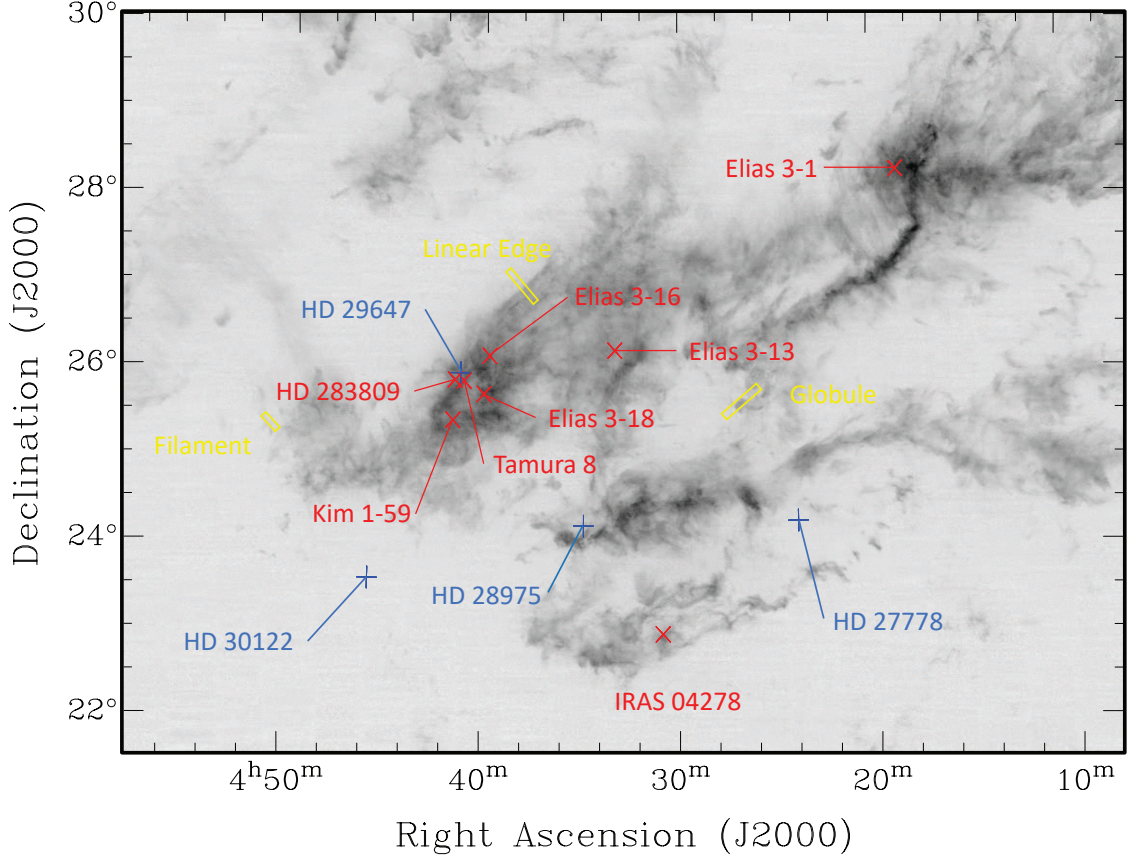}
\caption{A map of $^{13}$CO intensity for the Taurus Molecular Cloud from 
\citet{pin10} providing the positions of stars discussed in the text.  Our targets, 
HD~27778, HD~28975, HD~29647, and HD 30122, are shown in blue 
with plus signs, while those from \citet{lac17} are shown in red and marked with 
an x.  The regions labeled as the Filament, the Linear Edge, and the Globule 
(Goldsmith et al. 2010) are approximately represented by yellow rectangles.
\label{fig.1}}
\end{figure}

The outline for the remainder of the paper is the following.  Section 2 provides an 
overview of our extensive set of observations: absorption from Ca~{\small II} 
$\lambda$3933, Ca~{\small I} $\lambda$4226, K~{\small I} $\lambda$7698 
and from molecular bands CH [$B$--$X$ (0,~0) and $A$--$X$ (0,~0)], CH$^+$ 
[$A$--$X$ (0,~0) and (1,~0)], C$_2$ [$A$--$X$ (2,~0)], CN [$B$--$X$ 
(0,~0), (1,~0), and $A$--$X$ (2,~0)], NH [$A$--$X$ (0,~0)], and CO 
[2-0 rovibration].  The photometric data at infrared wavelengths are also 
described here.  Section 3 presents the results from the absorption-line 
measurements and the chemical and excitation analyses we perform on them.  
Maps of dust temperature derived from the far infrared (FIR) data in the vicinity 
of the four stars are presented in this section and the enhanced FIR emission 
toward HD~29647 is used to estimate the distance between the star and the cloud.  
The discussion in Section 4 focuses on interpreting 
our work in terms of the transition from diffuse atomic and molecular 
gas to material in a molecular cloud.  We attempt to integrate many of the 
previous efforts on molecular gas in Taurus into a self-consistent picture.  
Final remarks appear in Section 5.  Appendix A provides the total equivalent 
widths determined from our spectra with comparisons to other measurements, while 
Appendices B and C discuss an analysis of ionization balance for calcium and a 
description of the line of sight toward HD 26571 from available data.

\newpage
\section{Observations}

Relevant data for the four stars appear in Table 1, where equatorial 
positions, apparent magnitudes at $B$, $V$, and $K$, spectral types, 
$E$($B-V$), and distances derived from {\it Gaia} Data Release 2 (DR2) 
are given.   For the two stars with CO measurements at IR wavelengths, 
HD~28975 and HD~29647, the Table also provides $K$ magnitudes.  The coordinates 
come from the SIMBAD Database, operated at Centre de Donn\'{e}es 
Astronomiques de Strasbourg (CDS) \citep{wen00}, while the distances are 
based on the parallaxes in the database from {\it Gaia} DR2 discussed 
by \citet{bai18}\footnote[8]{http://gaia.ari.uni-heidelberg.de/tap.html}.  
Specific references for the other data are given in the table.  In what follows, 
we describe the spectroscopic and photometric measurements for our study.

\begin{deluxetable}{lcccccccc}[tbh]
\tablecaption{Stellar Data}
\tablehead{
\colhead{Star} & \colhead{R. A. (2000)\tablenotemark{a}} & 
\colhead{DEC (2000)\tablenotemark{a}} & \colhead{$B$} & 
\colhead{$V$} & \colhead{$K$} & \colhead{SpT} & 
\colhead{$E$($B-V$)} & \colhead{Distance\tablenotemark{b}} \\
\colhead{} & \colhead{($^h$:$^m$:$^s$)} & 
\colhead{($^{\circ}$:$^{\prime}$:$^{\prime\prime}$)} & \colhead{} & 
\colhead{} & \colhead{} & \colhead{} & \colhead{} & \colhead{(pc)}
}
\startdata
HD~27778 & 04:24:00 & $+$24:18:04 & 6.54\tablenotemark{c} & 
6.36\tablenotemark{c} & $\ldots$ & B3V\tablenotemark{d} & 
0.37\tablenotemark{e} & 224(2)\tablenotemark{f} \\
HD~28975 & 04:34:50 & $+$24:14:40 & 9.90\tablenotemark{c} & 
9.10\tablenotemark{c} & 7.11\tablenotemark{g} & 
A4III\tablenotemark{h} & 0.60\tablenotemark{h} & 194(2) \\
HD~29647 & 04:41:08 & $+$25:59:34 & 9.22\tablenotemark{h} & 
8.31\tablenotemark{h} & 5.36\tablenotemark{g} & 
B9III Hg-Mn\tablenotemark{d} & 1.09\tablenotemark{i} & 155(2) \\
HD~30122 & 04:45:42 & $+$23:37:41 & 6.41\tablenotemark{c} & 
6.34\tablenotemark{c} & $\ldots$ & B5III\tablenotemark{d} & 
0.23\tablenotemark{j} & 256(4) \\
\enddata
\tablenotetext{a}{SIMBAD -- Wenger et al. 2000.}
\tablenotetext{b}{{\it Gaia} DR2 -- Bailer-Jones et al. 2018.}
\tablenotetext{c}{TYCHO Catalog -- H{\o}g et al. 2000.}
\tablenotetext{d}{Mooley et al. 2013.}
\tablenotetext{e}{Jensen et al. 2007.}
\tablenotetext{f}{Numbers in parentheses are uncertainties in distance.}
\tablenotetext{g}{2MASS Catalog -- Cutri et al. 2003.}
\tablenotetext{h}{Ducati 2002.}
\tablenotetext{i}{Crutcher 1985.}
\tablenotetext{j}{Fitzpatrick \& Massa 2007.}
\end{deluxetable}

\subsection{Tull Spectrograph}

Observations of HD~28975 and HD~29647 were acquired with the 
2.7 m Harlan J.~Smith Telescope (HJST) at McDonald Observatory using the 
Tull (2dcoud\'{e}) Spectrograph \citep[TS,][]{tul95} in its high-resolution mode 
(TS21).  The TS data were obtained over the course of four nights in 2017 
September.  All of the observations employed the E1 grating with a 250 $\mu$m 
slit (Slit 3) and a $2048\times2048$ CCD (TK3). Three separate wavelength 
settings were utilized. The first one provided access to the CN $B$--$X$ (0,~0) 
band near 3874~\AA, the CH$^+$ $A$--$X$ (0,~0) transition at 4232~\AA, 
the CH $A$--$X$ (0,~0) transition at 4300~\AA, the Ca~{\small II}~K line at 
3933~\AA, and the Ca~{\small I} transition at 4226~\AA.  The second wavelength 
setting yielded data on the C$_2$ $A$--$X$ (2,~0) band near 8757~\AA\ and the 
$^S$R$_{21}$(0) line of the CN $A$--$X$ (2,~0) band at 7871.644 \AA. A 
third setting was necessary to cover the K~{\small I} line 
at 7698~\AA.  Multiple 30-minute exposures 
were taken of HD~28975 and HD~29647 at each wavelength setting.  A set of 10 
biases and 10 flats (per setting) were obtained each  night, while Th-Ar comparison 
spectra were recorded throughout the night at intervals of 2--3 hours. Four 
30-minute dark frames were acquired on the first night of the run.

The raw TS data were reduced using standard procedures within the Image 
Reduction and Analysis Facility (IRAF) environment.  The overscan region in each of 
the raw images was fitted with a low-order polynomial and the excess counts were 
removed.  An average bias was created for each night and was used to correct the 
darks, flats, stellar exposures, and comparison lamp frames.  No dark correction was 
necessary because the level of dark current was found to be insignificant after the bias 
was removed.  Cosmic rays were eliminated from the stellar and comparison lamp 
exposures.  Any cosmic rays present in individual flat lamp exposures were effectively 
removed by taking the median of all flats for a given night.  Scattered light was 
modeled in the dispersion and cross-dispersion directions and subtracted from the 
stellar exposures and from the median flat.  The flat was then normalized to unity and 
divided into the stellar and comparison lamp frames.  One-dimensional spectra were 
extracted from the processed images by summing across the width of each order.  
The extracted stellar spectra were calibrated in wavelength space after identifying 
emission lines in the Th-Ar comparison spectra.  Finally, the wavelength calibrated 
spectra were shifted to the reference frame of the Local Standard of Rest (LSR).   
For these sight lines in Taurus, the difference between heliocentric and LSR 
velocities is about $+10$ km s$^{-1}$.

The multiple exposures of HD~28975 and HD~29647 obtained at a particular 
wavelength setting were co-added to maximize the signal-to-noise (S/N) ratio 
achieved in the final spectra.  The co-added spectra were then normalized to 
the continuum by fitting low-order polynomials to regions free of interstellar 
absorption within small spectral windows surrounding the interstellar lines of 
interest.  From measurements of the widths of Th~{\small I} emission lines in 
the comparison spectra, we find that a resolving power of 
$R=\lambda/\Delta\lambda\approx125,000$ was achieved for the observations 
covering the short wavelength lines and the C$_2$ band.  For the observations 
covering the K~{\small I} line, which were acquired on the last 
night of the run, a somewhat lower resolving power of $R\approx110,000$ was 
attained.  The S/N ratios achieved in the final co-added spectra of HD~28975 
and HD~29647 were $\sim$30 in the vicinity of the CN $B$--$X$ and 
Ca~{\small II} lines, $\sim$80 near the CH, CH$^+$, and Ca~{\small I} 
transitions, $\sim$150 near the C$_2$ and K~{\small I} features, and $\sim$250 
near the CN $A$--$X$ transition at 7871 \AA.

A similar set of TS spectra for gas toward HD~30122 was 
acquired in 2019 September and December. These observations employed the short 
wavelength setting (covering the CN $B$--$X$, CH, CH$^+$, Ca~{\small I}, and 
Ca~{\small II} lines), and the K~{\small I} setting, but did not include the 
setting covering the C$_2$ band.  The procedures used for obtaining and reducing 
the data were the same as those described above. The resolving power of the 
spectrograph was essentially unchanged for these observing runs compared to the 
previous one.   The resulting S/N ratios were $\sim$100 and $\sim$140 near the 
CN $B$--$X$ and Ca~{\small II} lines, respectively, $\sim$170 near the CH, 
CH$^+$, and Ca~{\small I} transitions, and $\sim$40 near the K~{\small I} line.

Figure 2 displays TS spectra showing molecular absorption from CN 
$B$--$X$ (0,~0) R(0), CH $\lambda4300$, CH$^+$ $\lambda4232$, 
followed by atomic absorption from Ca~{\small I}, K~{\small I}, and 
Ca~{\small II}.  The lines toward HD~28975 appear on the left side, 
and those toward HD~29647 on the right.  Numerous stellar features 
complicate the analysis of the spectra obtained for HD~29647 (a HgMn star).  
For example, the interstellar CH $\lambda4300$ line is partially blended with 
a stellar Mn~{\small II} line that has a laboratory wavelength of 4300.25~\AA\ 
(see Figure 2). In addition, the Ca~{\small II} $\lambda3933$ line is most 
likely a complicated blend of stellar and interstellar absorption. (As a 
result, we make no attempt to derive a Ca~{\small II} column density toward 
HD~29647.) The poorer S/N ratio obtained for 
TS spectra acquired below 4000 \AA\ arises from decreased stellar flux caused 
by a combination of significant reddening by foreground dust and less 
sensitivity for the spectrograph.

\begin{figure}[tbh]
\plottwo{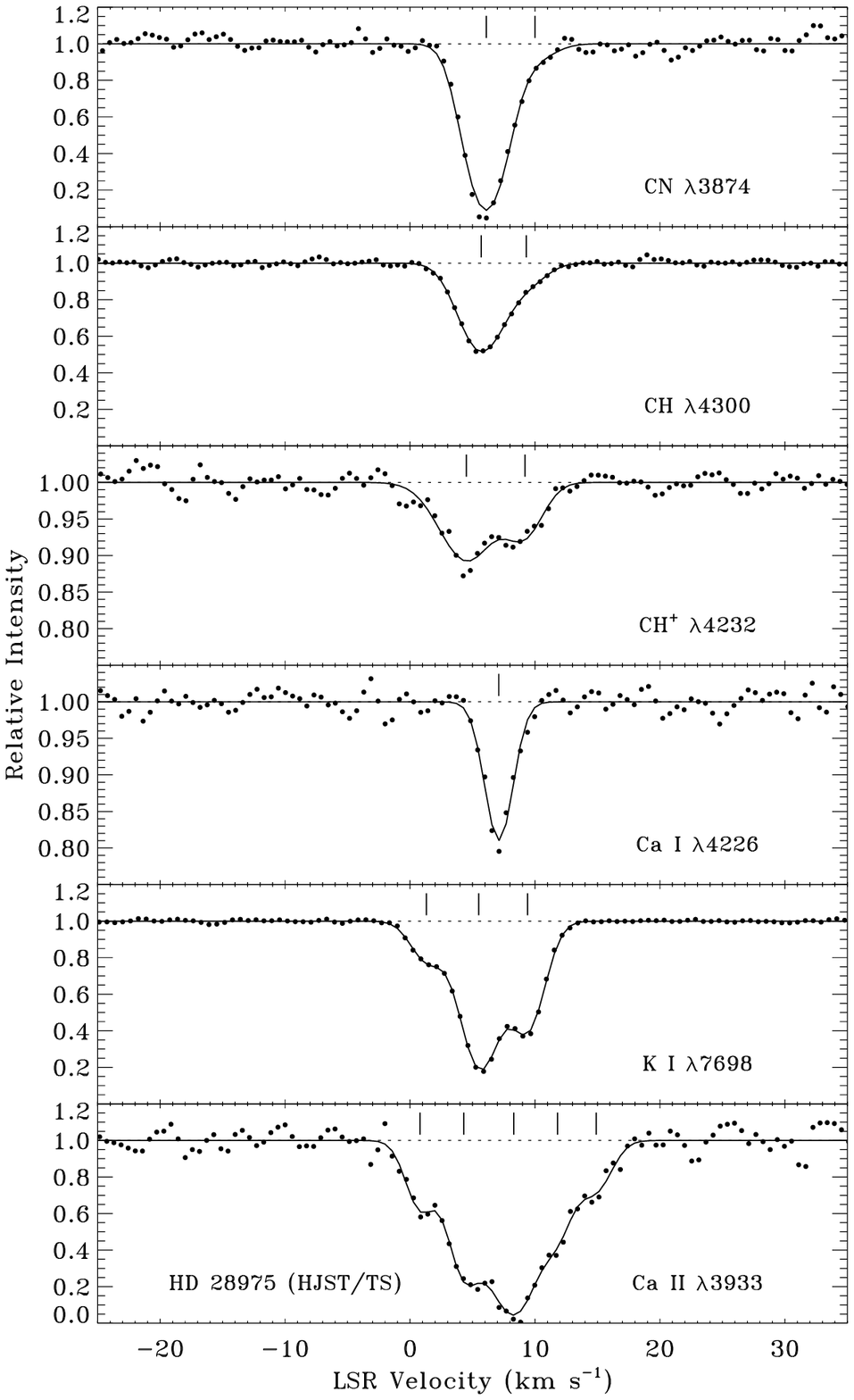}{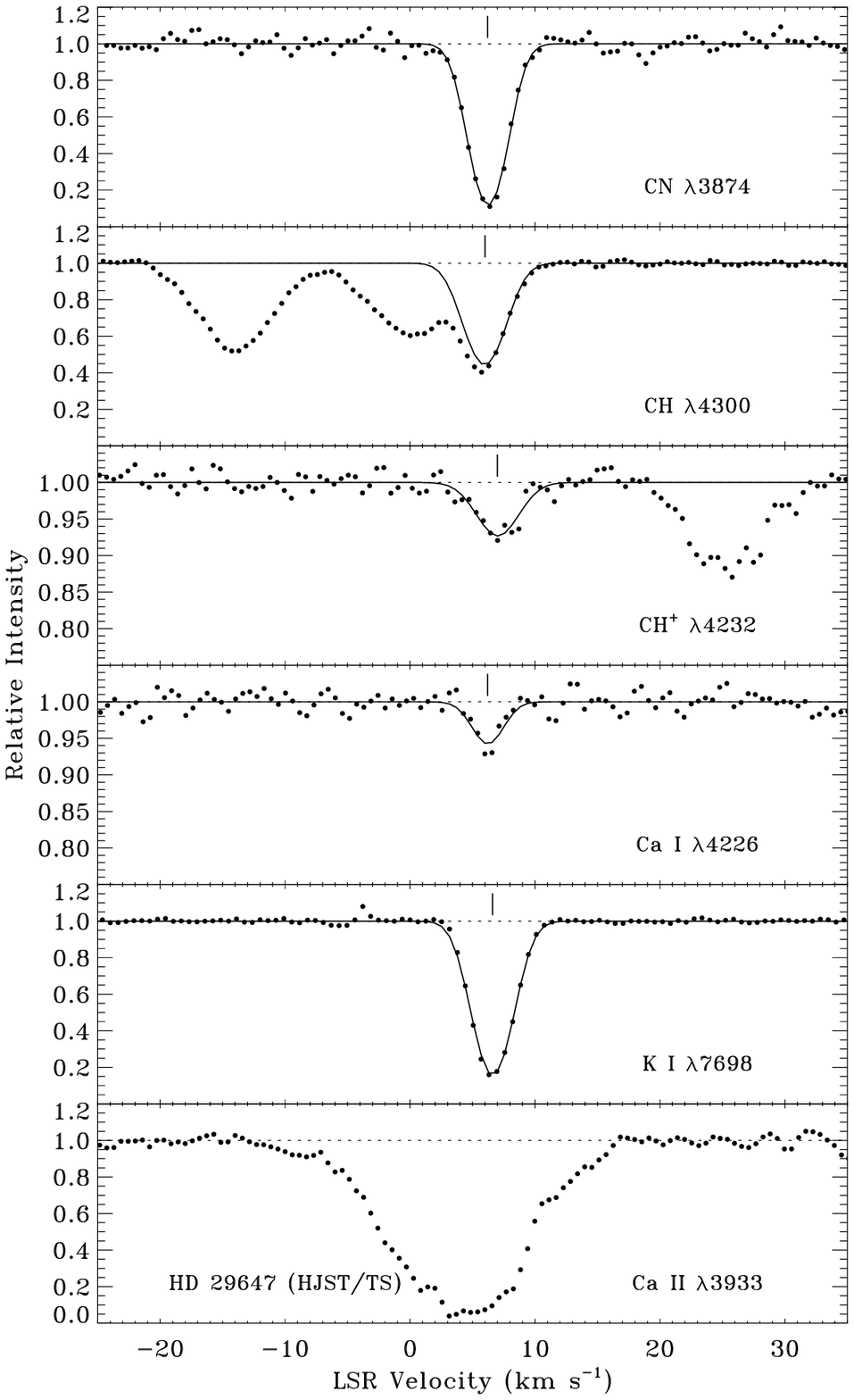}
\caption{TS spectra of interstellar absorption toward HD~28975 (left panels) 
and HD~29647 (right panels) at visible wavelengths.  Note that the 
vertical scales differ from panel to panel.  Many stellar features from 
HD~29647 are seen, especially affecting absorption from interstellar CH and 
Ca~{\sc II}.  The two stellar lines near CH are Ti~{\sc II} at 4300.042 \AA\ and 
Mn~{\sc II} at 4300.254 \AA.  In the case for Ca~{\sc II}, 
there is no clear evidence for absorption from interstellar material.  
The complex profile of the Ca~{\sc II} feature, which resembles that 
of the stellar Fe~{\sc II} line lying near the interstellar CH$^+$ line, makes 
it difficult to discern the presence of interstellar Ca~{\sc II}.  The vertical 
tick marks above the spectra indicate the velocities for each component.}
\end{figure}

\newpage
\subsection{Ultraviolet and Visual Echelle Spectrograph}

Archival spectra of HD~27778 and HD~29647, acquired using the 
Ultraviolet and Visual Echelle 
Spectrograph (UVES) of the Very Large Telescope (VLT), were obtained from the 
European Southern Observatory (ESO) Science Archive Facility. The VLT/UVES 
spectra of HD~29647 were acquired in 2008 August under program 
081.D-0498 (PI: S.~Hubrig) 
and cover nearly the entire visible spectrum from 3300 to 9500~\AA\ (with gaps 
between 4500 and 5700~\AA\ and between 7500 and 7700~\AA). Observations of 
HD~27778 were obtained in 2014 October and November under program 194.C-0833 
(PI: N.~Cox). These data (likewise) cover essentially the entire visible spectrum 
(with small gaps) from 3100 to 10400 \AA\ \citep[see][]{cox17}. In addition 
to providing duplicate observations of the lines already covered by the TS data 
(for HD~29647), the VLT/UVES spectra provide information on the NH $A$--$X$ (0,~0) 
band near 3358~\AA, the CN $B$--$X$ (1,~0) band near 3579~\AA, the 
CH$^+$ $A$--$X$ (1,~0) transition at 3957~\AA, the CH $B$--$X$ 
(0,~0) lines near 3886~\AA, the C$_3$ $A$--$X$ (000)$-$(000) band near 
4051~\AA, and the C$_2$ $A$--$X$ (3,~0) band near 7719~\AA.  Severe 
contamination from telluric features (and stellar lines in the case of HD~29647) 
prevented our use of features associated with the C$_3$ and C$_2$ (3,~0) 
bands.

After downloading the raw science and calibration frames from the ESO archive, the 
VLT/UVES data were reduced using the UVES pipeline software, which corrects 
for the bias in the data, subtracts scattered light, 
finds and extracts the echelle orders, 
flat fields the data, applies a dispersion solution, and then merges the orders to 
produce a final spectrum. The optimal extraction method was adopted for data 
obtained with the blue arm. However, for data obtained with the red arm, for which 
the default reduction procedures often lead to residual fringing and rippling in 
the reduced spectra, the average extraction method was used and the flat fields were 
divided into the stellar spectra pixel by pixel (rather than after extracting the spectra 
as is the default approach). The extracted and merged spectra were then shifted 
to the LSR frame of reference and small segments of spectra surrounding interstellar 
lines of interest were normalized to the continuum in the same way as for the 
TS data discussed above. The VLT/UVES spectra are 
characterized by higher S/N ratios than were achieved with the TS data 
(by factors of 3 to 8 for HD~29647), but the UVES data were acquired at lower 
spectral resolution. (The blue UVES data have $R\approx80,000$, while 
$R\approx100,000$ applies to the red data.)

\begin{figure}[tbh]
\plottwo{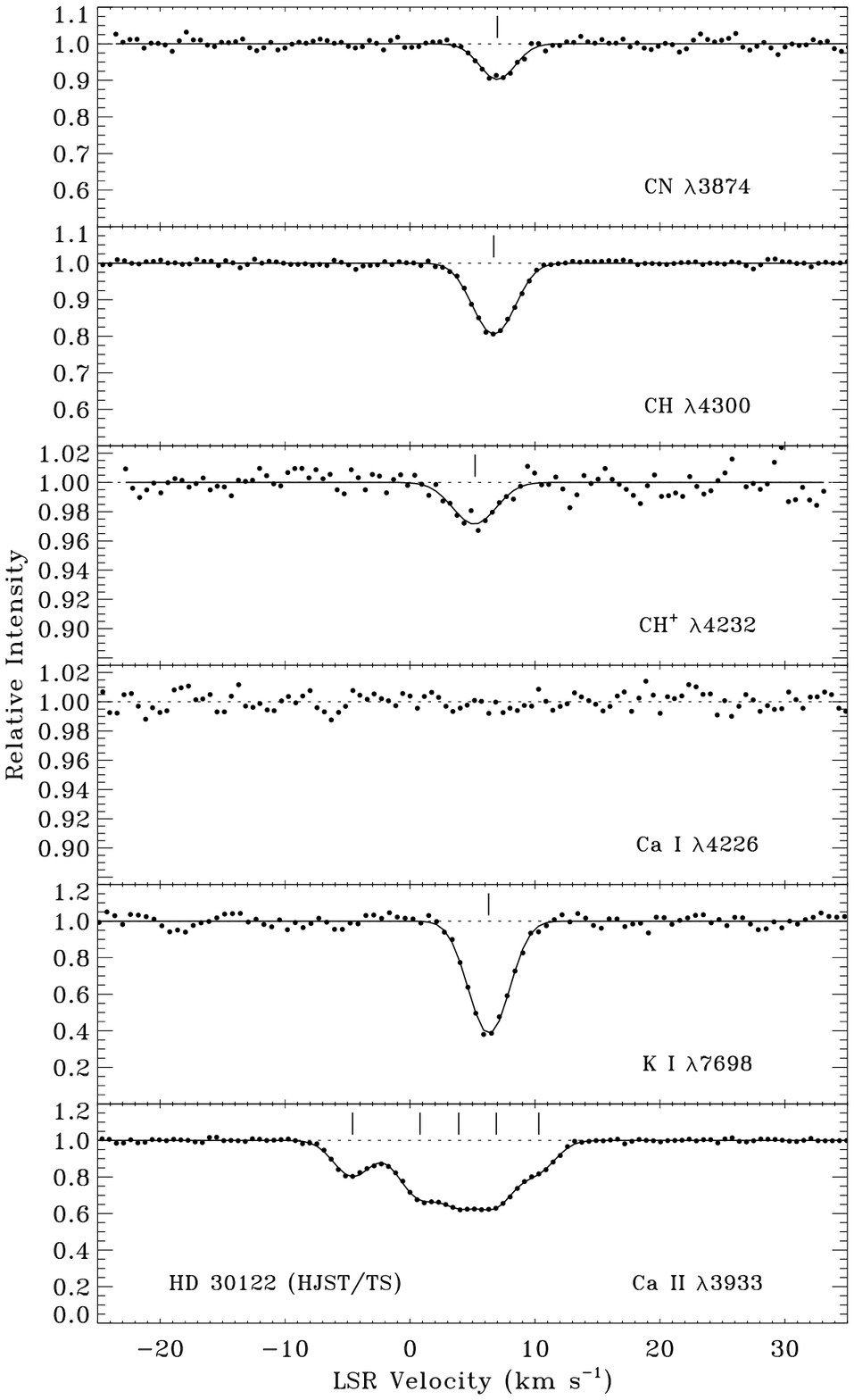}{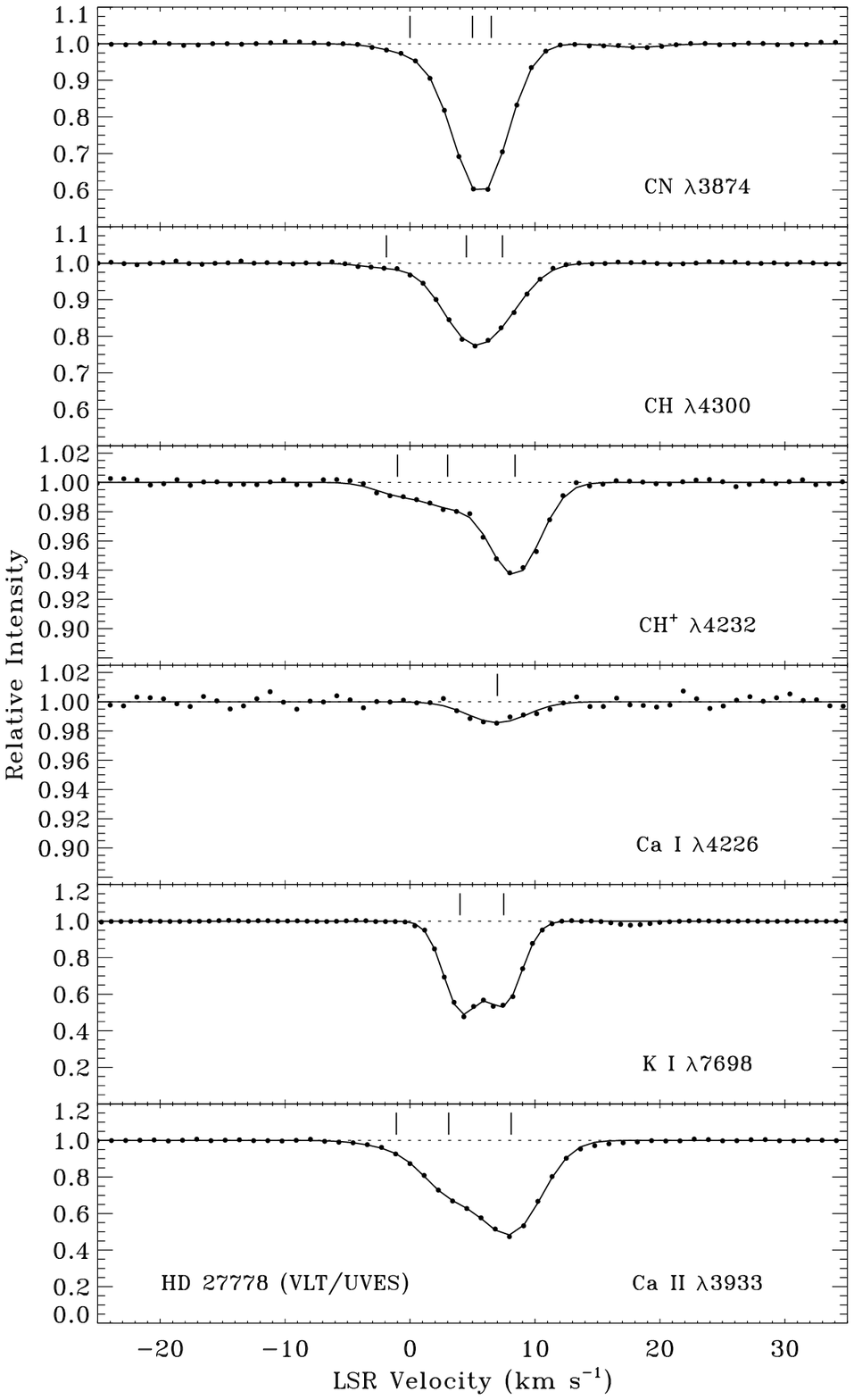}
\caption{Interstellar absorption toward HD~30122 with TS (left panels) and 
toward HD~27778 with UVES (right panels) at visible wavelengths.  In the spectrum for 
HD~30122, a broad stellar Fe~{\sc II} feature near the interstellar CH$^+$ line was 
removed, as was stellar absorption from Ca~{\sc II}, before fitting.  See 
Fig. 2 for additional details.\label{fig.3}}
\end{figure}

Additional examples of TS and UVES spectra for our targets appear in Figures 3 
and 4. Figure 3 displays TS spectra for HD~30122 (left panels) and UVES spectra for 
HD~27778 (right panels), adopting the same format as in Fig. 2.  (We note that a 
stellar feature affecting the Ca~{\small II} profile toward HD~30122 was fitted 
and removed from the spectrum during the continuum normalization process.)  
The rich spectra of the C$_2$ $A$--$X$ (2,~0) band toward HD~28975 and HD~29647 
appear in Figure 4.  The P-. Q-, and R-branches show detections up to 
$J^{\prime\prime}$ = 8 (and 10) toward HD~28975 (HD~29647).  The total 
equivalent widths of interstellar species seen in our spectra from TS and UVES 
are given in Appendix A, as is a comparison with other determinations.

\begin{figure}[tbh]
\gridline{\rotatefig{270}{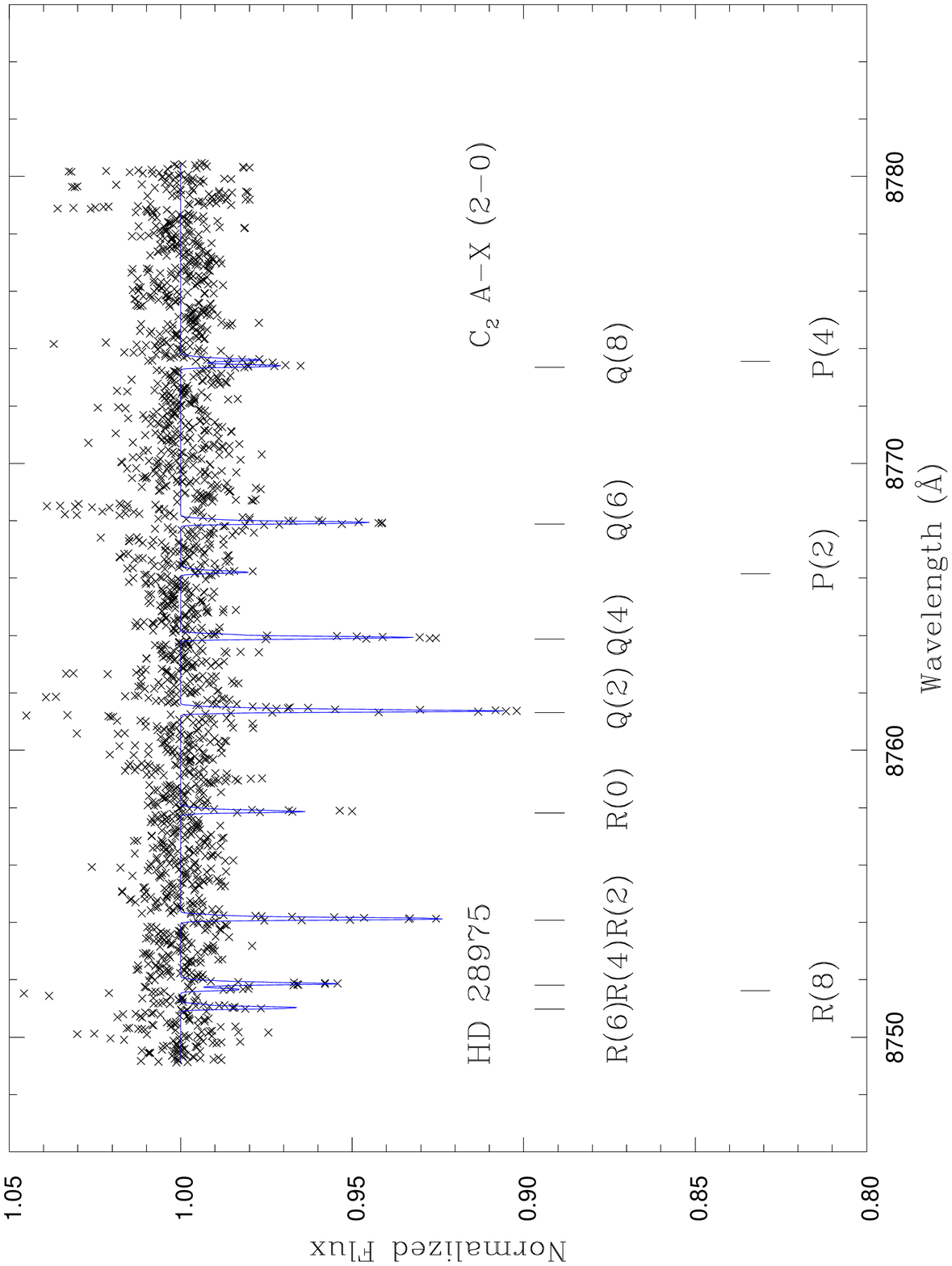}{0.36\textwidth}{(a)}
          \rotatefig{270}{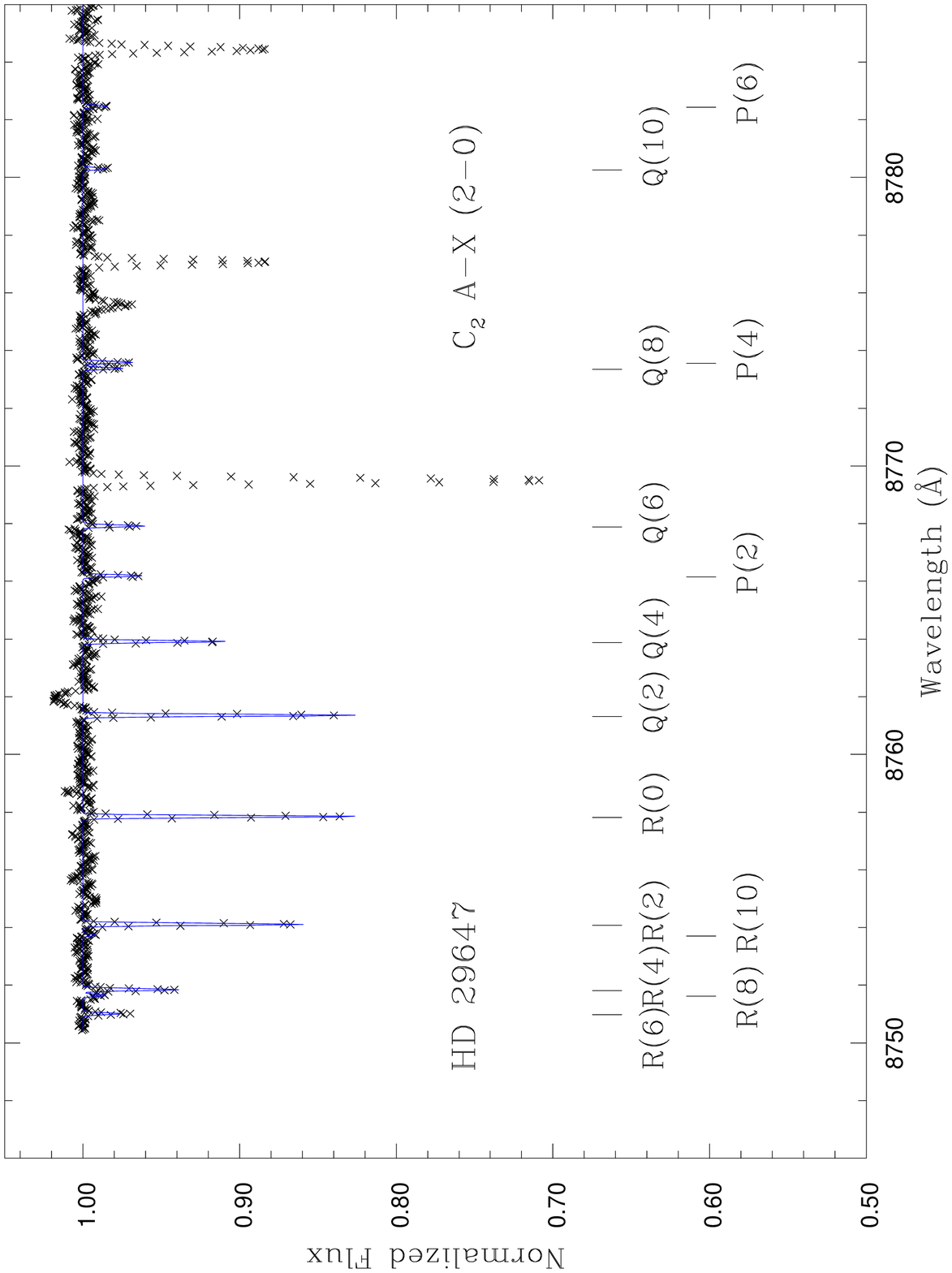}{0.36\textwidth}{(b)}}
\caption{Absorption from the C$_2$ $A$--$X$ (2,~0) band toward 
HD~28975 with TS (a) and toward HD~29647 with UVES 
(b).  The UVES spectrum for HD~29647 provides detections of 
the Q(10) and P(6) lines.  The bump near Q(2) in plot (b) is from the 
residuals remaining after removing the strong Paschen line of H~{\sc I} 
at 8750 \AA.  The other, unidentified lines are stellar absorption from 
Mn~{\sc II} at 8769.176 and 8784.127 \AA\ and Kr~{\sc I} at 8776.750 \AA.  
Observed (x) and synthesized (blue line) spectra are presented.  Details 
of the profile fitting are described in \textsection{3.1}.  Line 
identifications appear below the features.\label{fig.4}}
\end{figure}

\subsection{IGRINS}

High-resolution ($R\approx45,000$) near infrared (14,500 to 24,500~\AA) 
spectra were obtained with the Immersion GRating INfrared Spectrometer 
\citep[IGRINS;][]{yuk10,par14}. The spectrum of HD~29647 employed 
IGRINS on HJST at McDonald Observatory in 2016 January, while the 
spectrum of HD~28975 is from IGRINS when it was installed on the 4.3~m Lowell 
Discovery Telescope (LDT) in 2018 October
\citep{mac16, mac18}. During the LDT measurements, a mechanical anomaly 
degraded the spectral resolution, reducing $R$ to 35,000.  
On both telescopes the spectral format of IGRINS 
is unchanged. Targets were observed in two positions on the slit to facilitate 
sky subtraction, and the A0V star k~Tau was observed at a similar airmass 
after each science target to use for telluric correction. Spectra were extracted 
using the IGRINS pipeline 
\citep*{lee17}\footnote[9]{\url{https://github.com/igrins/plp/tree/v2.1-alpha.3}}, 
which performs flat-field correction, wavelength calibration with night sky OH 
emission and telluric absorption lines, and optimal extraction of the 
one-dimensional spectrum. Telluric absorption lines were corrected by dividing 
the science spectrum by the k~Tau spectrum, after the latter had been divided by 
the Vega model of \cite{kur79}.

The scientific goals of our project required spectra with S/N ratio 
of about 500.  We took care when removing the telluric lines so that residual 
absorption was less than 1\% of the target star's continuum.  This was 
accomplished by applying small corrections to the spectrum of k~Tau.  
First, we made an adjustment so that the airmasses for k~Tau and our 
targets were essentially the same; this step removed the telluric CO 
features in the wavelength intervals of interest to us.  However, variation in 
the water vapor during the observations caused residuals.  These were corrected 
by multiplying or dividing by a small power of a telluric water vapor absorption 
spectrum calculated from the HITRAN line list \citep{rot05} and the U.S. Standard 
Atmosphere, provided by JHL.

We sought absorption by interstellar CO and $^{13}$CO in their 2-0 rovibrational 
bands, with respective R(0) lines at 2.3453 and 2.3978 $\mu$m.  While the 
H$_2$ S(0) line from the 1-0 rovibrational band 
at 2.2232 $\mu$m may be observed in IGRINS spectra \citep{lac17}, 
the S/N ratios obtained by us and H$_2$ column densities less than 
10$^{22}$ cm$^{-2}$ toward HD~28975 and HD~29647 prevented 
our seeing H$_2$ absorption.  Normalized 
spectra showing CO toward our two targets appear in Figure 5.  
We did not detect $^{13}$CO absorption, but we discuss a 
meaningful upper limit toward HD~29647 in \textsection{4}. Appendix A also 
provides equivalent widths from the CO spectra.

\begin{figure}[tbh]
\gridline{\rotatefig{270}{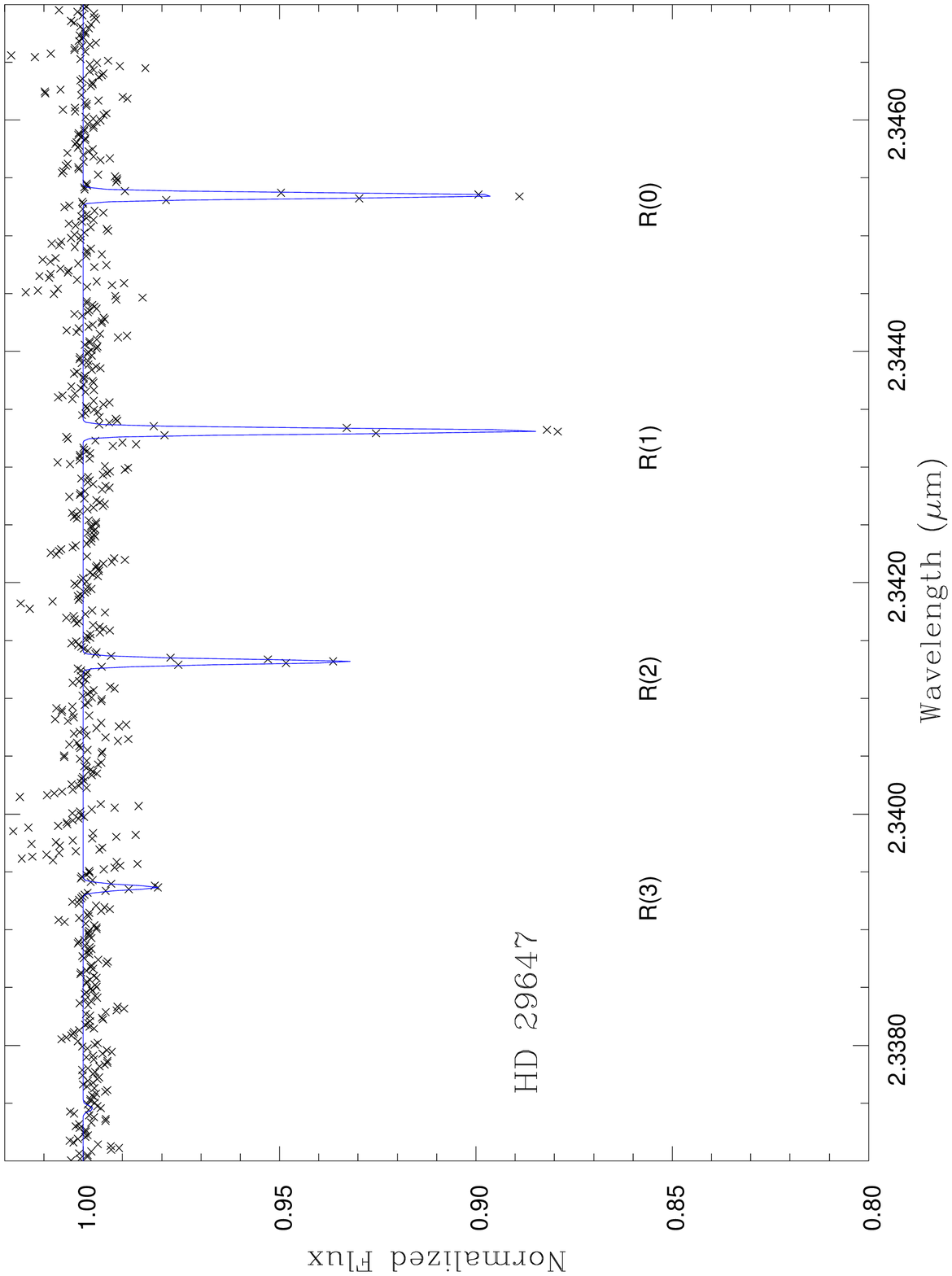}{0.36\textwidth}{(a)}
          \rotatefig{270}{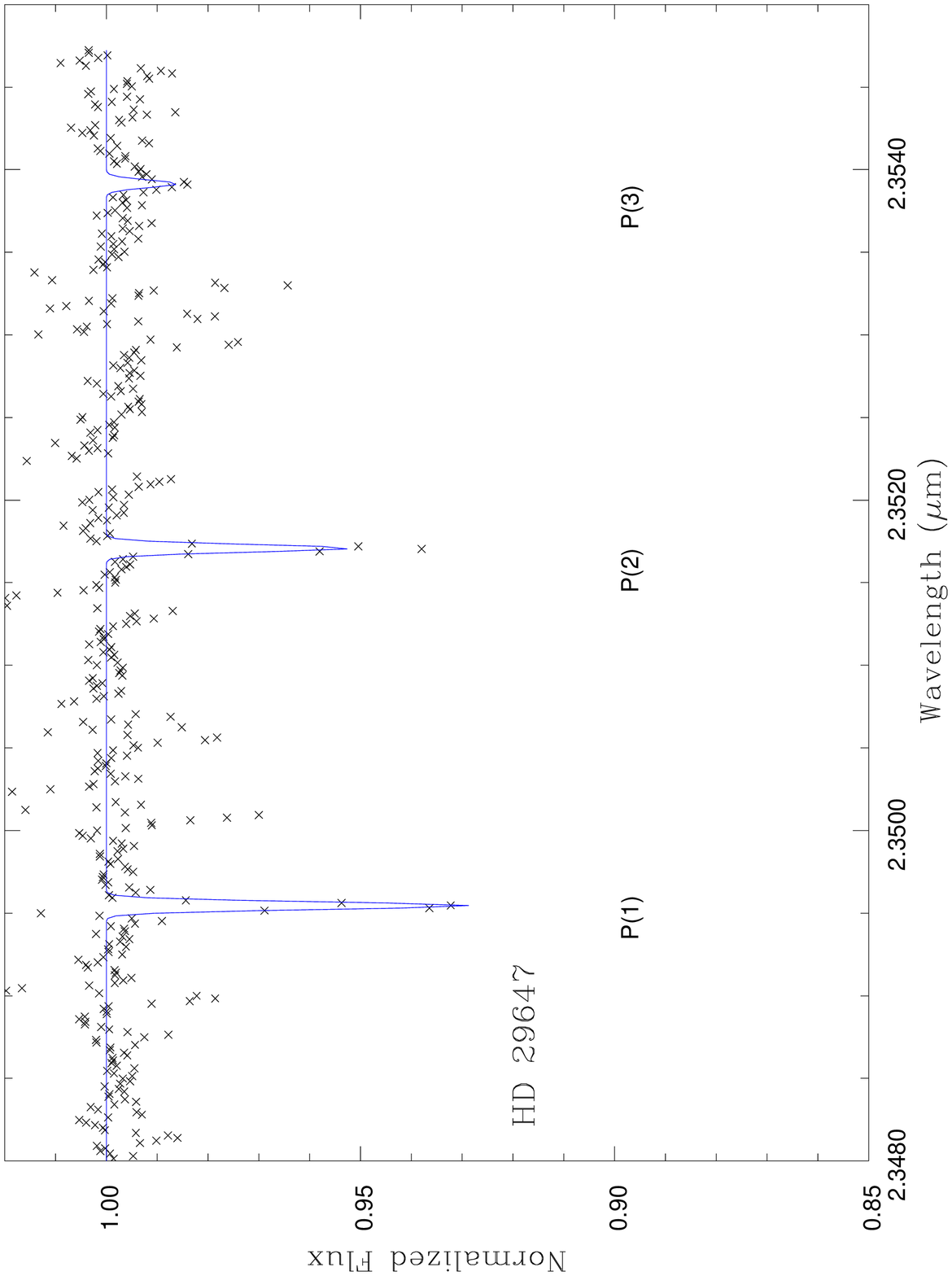}{0.36\textwidth}{(b)}}
\gridline{\rotatefig{270}{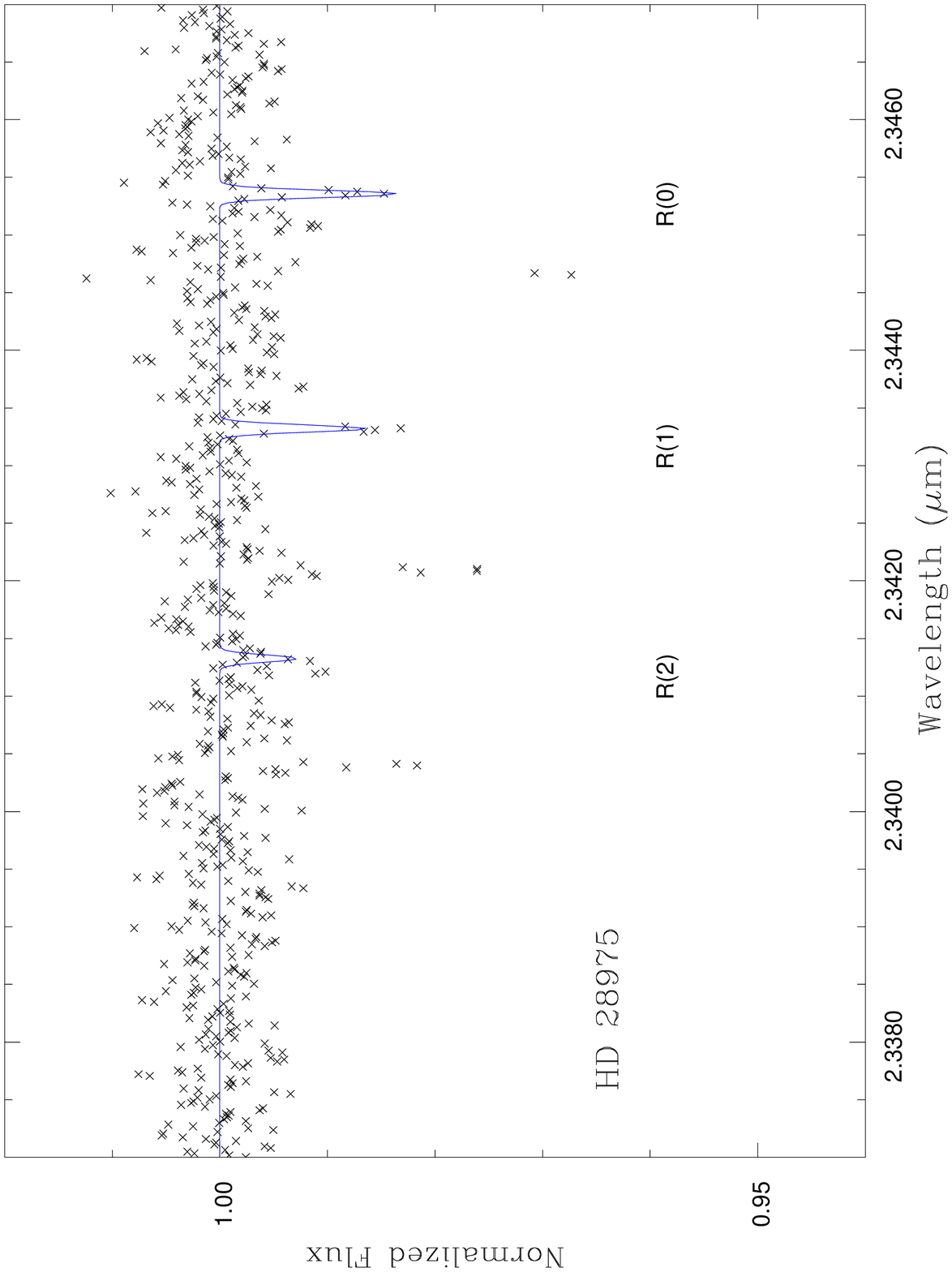}{0.36\textwidth}{(c)}}
\caption{IGRINS spectra revealing absorption from the (a) R- and (b) 
P-branches of interstellar CO toward HD~29647, as well as the R-branch toward 
HD~28975 (c).  Observed (x) and synthesized (blue line) spectra are presented.  
Details of the profile fitting are described in \textsection{3.1}.  Line 
identifications appear below the features.\label{fig.5}}
\end{figure}

\subsection{IR Photometry}

We use the mid- to far-IR images publicly available to describe the 
environment of each line of sight. Whenever possible, we use the images 
with the best angular resolution: Spitzer in the mid-IR 
\citep[3.6 to 24 microns from the Taurus Legacy Survey,][]{reb10}, 
far-IR \citep[160 microns,][]{fla09}, and Herschel in the far-IR 
\citep[70 to 500 microns from the Guaranteed Time Key 
Programme,][]{kir13}. For the line of sight toward HD~30122, which lies 
at the very edge of the Spitzer Taurus survey, we use the WISE all-sky 
images instead in the mid-IR \citep[3.4 to 22 microns,][]{wri10}. 
We also use the dust temperature map from \citet{fla09} derived from 
Spitzer 160 and IRIS 100 microns data \citep{miv05}. We search 
for previous mentions of the background stars in the literature to support our 
description.

\newpage
\section{Results and Analysis}

\subsection{Profile Syntheses}

We utilized modified versions of the FORTRAN program ISMOD developed 
by Y.~Sheffer \citep[e.g., ][]{she08} for our profile syntheses.  The program 
fits one or more Voigt components to an absorption profile and returns the 
column density $N$(X) for species X, $b$-value, and velocity $v_{\rm LSR}$ 
of each component after minimizing the root mean square in the residual 
spectrum.  Oscillator strengths ($f$-values) for atomic lines were taken from 
\citet{mor03}. For the CH and CH$^+$ transitions, we adopted $f$-values 
from \citet{gre93}.  For the $B$--$X$ and $A$--$X$ bands of $^{12}$CN 
and $^{13}$CN, we used oscillator strengths from \citet{bro14} and 
\citet{sne14}.  The adopted $f$-value for the NH transition came from 
\citet{fer18}.  As for the $A$--$X$ (2,~0) band of C$_2$, the necessary 
data were compiled by \citet{son07}, while for the (2,~0) rovibrational band 
of CO, they were from \citet{bla84}.  The large-scale calculations on 
the $f$-values for the C$_2$ $A$--$X$ (2,~0) band \citep{kok07, sch07} confirm 
the value adopted by \citet{son07}.

For many of the atomic and molecular transitions observed toward our targets, 
we fit each individual absorption profile separately.  For 
HD~29647, the column densities derived independently from the various CH and 
CH$^+$ transitions available from the UVES spectra allowed us to check the 
overall consistency in these column densities and compare them with those 
derived from the TS data.  However, for some species (e.g., CN, CO, and 
C$_2$) more sophisticated profile fitting procedures were required.  The 
R(0), R(1), and P(1) lines of the $B$--$X$ (0,~0) band of $^{12}$CN toward 
HD~28975 and HD~29647 are very strong and hence are likely severely 
affected by optical depth effects, which makes deriving accurate column densities 
difficult.  Since the UVES data toward HD~29647 provide access to both the 
(0,~0) and (1,~0) bands of the CN $B$--$X$ system, we take advantage of the 
factor-of-10 difference in $f$-value for the two bands and fit the R(0) lines of 
the bands simultaneously to derive a single value for the 
column density in the $N^{\prime\prime}$ = 0 level.  We then fit the R(1) 
and P(1) lines together to obtain a value for $N_{CN}$($N=1$).  We included 
fits to the corresponding $^{13}$CN features while synthesizing the R(0), R(1), 
and P(1) lines of the (0,~0) band.  The R(2) and P(2) lines are weak and were 
fitted independently.  The (1,~0) band was not covered by the TS 
observations (and the S/N is not nearly as good in the TS spectra as in 
the UVES data).  We therefore fit the R(0), R(1), and P(1) lines recorded by 
the TS observations of HD~28975 and HD~29647 independently, 
but supplemented these determinations with fits to the weak CN line at 
7871~\AA, which is part of the $A$--$X$ (2,~0) band and also probes the 
$N^{\prime\prime}$ = 0 level.  This line was available in one of the orders from 
the TS observations that provided the C$_2$ data.  The CN lines are significantly 
weaker toward HD~27778 and HD~30122.  Final CN column densities for these directions 
were derived from independent fits to the available lines.  Fits to the CN lines 
from the $B$--$X$ (0,~0) band toward all four targets appear in Figure 6. The 
component structures derived from these and other fits to the atomic and molecular 
lines observed in the four directions are given in Table 2, where 
$v_{\rm LSR}$, $N$(X), and $b$-value are listed for each component.

\begin{rotate}
\begin{deluxetable}{lccccccccccccccccccccccc}[tbh]
\tablecolumns{24}
\setlength{\tabcolsep}{0.02in}
\tablewidth{0pt}
\tabletypesize{\scriptsize}
\tablecaption{Component Structure}
\tablehead{
 \\
\colhead{Star} & \multicolumn{3}{c}{Ca~{\sc ii}} & \colhead{} & 
\multicolumn{3}{c}{K~{\sc i}} & \colhead{} & \multicolumn{3}{c}{Ca~{\sc i}} & 
\colhead{} & \multicolumn{3}{c}{CH$^+$} & \colhead{} & 
\multicolumn{3}{c}{CH} & \colhead{} & \multicolumn{3}{c}{CN} \\
\cline{2-4} \cline{6-8} \cline{10-12} \cline{14-16} \cline{18-20} \cline{22-24}
\colhead{} & \colhead{$v_{\rm LSR}$} & \colhead{$N/10^{11}$} & \colhead{$b$} & 
\colhead{} & \colhead{$v_{\rm LSR}$} & \colhead{$N/10^{11}$} & \colhead{$b$} & 
\colhead{} & \colhead{$v_{\rm LSR}$} & \colhead{$N/10^{10}$} & \colhead{$b$} & 
\colhead{} & \colhead{$v_{\rm LSR}$} & \colhead{$N/10^{12}$} & \colhead{$b$} & 
\colhead{} & \colhead{$v_{\rm LSR}$} & \colhead{$N/10^{12}$} & \colhead{$b$} & 
\colhead{} & \colhead{$v_{\rm LSR}$} & \colhead{$N/10^{12}$} & \colhead{$b$} \\
\colhead{} & \colhead{(km s$^{-1}$)} & \colhead{(cm$^{-2}$)} & 
\colhead{(km s$^{-1}$)} & \colhead{} & \colhead{(km s$^{-1}$)} & 
\colhead{(cm$^{-2}$)} & \colhead{(km s$^{-1}$)} & \colhead{} & 
\colhead{(km s$^{-1}$)} & \colhead{(cm$^{-2}$)} & \colhead{(km s$^{-1}$)} & 
\colhead{} & \colhead{(km s$^{-1}$)} & \colhead{(cm$^{-2}$)} & 
\colhead{(km s$^{-1}$)} & \colhead{} & \colhead{(km s$^{-1}$)} & 
\colhead{(cm$^{-2}$)} & \colhead{(km s$^{-1}$)} & \colhead{} & 
\colhead{(km s$^{-1}$)} & \colhead{(cm$^{-2}$)} & \colhead{(km s$^{-1}$)}}
\startdata
HD~27778 & $-$1.1 & 0.3 & 2.1 & & \ldots & \ldots & \ldots & & \ldots & \ldots & 
\ldots & & $-$1.0 & 0.6 & 0.5 & & $-$1.9 & 1.1 & 0.5 & & +0.0 & 0.5\tablenotemark{a} & 
1.6 \\
 & $+$3.1 & 2.6 & 1.5 & & $+$4.0 & 4.6 & 0.8 & & \ldots & \ldots & \ldots & & 
    $+$3.0 & 1.2 & 0.9 & & $+$4.5 & 19.1 & 1.3 & & $+$5.0 & 8.0 & 1.2 \\
  & $+$8.1 & 6.9 & 2.0 & & $+$7.5 & 3.8 & 0.9 & & $+$7.0 & 0.4 & 2.1 & & 
$+$8.4 & 5.6 & 1.3 & & $+$7.4 & 10.5 & 1.2 & & $+$6.5 & 5.9 & 0.9 \\
HD~28975 & $+$0.8 & 2.5 & 0.6 & & $+$1.3 & 1.3 & 0.7 & & \ldots & \ldots & 
\ldots & & \ldots & \ldots & \ldots & & \ldots & \ldots & \ldots & & \ldots & \ldots & 
\ldots \\
  & $+$4.3 & 10.0 & 1.0 & & $+$5.5 & 21.2 & 1.0 & & \ldots & \ldots & \ldots & & 
$+$4.5 & 9.9 & 2.6 & & $+$5.7 & 50.0 & 1.6 & & $+$6.1 & 57.8 & 1.4 \\
  & $+$8.3 & 39.4 & 1.3 & & $+$9.4 & 7.6 & 0.7 & & $+$7.1 & 3.3 & 0.6 & & 
$+$9.2 & 4.4 & 1.5 & & $+$9.3 & 7.7 & 1.1 & & $+$10.0 & 
1.7 & 1.6 \\
 & $+$11.8 & 3.7 & 0.9 & & \ldots & \ldots & \ldots & & \ldots & \ldots & \ldots & & 
\ldots & \ldots & \ldots & & \ldots & \ldots & \ldots & & \ldots & \ldots & \ldots \\
 & $+$14.9 & 1.6 & 1.0 & & \ldots & \ldots & \ldots & & \ldots & \ldots & \ldots & & 
\ldots & \ldots & \ldots & & \ldots & \ldots & \ldots & & \ldots & \ldots & \ldots \\
HD~29647 & \tablenotemark{b} & \tablenotemark{b} & \tablenotemark{b} & & 
$+$6.6 & 29.1 & 1.0 & & $+$6.2 & 1.0 & 1.0 & & 
$+$7.0 & 5.2 & 1.9 & & $+$6.0 & 58.7 & 1.3 & & $+$6.2 & 86.6 & 1.1 \\
HD~30122 & $-$4.6 & 1.3 & 1.4 & & \ldots & \ldots & \ldots & & \ldots & \ldots & 
\ldots & & \ldots & \ldots & \ldots & & \ldots & \ldots & \ldots & & \ldots & \ldots & 
\ldots \\
  & $+$0.8 & 2.4 & 1.8 & & \ldots & \ldots & \ldots & & \ldots & \ldots & \ldots & & 
\ldots & \ldots & \ldots & & \ldots & \ldots & \ldots & & \ldots & \ldots & 
\ldots \\
  & $+$3.9 & 1.9 & 1.3 & & \ldots & \ldots & \ldots & & \ldots & $<$ 0.2 & \ldots & & 
$+$5.2 & 2.1 & 2.0 & & \ldots & \ldots & \ldots & & \ldots & \ldots & \ldots \\
 & $+$6.9 & 2.4 & 1.5 & & $+$6.3 & 6.0 & 1.4 & & \ldots & $<$ 0.2 & \ldots & & 
\ldots & \ldots & \ldots & & $+$6.7 & 15.6 & 1.4 & & $+$7.0 & 1.6 & 1.3 \\
 & $+$10.3 & 1.0 & 1.3 & & \ldots & \ldots & \ldots & & \ldots & \ldots & \ldots & & 
\ldots & \ldots & \ldots & & \ldots & \ldots & \ldots & & \ldots & \ldots & \ldots \\
\enddata
\tablenotetext{a}{The column density for $N$ $=$ 0 in CN was multiplied by 1.4 to 
account for excited rotational levels populated by the cosmic background.}
\tablenotetext{b}{No attempt was made to fit the Ca~{\sc ii} line toward HD 29647 
since the line is predominantly stellar in nature.}
\end{deluxetable}
\end{rotate}

A different approach was taken when fitting the C$_2$ and CO lines toward 
HD~28975 and HD~29647.  It was a necessity for the CO lines because 
the data have coarser spectral resolution.  The initial guesses 
for component structure were based on 
the TS and UVES results for CN.  An additional constraint applied to the CO 
spectra was the rotational excitation temperature, $T_{0,J}$(CO).  Lines 
associated with the same $J^{\prime\prime}$ in the P-, Q-, and R-branches 
in C$_2$ and in the P- and R-branches in CO (for HD~29647) provided 
consistency checks during the fitting process.  The largest optical depth at line 
center approached 2 toward HD~29647.  Acceptable fits were not always 
possible, and we had to keep some of the parameters fixed.  For example, 
$T_{0,J}$(CO) was kept at 11.5 K toward HD~28975, with an allowed 
range of $\pm2.5$ K based on the root-mean-square outcomes.  The resulting 
fits are provided in Figures 4 and 5.  The total column densities for the two lines 
of sight appear in Table 3.

\begin{deluxetable}{lcc}[tbh]
\tablecaption{Column Densities for C$_2$ and CO Rotational Levels}
\setlength{\tabcolsep}{0.5in}
\tablehead{
 \\
\colhead{Line} & \colhead{HD~28975} & \colhead{HD~29647}
}
\startdata
\multicolumn{3}{c}{C$_2$ ($10^{12}$ cm$^{-2}$)} \\
 & \tablenotemark{a} &  \tablenotemark{b}$^,$\tablenotemark{c} \\ \hline
$N_{C_2}$(0) & 4.9(2.0)\tablenotemark{d} & 20.8(0.6)/19.7(0.4) \\
$N_{C_2}$(2) & 26.4(2.1) & 38.9(0.6)/38.5(0.4) \\
$N_{C_2}$(4) & 18.7(2.0) & 18.0(0.6)/18.8(0.4) \\
$N_{C_2}$(6) & 15.0(2.5) & 10.3(0.7)/7.8(0.4) \\
$N_{C_2}$(8) & 7.8(2.6) & 5.8(0.7)/5.0(0.5) \\
$N_{C_2}$(10) & $\ldots$ & $\ldots$/3.0(0.5) \\
$N_{tot}$(C$_2$) & 72.8 & 93.8/92.8 \\ \hline
\multicolumn{3}{c}{CO ($10^{16}$ cm$^{-2}$)} \\
 & \tablenotemark{e} & \tablenotemark{f} \\ \hline
$N_{CO}$(0) & 3.6(1.0) & 25.4(1.7) \\
$N_{CO}$(1) & 6.7(1.8) & 42.6(1.7) \\
$N_{CO}$(2) & 4.2(1.2) & 22.2(1.4) \\
$N_{CO}$(3) & $\ldots$ & 5.4(1.2) \\
$N_{CO}$(4) & $\ldots$ & $\le~4.2$\tablenotemark{g} \\
$N_{tot}$(CO) & 14.5 & 95.6 \\
\enddata
\tablenotetext{a}{Line parameters: $v_{\rm LSR}$ are $+6.1$ and $+9.7$ km 
s$^{-1}$; $b$ are 1.5 and 1.6 km s$^{-1}$, with relative fractions of 0.85 
and 0.15.}
\tablenotetext{b}{The first entry provides TS results, and the second UVES.}
\tablenotetext{c}{Line parameters: TS -- $v_{\rm LSR}$ is $+5.8$ km s$^{-1}$ 
and $b$ is 0.9 km s$^{-1}$; UVES -- $v_{\rm LSR}$ is $+5.6$ km s$^{-1}$ 
and $b$ is 0.9 km s$^{-1}$.}
\tablenotetext{d}{The uncertainties are given in parentheses.}
\tablenotetext{e}{Line parameters: $v_{\rm LSR}$ are $+6.1$ and $+9.7$ km 
s$^{-1}$; $b$ are 1.2 and 1.6 km s$^{-1}$, with relative fractions of 0.85 
and 0.15.}
\tablenotetext{f}{Line parameters: $v_{\rm LSR}$ is $+5.4$ km s$^{-1}$ 
and $b$ is 0.5 km s$^{-1}$.}
\tablenotetext{g}{A meaningful 2-$\sigma$ upper limit was possible in this case.}
\end{deluxetable}

\subsection{Rotational Excitation}

Absorption from more than one rotational level in C$_2$, CO, and CN was 
detected, allowing us to infer the physical conditions (gas density, gas 
temperature, and strength of the interstellar radiation field) from the amount of 
excitation above that due to the cosmic microwave background 
(CMB) at a temperature of $T_{\rm CMB}$ $=$ 2.725 K \citep{fix09}.  
According to \citet{pan05}, the three molecules occupy similar 
regions along the line of sight to the background star.  Because the processes 
leading to excitation differ, the results from C$_2$, CO, and CN are 
complementary.  In particular, C$_2$ excitation by atomic and molecular 
hydrogen and optical pumping \citep{van82} from low lying levels provides 
information on gas temperature and from levels greater than $J$ $=$ 4 on the 
ratio of gas density to strength of the infrared radiation field.  Because 
C$_2$ is a homonuclear molecule, rotational transitions with $\Delta J$ 
equal 1 are forbidden, unlike the cases for CO and CN.  The IR 
radiation field is assumed to have a strength comparable to that of the 
average interstellar radiation field, unless evidence suggests otherwise.  
Analysis of CO excitation \citep[e.g.,][]{gol13} yields an estimate for the 
density of H$_2$, provided that the gas temperature is available from H$_2$ 
or C$_2$ measurements.  In diffuse molecular gas, with an ionization 
fraction $x$(e) $\approx10^{-4}$, the dominant collision partner for 
CN excitation is electrons \citep{bla91}.  If the ionization fraction is 
known, the resulting electron density can be converted to a total  
proton density, $n_{tot}$(H) = $n$(H) + 2 $n$(H$_2$).  In molecular 
clouds, where the ionization fraction is very low (i.e., $\lesssim10^{-5}$) 
and gas densities are high ($\gtrsim$~10$^4$ cm$^{-3}$), collisions 
involving molecular hydrogen and atomic helium begin to become important.  
We consider this possibility when attempting to merge the results from our 
analyses from a diffuse cloud perspective with the results of radio emission 
from CN toward HD~29647 \citep{cru85}.  Although it is common to 
infer the density of collision partners, $n$(H) + $n$(H$_2$), in analyses of 
rotational excitation, we determine total proton densities for ease of 
comparison with analyses based on electron density or chemical 
considerations.  For the molecule-rich diffuse material in our work, we assume 
that $n$(H) = $n$(H$_2$).  Throughout the paper, the subscript {\it tot} 
refers to {\it total} proton gas and hydrogen column density.

\subsubsection{C$_2$}

For C$_2$ excitation, the discussion in \citet{hup12} forms the basis for 
our analysis.  In particular, collisional cross sections 
\citep{lav91, rob92, naj08, naj09} and the $f$-value for the $A$--$X$ (2,~0) 
band \citep{erm95, kok07, sch07} are used.  The line-of-sight column densities 
and their uncertainties for the gas toward HD~28975 and HD~29647 appear in 
Table 3.  For HD~27778, the results of \citet{son07} are adopted.  Results 
for C$_2$ toward HD~30122 are not available.  We obtain a gas temperature, 
$T$(C$_2$), and total proton density, $n_{tot}^{ex}$(C$_2$), of 10 K and 
greater than $\sim350$ cm$^{-3}$ toward HD~29647.  A remarkable outcome of 
this analysis is that only a value of 10 K for 
$T$(C$_2$) is possible although temperatures up to 100 K were considered.  
For the gas toward HD~28975 and HD~27778, values of 
$T$(C$_2$) and $n_{tot}^{ex}$(C$_2$) are inferred to be 30 K/250 cm$^{-3}$ and 
50 K/150 cm$^{-3}$, respectively.  For comparison when adopting the same 
cross section and $f$-value, the values for the gas toward HD~27778 from 
\citet{son07} are $50~\pm~10$ K and $140~\pm~20$ cm$^{-3}$.  As discussed 
below, there is evidence for a weaker interstellar radiation field penetrating 
the TMC; adopting a value for $I_{IR}$ (the strength of the adopted IR 
radiation field relative to the interstellar value) of 0.5 in this analysis 
leads to a halving of $n_{tot}^{ex}$(C$_2$).

\newpage
\subsubsection{CO}

The effort by \citet{gol13}, where the cross sections from \citet{yan10} are 
used, is the basis for our analysis of CO excitation.  The analysis, however, 
uses a finer grid of gas temperatures (20, 40, 60, 80, 100 K) and adopts 
values from observations of H$_2$ or C$_2$ instead of generic ones.  Once 
the gas temperature is specified, the CO excitation temperature determines 
the density of collision partners.  As noted in \citet{gol13}, collisions with 
H$_2$ dominate over those involving atomic hydrogen.  We multiply 
this density, $n$(H$_2$), by a factor to convert it to total proton density 
from CO excitation, $n_{tot}^{ex}$(CO).  The adopted factor depends on properties 
revealed by our analyses.  For diffuse molecular material like that toward 
HD~27778 and HD~30122, a factor of 3 is adopted, since $n$(H) and 
$n$(H$_2$) are comparable.  As discussed in $\S{4.1.2}$, the sight line 
toward HD~29647 is dominated by molecular gas, while that toward 
HD~28975 seems to represent an intermediate case.  As a result, the 
conversion factors used are 2 and 2.5, respectively.  The excitation 
temperatures needed for the analysis are $T_{10}$(CO) and when 
available $T_{21}$(CO) and $T_{32}$(CO).  Since the fitting of the 
CO profiles provides a different set of excitation temperatures, 
$T_{01}$(CO), $T_{02}$(CO), and $T_{03}$(CO), as was done in 
\citet{she08} for instance, we first transformed the later set into the 
former ones through the use of equations (2) to (4) from \citet{gol13}.

All four sight lines have information on CO excitation.  For HD~29647 
and HD~28975, the IGRINS results presented here are adopted, while for 
HD~27778 and HD~30122, the HST results from \citet{she08} are 
used.  Our fitting of the CO spectrum toward HD~29647 yielded the same 
values for $T_{10}$(CO), $T_{21}$(CO), and $T_{32}$(CO) of 
9.5 K considering the uncertainty in each determination of 1.0 K.  
\citet{cru85} obtains CO excitation temperatures of 9.2 K ($+5.1$ km 
s$^{-1}$) and 7.5 K ($+6.5$ km s$^{-1}$), where the LSR velocity of the 
component is in parentheses, from his radio observations.  For 
$T_{10}$(CO) of 9.5 K and $T_k$ $<$~20 K from $T$(C$_2$), 
$n_{tot}^{ex}$(CO) of about 1800 cm$^{-3}$ is inferred.  Even higher 
densities are suggested from $T_{21}$(CO) and $T_{32}$(CO).  A comparable 
total proton density is obtained for the gas toward HD~28975 when 
considering $T_{10}$(CO) $=$ 11.5 K and $T$(C$_2$) $=$ 30 K.  
The analysis of CO toward HD~27778 and HD~30122 by \citet{she08} 
yielded respective values for $T_{10}$(CO) of 5.3 and 3.8 K.  They also 
derived gas temperatures from H$_2$ of 51 and 61 K, from which 
we find values for $n_{tot}^{ex}$(CO) of 975 and 450 cm$^{-3}$.

The radiation temperature, $T_R$, obtained from the microwave data 
of \citet{cru85}, as well as from the measurements by \citet{hey87}, 
\citet{van91}, and \citet{lis08}, can be used 
to check for consistency in the 
physical conditions for the direction toward HD~29647.  \citet{cru85} 
used the 36-foot NRAO telescope at Kitt Peak for the $J$ $=$ 
$1\rightarrow0$ line and the 4.9 m University of Texas telescope at the 
Millimeter Wave Observatory for the $J$ $=$ $2\rightarrow1$ transition.  
The respective beam sizes and spectral resolutions were 
60$^{\prime\prime}$/0.52 km s$^{-1}$ and 
65$^{\prime\prime}$/0.16 km s$^{-1}$.  Emission was seen at $+5.1$ 
and $+6.5$ km s$^{-1}$ with line widths of about 1.0 km s$^{-1}$ in 
both lines.  The pairs representing these radial velocities 
for $T_R$ were 5.8/4.2 K and 3.6/2.0 K for 
$J$ $=$ $1\rightarrow0$ and $J$ $=$ $2\rightarrow1$, respectively.  The 
14 m Five College Radio Observatory telescope was used by \citet{hey87} 
for their measurements of $J$ $=$ $1\rightarrow0$ emission.  These 
observations had a beam size of 45$^{\prime\prime}$, a spectral resolution 
of 0.52 km s$^{-1}$, and a spacing of 2$^{\prime}$.  The coarse spacing 
revealed a single component with $T_R^*$ of 5.9 K at $+5.9$ km 
s$^{-1}$ with a line width of 2.8 km s$^{-1}$.  These results are similar 
to a single component at $+5.7$ km s$^{-1}$ with emission 
weighted by $T_R$ found by Crutcher.  It is worth noting that our 
measurements of CH, C$_2$, CN, and CO show absorption at about 
$+6.0$ km s$^{-1}$.  The study by \citet{van91} was based on data 
obtained with the 15 m Swedish-ESO Submillimeter Telescope for the 
$J$ $=$ $1\rightarrow0$ lines and the Caltech Submillimeter Observatory 
for the $J$ $=$ $2\rightarrow1$ and $J$ $=$ $3\rightarrow2$ lines.  The 
respective beam sizes and spectral resolutions were 
44$^{\prime\prime}$/0.13 km s$^{-1}$, 
32$^{\prime\prime}$/0.043 km s$^{-1}$, and 
20$^{\prime\prime}$/0.043 km s$^{-1}$, making these the observations 
with the highest spatial and spectral resolutions.  \citet{van91} provided 
beam efficiencies so that we could convert their tabulated values for $T_A^*$ 
into $T_R^*$ for comparison with the earlier studies.  Emission was seen at 
$+5.2$ and $+7.1$ km s$^{-1}$ and had typical line widths of about 2.0 
km s$^{-1}$.  The intensities described by $T_R^*$ were somewhat lower 
than the earlier measurements and those of \citet{lis08}; they found 4.1 and 
2.9 K for $J$ $=$ $1\rightarrow0$, 2.2 and 1.6 K for $J$ $=$ 
$2\rightarrow1$, and 2.2 and 1.1 K for $J$ $=$ $3\rightarrow2$.  \citet{lis08} 
used the 12 m Arizona Radio Observatory for measurements of CO 
$J$ $=$ $1\rightarrow0$ emission; the beam size and spectral resolution 
were 65$^{\prime\prime}$ and 0.13 km s$^{-1}$.  The emission, occurring 
at an average velocity of $+$6.14 km s$^{-1}$ with $T_R^*$ of 5.84 K, had 
a peak near $+$5.0 km s$^{-1}$ and a shoulder centered around $+$7.0 km 
s$^{-1}$; the accompanying spectrum of $^{13}$CO emission clearly 
reveals the two components.

The excitation calculations for CO emission were carried out using the RADEX 
code \citep{van07}.  As with the analysis noted above, the rate coefficients 
for collisions with H$_2$ were from \citet{yan10}.  We assumed equal 
abundances of ortho- and para-H$_2$, but the dependence on the spin state of 
H$_2$ is small for diffuse cloud temperatures, typically $\pm~<10$\% from the 
average at 10 K, and much less at higher temperatures.  For the 
$J$ $=$ $1\rightarrow0$ CO transition, the rate coefficients for para-H$_2$ 
collisions on CO at 10(100) K are 
$3.3\times10^{-11}$($3.5\times10^{-11}$) cm$^3$ s$^{-1}$, 
and those for ortho-H$_2$ on CO are 
$3.8\times10^{-11}$($3.5\times10^{-11}$) cm$^3$ s$^{-1}$.  An expanding 
spherical cloud with $v$ proportional to radius or a plane-parallel slab are 
invoked to handle radiative transfer.  The models 
are based on the large velocity gradient approximation.  Optical depths 
approach values of 100 for $J$ $=$ $1\rightarrow0$ and 
$J$ $=$ $2\rightarrow1$ when considering column densities of 10$^{18}$ 
cm$^{-2}$, as seen toward HD~29647.

We considered cases with $T_k$ equal to 10 and 20 K and line widths of  2.0 
km s$^{-1}$ and sought agreement with the excitation temperatures found in 
our CO measurements and with the values of $T_R$ from CO emission 
\citep{cru85, hey87, van91}.  Separate models were run with densities differing 
by 0.25 in the log.  These densities of $n$(H$_2$) were multiplied by 2 to yield 
total proton densities, as noted above for HD~29647.  
For a given density, the results from the 
plane-parallel slab yielded slighted higher excitation and radiation temperatures.  
The calculations with $T_k$ of 20 K produced excitation temperatures larger 
than observed, for values of $n_{tot}^{ex}$(CO) greater than 200 cm$^{-3}$ as 
found in our other analyses.  Moreover, the fact that $T_k$ needs to be 10 K 
confirms our results from C$_2$ excitation.  Total proton densities of about 350 
to 600 cm$^{-3}$ best match the observations, within a factor of 2 or so of our 
other determinations.

We also performed calculations to predict line intensities for CO emission 
toward HD~28975.  We considered $T_k$ of 20 and 30 K from our C$_2$ 
analysis and line widths of 2.0 and 3.0 km s$^{-1}$, seeking agreement with the 
excitation temperatures found from the analysis of the IR spectra.  A value of 
$1.78\times10^{17}$ cm$^{-2}$ was adopted for $N_{tot}$(CO), close to 
the value found from IGRINS spectra.  The most consistent results using RADEX 
were $n_{tot}^{ex}$(CO) between 180 and 320 cm$^{-3}$ with values of $T_R$ about 
10(7) K for the $J$ $=$ $1\rightarrow0$($2\rightarrow1$) line.  The plane-parallel 
model produced satisfactory results for $T_k$ of 20 K.  We multiply 
$n$(H$_2$) by 2.5 to infer a total proton density of about 450 to 800 cm$^{-3}$ 
for $n_{tot}^{ex}$(CO), as suggested by the discussion in $\S{4.1.2}$.  It is not 
surprising that the radiation temperatures are higher in this case in light of the 
higher kinetic and excitation temperatures but comparable total density.  This 
range in $n_{tot}^{ex}$(CO) is consistent with the chemical results discussed 
below and within a factor of 3 or so of those from C$_2$ excitation.

\subsubsection{CN}

The degree of excitation observed in CN in diffuse molecular gas differs 
from the situation for C$_2$ and CO, the other molecules examined here.  
When local sources of excitation are present, the CN excitation temperature 
that corresponds to transitions between the $N$ $=$ 0 and $N$ $=$ 1 levels, 
$T_{01}$(CN), may be somewhat higher than the temperature of the CMB, 
and only measurements with high signal to noise are able to discern the effect 
of local sources.  
The difficulty arises because CN has a much larger dipole moment than CO; 
the homonuclear C$_2$ molecule has no dipole moment.  Excitation of the 
$N$ $=$ 2 level from $T_{12}$(CN) is even harder to detect in diffuse gas.  
For typical values of the ionization fraction, electron impacts will always 
dominate over collisions with neutral species in populating the upper rotational 
levels of CN.  The results on column density for rotational levels in CN that are 
used for the analysis of its excitation appear in Table 4; fits to the profiles 
are provided in Figure 6.

\begin{deluxetable}{lcccccc}[tbh]
\tablecolumns{7}
\tablewidth{0pt}
\tabletypesize{\small}
\tablecaption{CN Rotational Column Densities and Excitation Temperatures}
\tablehead{\colhead{Star} &  \colhead{$v_{\rm LSR}$} & \colhead{$N$ $=$ 0} & 
\colhead{$N$ $=$ 1} & \colhead{$N$ $=$ 2} & \colhead{$T_{01}$(CN)} & 
\colhead{$T_{12}$(CN)} \\
\colhead{} & \colhead{(km s$^{-1}$)} & \colhead{(10$^{12}$ cm$^{-2}$)} & 
\colhead{(10$^{12}$ cm$^{-2}$)} & \colhead{(10$^{12}$ cm$^{-2}$)} & 
\colhead{(K)} & \colhead{(K)}}
\startdata
HD~27778\tablenotemark{a} & $+$5.6 & $9.46\pm0.47$ & $4.44 \pm 0.08$ & $\ldots$ & 
$2.934 \pm 0.084$ & $\ldots$ \\
HD~28975 & $+$6.1 & $42.02 \pm 4.08$ & $15.78 \pm 0.69$ & $\ldots$ &
$2.617\pm0.134$ & $\ldots$ \\
HD~29647\tablenotemark{b} & $+$6.2 & $60.36 \pm 3.76$ & $26.25 \pm 1.61$ & 
$\ldots$ & $2.816 \pm 0.127$ & $\ldots$ \\
 & $+$6.2 & $66.14 \pm 3.89$ & $26.90 \pm 1.25$ & $0.98 \pm 0.08$ & 
$2.722 \pm 0.102$ & $2.847 \pm 0.068$ \\
HD~30122 & $+$7.0 & $1.06 \pm 0.11$ & $0.48 \pm 0.15$ & $\ldots$ & 
$2.940 \pm 0.448$ \\
\enddata
\tablenotetext{a}{Results are based on the sum of the two main CN components; 
velocity listed is the column-density weighted mean velocity.}
\tablenotetext{b}{First line gives results from TS data; second line lists values 
derived from UVES spectra.}
\end{deluxetable}

\begin{figure}[tbh]
\centering
\includegraphics[width=6in]{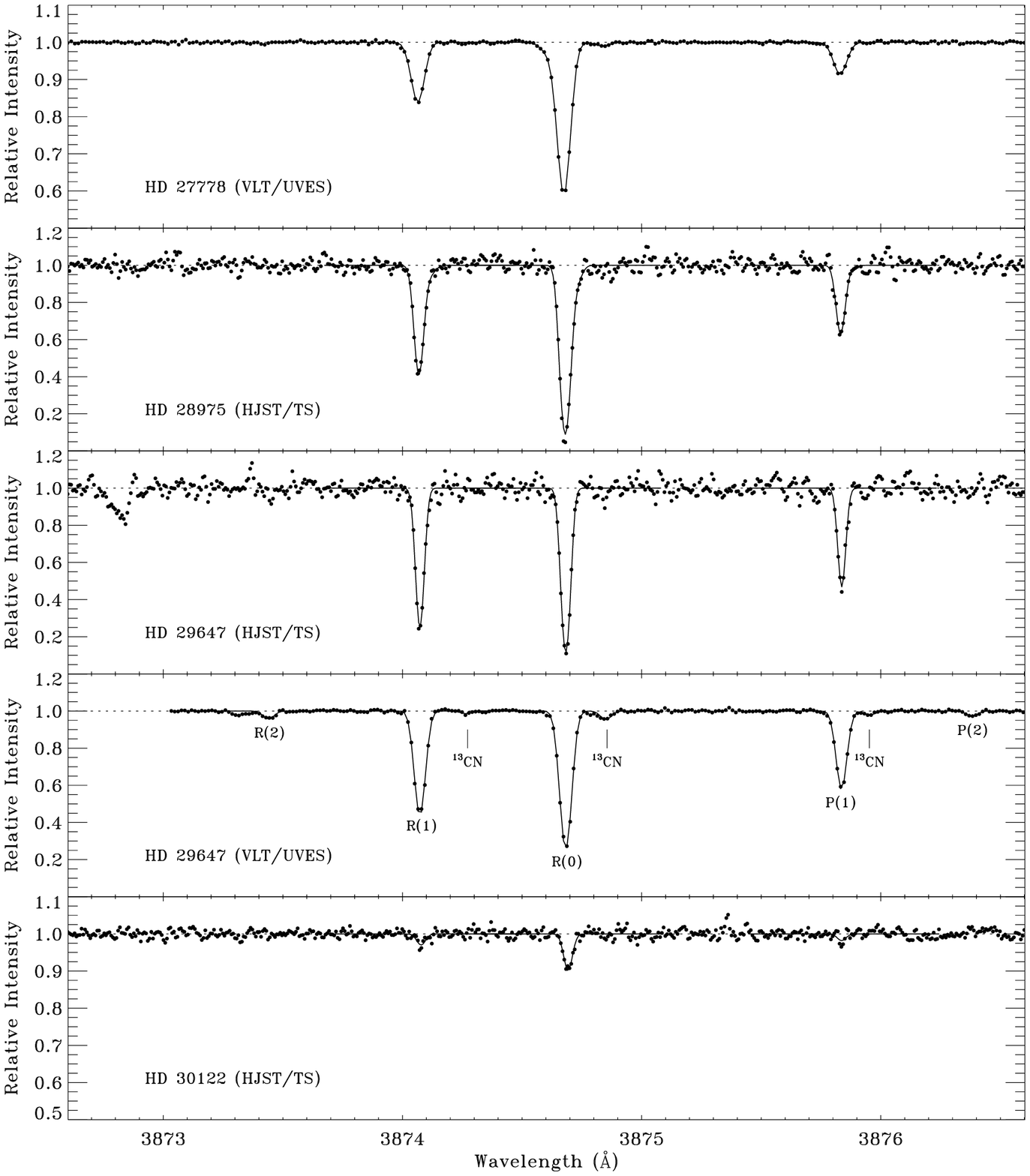}
\caption{Profile synthesis fits to the $B$--$X$ (0,~0) band of CN toward HD~27778 
(from UVES data), HD~28975 and HD~29647 (from TS data), HD~29647 (from UVES spectra), 
and HD~30122 (from TS data).  The difference in resolution and S/N between the 
TS and UVES spectra is evident in the figure.  In the panel containing the UVES 
spectrum of HD~29647, we include labels for the individual $^{12}$CN lines, and tick 
marks give the expected positions of the corresponding $^{13}$CN features.  Solid 
curves show the synthetic spectra obtained through profile fitting.}
\end{figure}

In our analysis of CN excitation toward stars in the Taurus region, we consider 
local excitation by collisions with electrons with rate coefficients from 
\citet{har13} and, for completeness, the effects of collisions from ortho and para H$_2$ 
molecules \citep{kal15} and neutral He atoms \citep{liq10}.  The analysis 
yields the electron density that corresponds to the measured excitation 
temperature in cases where $T_{01}$(CN) is statistically greater 
than $T_{\rm CMB}$.  From our McDonald observations of CN absorption 
toward HD~29647, we find an excitation temperature of 
$T_{01}$(CN) $=$ $2.816~\pm~0.127$ K.  If the kinetic temperature of the 
gas in this direction $T_k$ equals $T$(C$_2$) and $T$(CO) of 
$\approx10$ K, as indicated by the analyses of C$_2$ and CO excitation, 
then the small excess in the CN rotational temperature would indicate that 
$n$(e) is about 0.19 cm$^{-3}$.  For $n_{tot}$(H) of about 1500 
cm$^{-3}$, as suggested by the analyses of CO excitation above and CN 
chemistry below, the electron fraction would be $x$(e) $\sim$ $1.3\times10^{-4}$, 
similar to the value expected for low density diffuse gas.  
These calculations assume that the H$_2$ ortho-to-para ratio is that determined 
by the kinetic temperature, which for the gas toward HD~29647 is assumed to be 
10 K. However, even if the ortho-to-para ratio were larger than this, due to 
turbulent dissipation, for example, the results on $n$(e) and $x$(e) would not 
be significantly different. Our analysis of the UVES data on 
CN toward HD~29647 yielded 
$T_{01}$(CN) of $2.722~\pm~0.102$ K, which is consistent with no additional 
excitation over that from the CMB.  However, at the 2$\sigma$ level, this 
measurement is consistent with $T_{01}$(CN) $<$ 2.926 K, which would 
indicate that $n$(e) $<$ 0.45 cm$^{-3}$, in agreement with the McDonald 
result.  The weighted average of the two measurements yields $2.759~\pm~0.080$ 
K for a conservative 2$\sigma$ limit of 2.919 K and a limit on $x$(e) of 
$2.9\times10^{-4}$.  Similarly, the CN excitation temperature 
that we find toward HD~28975, $T_{01}$(CN) $=$ $2.617~\pm~0.134$ 
K is consistent with $T_{\rm CMB}$.  At the 2$\sigma$ level, the upper limit of 
$T_{01}$(CN) is 2.885 K for this sight line and yields $n$(e) less than 0.25 
cm$^{-3}$ for $T_k$ of 30 K from C$_2$ excitation.  The values of 
$n_{tot}$(H) inferred from CO excitation (1900 cm$^{-3}$) and CN chemistry 
(1200 cm$^{-3}$) suggest that $x$(e) $<$ $2.1\times10^{-4}$.

We can perform the same analyses on the UVES data for HD~27778 shown in 
Table 4 and compare them to the results by \citet{rot95}.  We obtained 
column densities for the $N$ $=$ 0 and 1 levels of $(9.46\pm0.47)\times10^{12}$ 
and $(4.44\pm0.08)\times10^{12}$ cm$^{-2}$, while \citet{rot95} found 
values of $(10.87^{+0.74}_{-0.69})\times10^{12}$ and 
$(4.57^{+0.083}_{-0.086})\times10^{12}$ cm$^{-2}$, respectively.  The two sets of 
results agree within about 1.2$\sigma$ considering the mutual uncertainties.  The 
corresponding values for $T_{01}$(CN) are $2.934\pm0.084$ K (present) and 
$2.747^{+0.096}_{-0.101}$ K (Roth \& Meyer) are consistent at the combined 
1.0$\sigma$ level; our determination and their 2$\sigma$ limit are indistinguishable.  
Our CN excitation temperature suggests an electron density of 0.34 cm$^{-3}$ 
when adopting a value for $T_k$ of 50 K from H$_2$ and C$_2$ observations.  
With $n_{tot}$(H) of about 700 cm$^{-3}$ from our analyses of CO 
excitation and CN chemistry, an ionization fraction of 4.9$\times10^{-4}$ is 
inferred.  The weighted average of the two excitation temperatures is 
$2.856~\pm~0.064$ K, indicating values for $n$(e) and $x$(e) of 0.21 
cm$^{-3}$ and 3.0$\times10^{-4}$.

These values for $x$(e) can be compared with the results for observed C$^+$ 
abundances in diffuse clouds of about $1.4\times10^{-4}$ \citep{sof97} and the 
upper limit for the gas toward HD~27778 of $1.1\times10^{-4}$ \citep{sof04}; 
C$^+$ is expected to be the dominant source of electrons in this material.  
Further discussion with respect to the carbon budget appears in $\S{4.1.2}$.  The 
results in Table 4 for HD~30122, where the CN column density is much less 
but the relative uncertainties in $N_{CN}$(0) and $N_{CN}$(1) are greater, yield 
an excitation temperature that is not distinguishable from $T_{\rm CMB}$.

\citet{cru85} also detected the ($N$,~$J$,~$F$) $=$ 
(1,~3/2,~5/2)--(0,~1/2,~3/2) and (1,~3/2,~3/2)--(0,~1/2,~1/2) lines of CN 
toward HD~29647 with the NRAO 36-foot telescope with a beam width of about 
$1^{\prime}$.  Each line has a velocity component at $v_{LSR}$ of $+5.1$ 
km s$^{-1}$ and the stronger $F$ $=$ 5/2--3/2 transition shows weak emission 
extending to about $+7$ km s$^{-1}$.  Thus, the component structure is similar to 
the $^{12}$CO emission lines.  According to Crutcher, the two CN lines have $T_R$ 
of 0.059 and 0.042 K and line widths of 1.0 km s$^{-1}$.  We also modeled the 
emission from these lines with RADEX, using the collisional data from \citet{liq10}.  
These results for He were scaled by 1.37 to approximate collisions with 
H$_2$.  Like the models of $^{12}$CO emission discussed above, we considered gas 
temperatures of 10 and 20 K, where the scaling is most likely to apply.  Because 
CN has a larger dipole moment than CO and as a consequence a larger critical 
density, we only chose a line width of 1 km s$^{-1}$, 
consistent with the observations for CN \citep{cru85} and CO 
\citep{van91}. In light of the RADEX results for CO, we only considered 
geometries representing an expanding sphere.  The models 
most consistent with the measurements by \citet{cru85} suggest H$_2$ densities 
of 1000 cm$^{-3}$ for 10 K and about 600 cm$^{-3}$ for 20 K, corresponding to 
total proton densities of 2000 and 1200 cm$^{-3}$, respectively.  We favor the 
10 K results because this temperature was required to reproduce the results for 
C$_2$ and CO excitation above.  This leads to comparable total proton 
densities for CO and CN excitation.  For completeness, we note that the optical 
depths for the two CN emission lines are about 5 and 10 (i.e., weaker versus 
stronger emission) and that $T_{ex}$(CN) is about 2.8 K, 
like we found from the observations at visible wavelengths.

\subsection{$^{13}$C$^{16}$O and $^{13}$C$^{14}$N}

Carbon monoxide is the second most abundant molecule after H$_2$.  As a 
result, it is the molecular species that contributes the greatest amount to the carbon 
budget.  This in turn affects the relative abundances of carbon isotopologues among 
molecular species.  For example, if the $^{12}$C$^{16}$O/$^{13}$C$^{16}$O 
ratio is greater than the ambient $^{12}$C/$^{13}$C ratio, the atomic carbon 
reservoir is depleted in $^{12}$C (and vice versa).  Thus, carbon-bearing molecules 
synthesized from the depleted reservoir, such as CN, are expected to show an 
enhancement in $^{13}$C (and vice versa).  \citet{rit11} provided evidence for 
this inverse relationship for CO and CN in diffuse molecular clouds relative to an 
ambient $^{12}$C/$^{13}$C ratio of about 70 extracted from their sample.  Two 
sight lines in the present study, those toward HD~27778 and HD~29647, allow us 
to extend this analysis.

The relative abundances of CO isotopologues are affected by two photochemical 
processes.  (In what follows, when no isotope is given, the most abundant isotope is 
assumed, $^{12}$C, $^{14}$N, or $^{16}$O.)  Isotope charge exchange, 
$^{13}$C$^+$ $+$ $^{12}$CO $\rightarrow$ $^{12}$C$^+$ $+$ $^{13}$CO, 
is favored over the reverse process because $^{13}$CO has a lower zero-point 
energy equivalent to 35 K in temperature units \citep{wat76}.  On the other hand, 
isotope selective photodissociation \citep[e.g.,][]{bal82,chu83,van88,vis09} 
favors the more abundant isotopic variant, $^{12}$CO.  For carbon monoxide, 
photodissociation is a line process and the more abundant variant has lines that are 
more optically thick, thereby shielding the CO from further photodissociation.  The 
relative mix of the two processes determines the amount of chemical fractionation 
present.

The data on column densities for CO and CN isotopologues come from \citet{she07} 
for CO toward HD~27778 and from the present study for CN.  For HD~27778, the 
results are $67~\pm~10$ for $^{12}$CO/$^{13}$CO and $63~\pm~25$ for 
$^{12}$CN/$^{13}$CN.  These isotopologue ratios are indistinguishable from the 
ambient value of 70, suggesting that no fractionation among carbon isotopes is present 
along this sight line.  However, in a plot of $N$($^{12}$CO)/$N$($^{13}$CO) vs. 
log[$N$($^{12}$CO)], the data for HD~27778 lie in a region where self shielding may 
be occurring in both isotopologues \citep{ric18}, returning the ratio to the 
ambient value for the atomic reservoir.

For the gas toward HD~29647, information on the $^{12}$CO/$^{13}$CO and the 
$^{12}$CN/$^{13}$CN ratios comes from our spectra.  For $^{13}$CO, we can set 
an upper limit of $1.06\times10^{17}$ cm$^{-2}$ for a ratio greater than 10.  The 
UVES spectrum provides a $^{12}$CN/$^{13}$CN ratio of $109.5~\pm~9.6$.  While a 
factor of 5 or so improvement in the IR data for the CO band is required for a 
definitive conclusion, it appears the $^{13}$C reservoir in CO may be enhanced toward 
HD~29647.  Because the $^{12}$CO column is so large, the self-shielding factor 
\citep[see][]{van88,vis09} is much smaller than 1, the value appropriate for unshielded 
gas (see Fig. 13 in \citet{she08} for an illustration of how the factor varies).  When 
combined with the large amount of dust attenuation, one would not expect isotope 
selective photodissociation to be operating along this direction.  That leaves isotope 
charge exchange, which can be described \citep{lam94} by 

\begin{equation}
(^{12}{\rm CO}/^{13}{\rm CO}) = (^{12}{\rm C}/^{13}{\rm C}) 
{\rm exp}(-35/T_{k}).
\end{equation}

\noindent For ($^{12}$C/$^{13}$C) of 70 and $T_k$ of 10 K, we obtain 2, much 
smaller than our lower limit.  A likely solution to this apparent inconsistency is that 
isotope charge exchange is not operating either, because the C$^+$ abundance is very 
low, a consequence of the limited ionization possible with so much grain attenuation.  
Since there must be some C$^+$ along the line of sight to explain the presence of 
observable amounts of CH$^+$ (though small), as well as CH and C$_2$ (see next 
Section), C$^+$ must occupy a limited region where sufficient UV radiation 
penetrates.  These points are considered further in the Discussion.

The integrated intensities for CO and $^{13}$CO emission toward HD~27778 and 
HD~29647 are available \citep{van91, lis08}.  Both studies provide results for the 
$J$ $=$ $1\rightarrow0$ transition.  \citet{van91} give ratios for each velocity 
component, but we consider the sum for comparison with \citet{lis08}.  For HD~27778, 
\citet{lis08} obtained an intensity ratio of 16.7, and both studies 
reveal a ratio of 2.8 toward HD~29647.  The focus of these efforts 
was not on estimates for the ambient $^{12}$CO/$^{13}$CO ratios, and so we can 
only discuss the results in general terms.  It is not surprising that the ratio is 
much larger toward HD~27778 because $N$(CO) is a factor of 80 smaller.  However, 
it is not clear whether the ratio of 2.8 toward HD~29647 is mainly a consequence of 
large optical depths or severe fractionation.  At the present time, the IGRINS data 
only yield an $N$($^{12}$CO)/$N$($^{13}$CO) ratio greater than 10.

\subsection{CN Chemistry}

The chemical network involving CH~$\rightarrow$~C$_2$~$\rightarrow$~CN 
and NH~$\rightarrow$~CN is described in \citet{fed94} 
and updated in subsequent papers \citep[e.g.,][]{pan01}.  
There are two updates to the chemical model used in the current version.  First, many 
rate coefficients have a measured temperature dependence, along with theoretical 
confirmation \citep[see][]{mce13}.  Second, new photodissociation rates for CN are 
available \citep{elq13,hea17}; we adopted the rate of $5.2\times10^{-10}$ s$^{-1}$ 
from \citet{hea17}, which is about a factor of 2 larger than the rate in earlier 
versions.  While we previously modeled the molecular material toward the four sight 
lines discussed in the present paper \citep{fed94,she08}, revisions to $E$($B-V$) 
(see Table 1), CH, C$_2$, and CN column densities (see Tables 2 and 3), and 
$T_k$ (see \S{3.2.1)} were made.  The value for 
$N_{tot}$(C$_2$) toward HD~27778 comes 
from \citet{son07} and \citet{hup12}.  Finally, our values for $N$(NH) from UVES 
spectra are utilized: $2.0\times10^{12}$/$0.7\times10^{12}$ cm$^{-2}$ for the 
components at $+$5.0/$+$6.5 km s$^{-1}$ toward HD~27778; and $8.0\times10^{12}$ 
cm$^{-2}$ toward HD29647.  Changes to $E$($B-V$) lead to new values for the 
parameter $\tau_{UV}$, the optical depth resulting from grain attenuation at UV 
wavelengths.  A particularly important change is the extinction curve for HD~27778 
\citep{fit07} with its rise at short wavelengths.

\begin{rotate}
\begin{deluxetable}{lccccccccccc}[tbh]
\tablecolumns{12}
\tablewidth{0pt}
\tabletypesize{\small}
\tablecaption{Chemical Results}
\tablehead{\colhead{Star} &  \colhead{$v_{\rm LSR}$} & \colhead{$\tau_{UV}$} & 
\colhead{$I_{UV}$} & \colhead{$T_k$} & \colhead{$N_o$(CH)} & 
\colhead{$N_o$(NH)} & \colhead{$N_o$(C$_2$)} & \colhead{$N_p$(C$_2$)} & 
\colhead{$N_o$(CN)} & \colhead{$N_p$(CN)} & \colhead{$n_{tot}$(Chem)} \\
\colhead{} & \colhead{(km s$^{-1}$)} & \colhead{} & \colhead{} & \colhead{(K)} & 
\colhead{(10$^{12}$ cm$^{-2}$)} & \colhead{(10$^{12}$ cm$^{-2}$)} & 
\colhead{(10$^{12}$ cm$^{-2}$)} & \colhead{(10$^{12}$ cm$^{-2}$)} & 
\colhead{(10$^{12}$ cm$^{-2}$)} & \colhead{(10$^{12}$ cm$^{-2}$)} & 
\colhead{(cm$^{-3}$)}}
\startdata
HD~27778 & $+$0.0 & 2.97 & 0.5 & 65 & 1.1 & $\ldots$ & $\ldots$ & 1.6 & 0.5 & 0.5 & 
675 \\
 & $+$0.0 & 2.97 & 1.0 & 65 & 1.1 & $\ldots$ & $\ldots$ & 1.7 & 0.5 & 0.5 & 
$\sim$1400 \\
 & $+$5.0 & 2.97 & 0.5 & 50 & 19.1 & 2.0 & 16.1 & 20.8 & 8.0 & 4.4 & 
300 \\
 & $+$5.0 & 2.97 & 1.0 & 50 & 19.1 & 2.0 & 16.1 & 20.2 & 8.0 & 4.3 & 
575 \\
 & $+$6.5 & 2.97 & 0.5 & 50 & 10.5 & 0.7 & 11.9 & 15.3 & 5.9 & 3.2 & 500 \\
 & $+$6.5 & 2.97 & 1.0 & 50 & 10.5 & 0.7 & 11.9 & 15.3 & 5.9 & 3.2 & 1000 \\
HD~28975 & $+$6.1 & 3.72 & 0.5 & 30 & 50.0 & $\ldots$ & 61.9 & 79.0 & 57.8 & 
36.6 & 475 \\
 & $+$6.1 & 3.72 & 1.0 & 30 & 50.0 & $\ldots$ & 61.9 & 84.4 & 57.8 & 
30.1 & 1200 \\
 & $+$10.0 & 3.72 & 0.5 & 30 & 7.7 & $\ldots$ & 10.9 & 7.0 & 1.7 & 1.8 & 150 \\
 & $+$10.0 & 3.72 & 1.0 & 30 & 7.7 & $\ldots$ & 10.9 & 7.4 & 1.7 & 2.0 & 325 \\
 & $+$10.0 & 3.72 & 0.5 & 50 & 7.7 & $\ldots$ & 10.9 & 6.1 & 1.7 & 1.9 & 
 175\\
 & $+$10.0 & 3.72 & 1.0 & 50 & 7.7 & $\ldots$ & 10.9 & 6.3 & 1.7 & 2.0 & 375 \\
HD~29647 & $+$6.2 & 6.76 & 0.5 & 10 & 58.7 & 8.0 & 93.8 & 88.3 & 86.6 & 
96.4 & $\sim$1000 \\
 & $+$6.2 & 6.76 & 1.0 & 10 & 58.7 & 8.0 & 93.8 & 78.4 & 86.6 & 
91.5 & $\sim$1000 \\
HD~30122 & $+$7.0 & 1.43 & 0.5 & 65 & 15.6 & $\ldots$ & $\ldots$ & 15.4 & 
1.6 & 1.7 & 275 \\
 & $+$7.0 & 1.43 & 1.0 & 65 & 15.6 & $\ldots$ & $\ldots$ & 14.8 & 
1.6 & 1.6 & 525 \\
\enddata
\end{deluxetable}
\end{rotate}

We sought factor-of-two agreement between the observed and predicted column 
densities for C$_2$ and CN, and we suggest a similar level of precision for the gas 
densities.  Our latest results appear in Table 5, where the star, the velocity 
component, $\tau_{UV}$, $I_{UV}$ (the relative enhancement 
of the UV radiation field over the average interstellar value), $T_k$, 
observed (o) and predicted (p) molecular column densities, and the $n_{tot}$(H) 
derived from the chemical model, $n_{tot}$(Chem), are listed.  There are 
multiple components with measures of $N$(CN) toward HD~27778 and HD~28975, and 
calculations were performed for each.  Since the velocity spread is small, we used the 
value for $\tau_{UV}$ for each component.  Because one molecular component 
toward HD~28975 has much smaller column densities, we considered two gas 
temperatures, 30 and 50 K, for it.  For the gas toward HD~27778, we previously 
adopted $I_{UV}$ $=$ 0.5 in light of the high Galactic latitude for the TMC 
\citep{fed94,she08}.  In their studies of the TMC, both \citet{fla09} and \citet{pin10} 
suggest that the interstellar radiation field is about 
50\% of the average value throughout
the cloud.  Making this change with the current version of the model for all four lines 
sight results in a gas density a factor of about 2 smaller toward HD~27778, HD~28975, 
and HD~30122, indicating that the C$_2$ and CN column densities are proportional to 
$n_{tot}$(Chem)/$I_{UV}$.  In other words, molecular destruction in the gas toward these 
stars mainly occurs through photodissociation.  
The difference is not necessarily 2 because 
any change in $N_p$(C$_2$) is propagated through the CN rate equations.  On the other 
hand, $\tau_{UV}$ is so large toward HD~29647 that photodissociation plays a limited 
role.  Then since collisional terms dominate both production and destruction, the density 
terms in the numerator and denominator of the steady-state rate equations cancel.  The 
results for this sight line in Table 5 come from the middle of the acceptable 
values.  In general, the predictions with a reduced $I_{UV}$ are in better agreement 
with observed column densities.  The 
results for $n_{tot}$(Chem) toward HD~30122 are lower than those in \citet{she08} 
because the CN photodissociation rate is larger and $\tau_{UV}$ is smaller.  
Overall, the values for $n_{tot}$(Chem) from the present study -- 100 to 1000 
cm$^{-3}$ -- are typical of CN and CO rich gas.  Since the column 
densities are so different for the two components toward HD~28975, the gas density 
of 475 cm$^{-3}$ for the one with larger $N$(CN) applies to the line-of-sight 
average as well.  For a similar reason, the result for the $+$0.0 km s$^{-1}$ component 
toward HD~27778 does not affect the average density.  
The results for gas density from our analyses of excitation 
and chemistry for each direction are discussed further in $\S{4.3}$.

\subsection{Dust Temperature}

\subsubsection{HD~27778 and HD~30122}

\begin{figure}
\includegraphics[width=\textwidth]{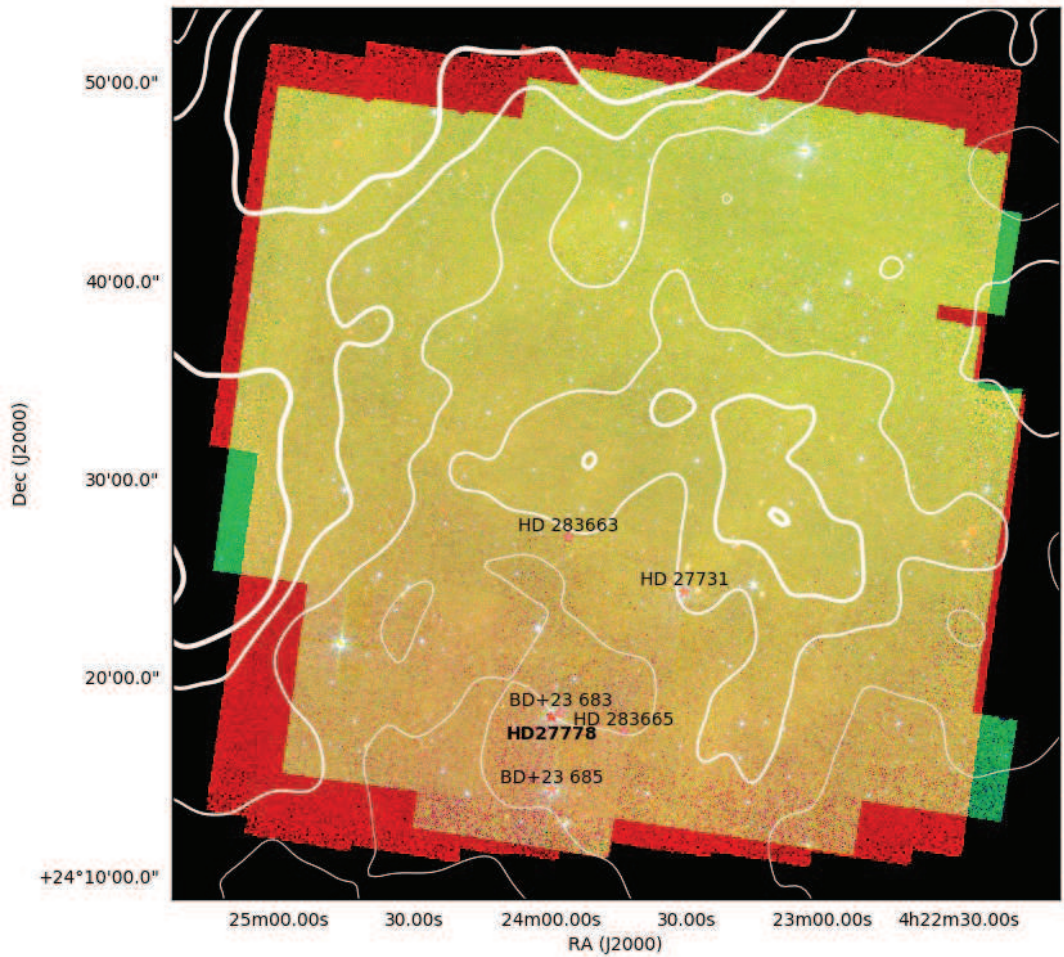}
\caption{Three color image (blue is 4.5~microns, green is 8.0~microns, 
and red is 24~microns from Spitzer) 
of the surroundings of HD~27778.  The overlaid contours are from MIPS~160~microns, 
smoothed by a 3-pixel wide Gaussian.  Contours are increasing in thickness from 40 to 
55 MJy sr$^{-1}$ in steps of 5 MJy sr$^{-1}$.  The location of HD~27778 is shown by 
a red star.  Other stars in the vicinity of HD~27778 are also shown.}
\label{fig:HD27778_IR}
\end{figure}

\begin{figure}
\includegraphics[width=\textwidth]{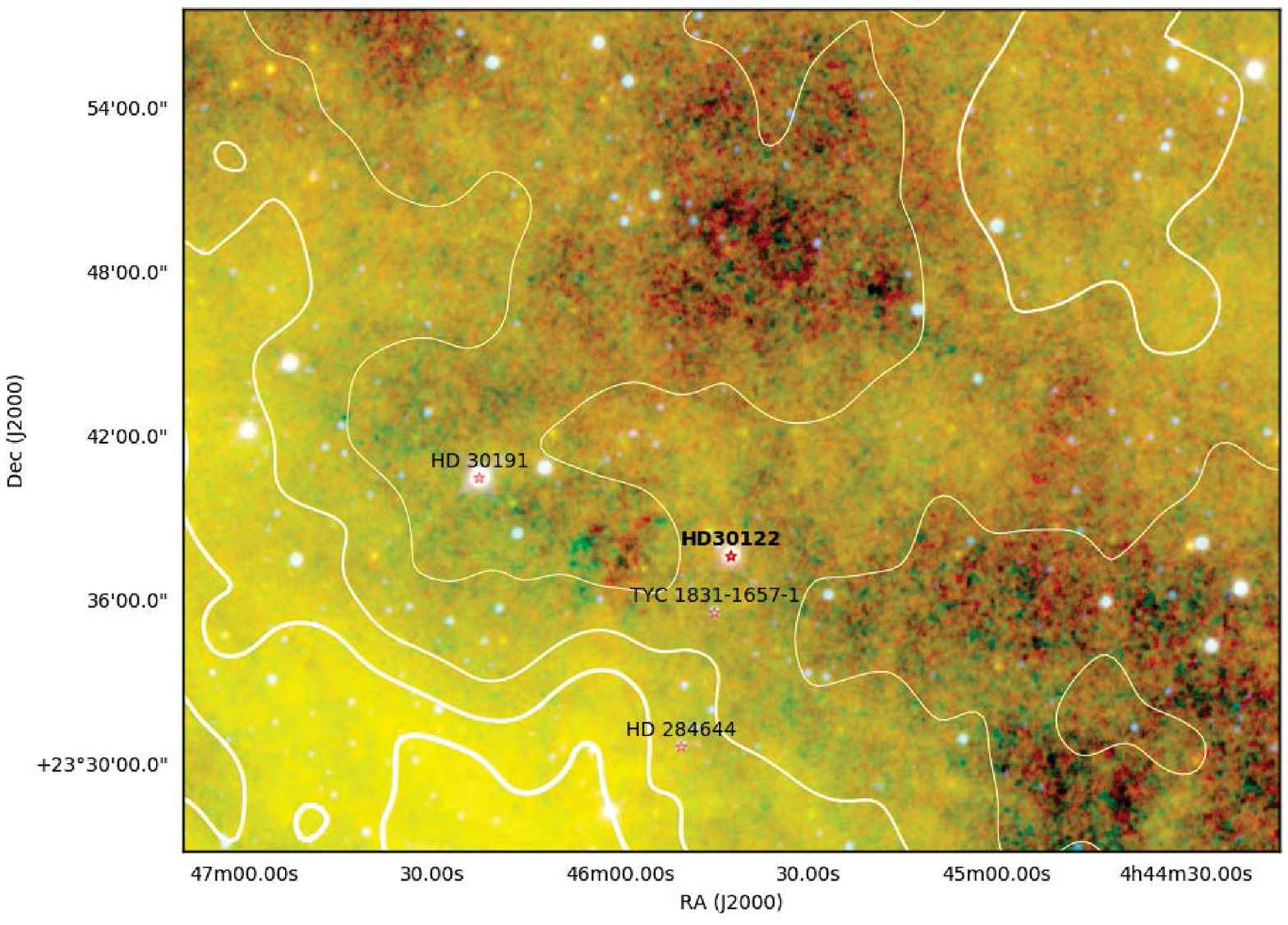}
\caption{Three color image (blue is 4.6~microns, green is 12~microns, 
and red is 22~microns from WISE) of 
the surroundings of HD~30122.  The overlaid contours are from MIPS 160 microns, 
smoothed by a 3-pixel wide Gaussian.  Contours are increasing in thickness from 50 to 65 
MJy sr$^{-1}$ in steps of 5 MJy sr$^{-1}$.  The location of HD~30122 is shown by a 
red star. Other stars in the vicinity of HD~30122 are also shown.}
\label{fig:HD30122_IR}
\end{figure}

The environments of HD~27778 and HD~30122 seem to be dominated by the diffuse 
interstellar medium.  No bright filament is visible in the far-IR, and the mid-IR does 
not reveal any dark cloud either.  The dust temperature map is quite homogeneous in the 
area surrounding both stars with values between 14.5 and 15 K.  The dust temperature 
maps from \citet{fla09} indicate values of 14.7 K (averaged within a 3 arcmin radius 
region centered on HD~27778) and 14.9 K (averaged within a 3 arcmin radius region 
centered on HD~30122).  Figures \ref{fig:HD27778_IR} and \ref{fig:HD30122_IR} 
show the mid-IR and far-IR emission in the environment of the two stars.  The distances 
to HD~27778 and HD~30122 are $224\pm2$ pc and $256\pm4$ pc, respectively, 
according to the $Gaia$ DR2 (see Table 1), locating them both well beyond the TMC, 
which is about 150 pc away as discussed in more detail below.

\subsubsection{HD~28975}

\begin{figure}
\includegraphics[width=\textwidth]{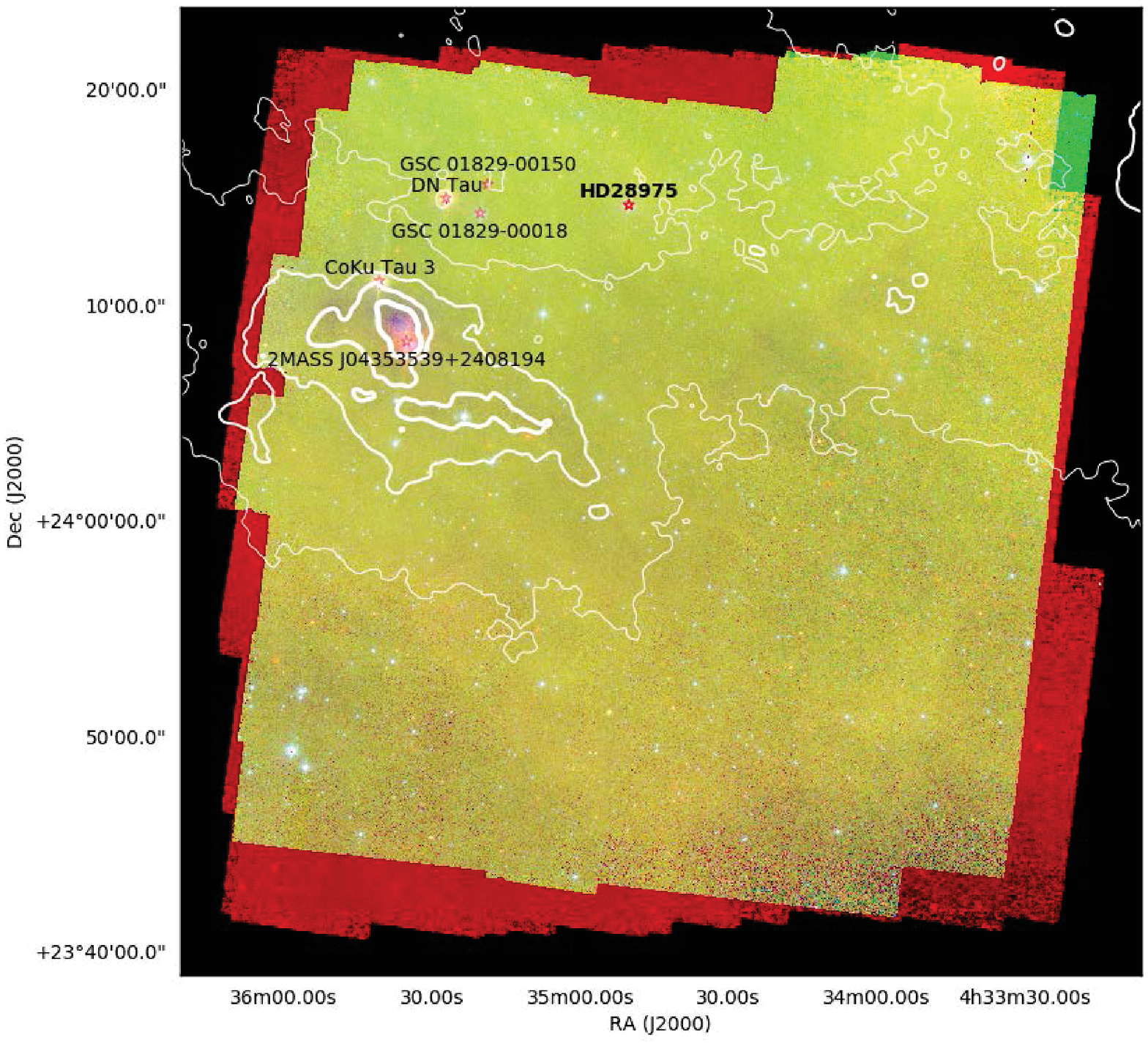}
\caption{Three color image (blue is 4.5~microns, green is 8.0~microns, and red is 
24~microns from Spitzer) 
of the surroundings of HD~28975.  The overlaid contours are from PACS 160 microns, 
smoothed by a 3-pixel wide Gaussian.  Contours are increasing in thickness from 40 to 160 
MJy sr$^{-1}$ in steps of 40 MJy sr$^{-1}$. The location of HD~28975 is shown by a red 
star.  Other stars in the vicinity of HD~28975 are also shown.}
\label{fig:HD28975_IR}
\end{figure}

The environment of HD~28795 is dominated by the presence of a dark cloud about 10 
arcmin to the SE.  A filamentary structure visible in the far-IR emission extends from the 
dark cloud south of HD~28975, though the emission levels remain low directly toward the 
star (see Figure \ref{fig:HD28975_IR}).  The presence of a nearby dark cloud might be 
the cause of the high $T_{ex}$(CO) found from IGRINS spectra.  The dust 
temperature averaged within a 3 arcmin radius region centered on HD~28975 is 14.0 K, 
which indicates that the line of sight might be dominated by somewhat colder gas 
than toward HD~27778 and HD~30122.  The distance to HD~28795 is $194\pm2$ pc according 
to the $Gaia$ DR2 (Table 1), indicating this star is also well beyond the TMC.

\subsubsection{HD~29647}

The environment of HD~29647 is the most intriguing of all background stars in this paper. 
This star is located within the Heiles Cloud-2 (HCL-2) 
and was previously associated with an IRAS source 
(IRAS 04380$+$2553) by \citet{whi04}.  Its distance was initially measured by $Hipparcos$ 
at about 180 pc, which would place the star a few tens of parsecs behind the TMC.  
\citet{whi04} used extinction curves toward this star and other nearby stars as well as IRAS 
images to conclude that (1) HD~29647 is ``embedded in or in very close proximity to dust that 
is being warmed, at least in part, by radiation from the star itself'' 
and (2) it is located behind TMC-1 and half-way within a diffuse screen 
that contributes about 3.6 mag of visual extinction.  
At higher angular resolution, an infrared nebula centered on HD~29647 and 5 arcminutes in 
radius, is clearly detected in the mid- to far-IR (3.6 to 160 microns).  The nebula is 
IRAS 04380$+$2553, and it is not detected at shorter wavelengths because visual extinction 
along the line of sight is so large.

In the mid-IR, two dark filaments are located to the ENE and S of HD~29647.  In the far-IR 
(160 microns and beyond), these filaments appear in emission, as they trace cold and large dust 
grains.  The dust temperature map of \citet{fla09} tells a similar story with cold filaments 
surrounding a ``warm spot'' at the location of HD~29647.  The temperature they derive toward 
the star is 15.4 K (averaged within a 3 arcmin radius region centered on the peak in dust 
temperature, 42 arcsec away from HD 29647), while it is about 13.5 K in the filaments and 
about 14.5 K in the surrounding diffuse medium.

\begin{figure}
\includegraphics[width=\textwidth]{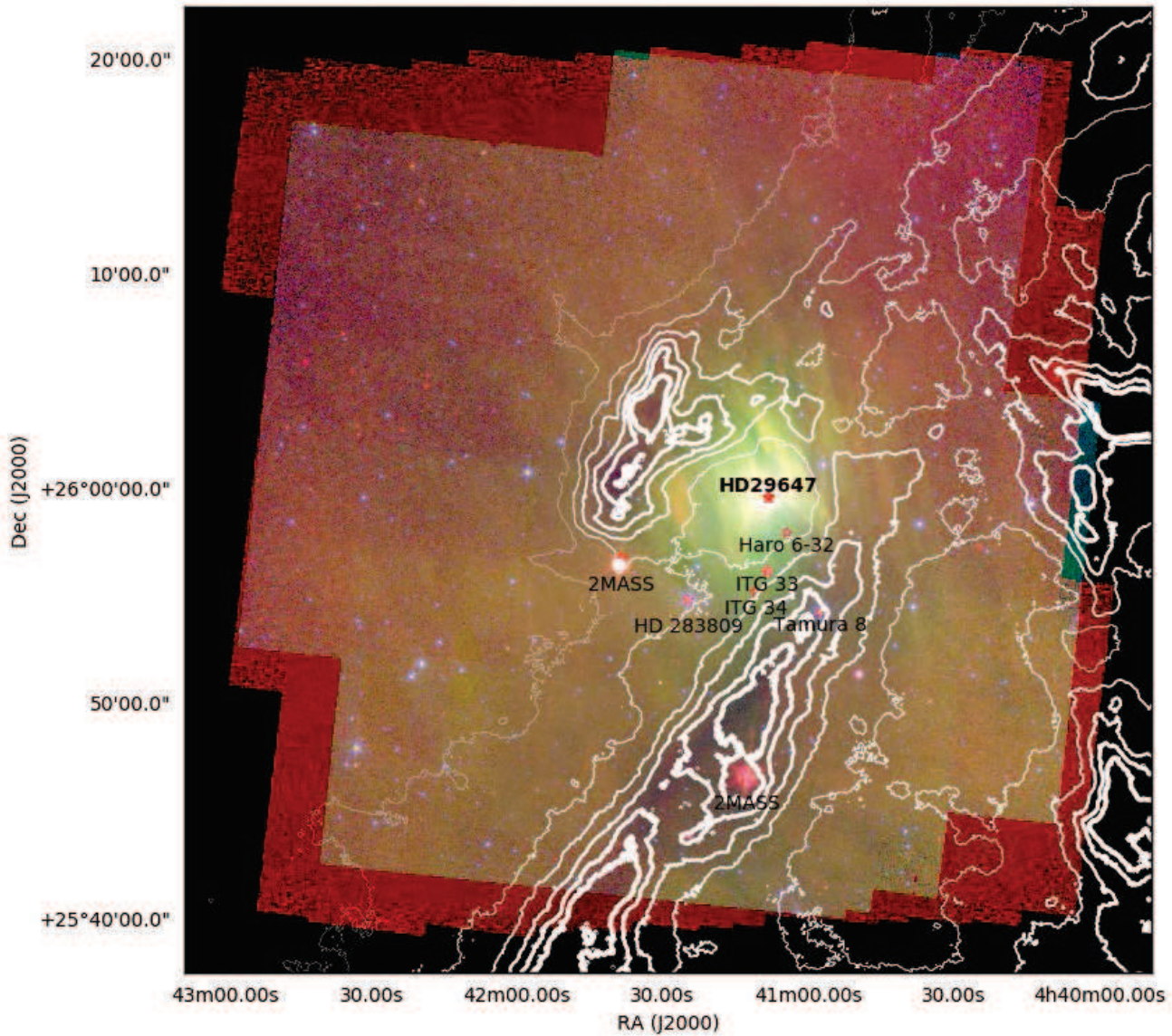}
\caption{Three color image (blue is 4.5~microns, green is 8.0~microns, 
and red is 24~microns from Spitzer) of the surroundings of HD~29647.  
The overlaid contours are from SPIRE 250 microns. Contours are increasing in thickness 
from 10 to 160 MJy sr$^{-1}$ in steps of 30 MJy sr$^{-1}$. The 
location of HD~29647 is shown by a red star, as well as for other objects, including stars in 
the survey by \citet{lac17}.}
\label{fig:HD29647_IR}
\end{figure}

Figure \ref{fig:HD29647_IR} shows a three color mid-IR image of the HD~29647 surroundings 
(4.5, 8.0, and 24~microns) with far-IR contours (250~microns).  The infrared nebula seems to 
be located exactly in a ``hole'' delineated by the filamentary structure observed in the far-IR 
which, at least visually, could be an indication that the two are related, though a fortuitous 
alignment cannot be ruled out.  We also note that the distance to HD~29647 has been revised 
by $Gaia$ DR2 measurements and now is $155\pm2$ pc, which would put the star somewhat 
closer to TMC-1.

We estimate the distance between HD~29647 and the IR nebula, adopting the method used by 
\citet{fed91} based on the prescription of \citet{dra79}.  In particular, the distance $r$ to a 
point source whose luminosity is $L$ and with dust temperature $T_d$ of the cloud is given 
by 

\begin{equation}
r = (LQ_{\star}/16{\pi}{\sigma}T_d^4Q_{ir})^{1/2},
\end{equation}

\noindent where $Q_{\star}$ is the effective absorption efficiency for the source spectrum, 
$Q_{ir}$ is the Planck-averaged emissivity at $T_d$, and $\sigma$ is the Stefan-Boltzmann 
constant.  We determined $Q_{\star}$ from optical properties for silicates \citep{dra85} and 
$Q_{ir}$ from \citet{dra84}.  The grain radius was taken to be 0.1 $\mu$m.  Use of a 
spectral type of B9III~Hg-Mn \citep{moo13} and $T_{\star}$ of 10,550 K \citep{cox00} 
resulted in a value of 0.330 for $Q_{\star}$.  Selecting a $T_{\star}$ of 11,500 K 
\citep{moo13} with or without an estimate for the effect of line blanketing in this metal-rich 
star led to $Q_{\star}$ values differing by only 15\%.  The adopted values for $R_{\star}$ and 
$Q_{ir}$ were $3.9\times10^{11}$ cm (5.6$R_{\sun}$) \citep{cox00} and 
$2.51\times10^{-4}$.  With $T_d$ of 15.4 K, the estimate for the distance between 
HD~29647 and the TMC is about 1 pc.

The final step in this analysis is to determine the likelihood that HD~29647 lies in the 
diffuse gas on the far side of the TMC, as suggested by \citet{whi04}.  We infer the 
extent of the diffuse gas surrounding the TMC by deriving the distance between 
HD~30122 and the nearby dark cloud, L1538, based on the map (Fig. 4) of 
\citet{gal19}.  The Galactic coordinates [$l$, $b$] for HD~30122 and the cloud are 
[$176.62^{\circ}$,~$-14.03^{\circ}$] and [$175.50^{\circ}$,~$-13.05^{\circ}$], 
yielding 1.5$^{\circ}$ on the sky or 4 pc for a distance of 150 pc.  If the total extent of 
the diffuse envelope surrounding the TMC is about twice this distance (10 pc), a gas 
density of about 100 cm$^{-3}$ is obtained when the value of $N_{tot}$(H) from 
$N$(K~{\small I}) in the direction of HD~29647 is adopted (see $\S{4.1.1}$).  
Such a value is very typical of diffuse gas.  Therefore, it appears that HD~29647 
lies within the diffuse material behind the TMC.

Three groups used $Gaia$ DR2 parallaxes to study the distance to the Taurus 
star-forming region and associated molecular material.  \citet{yan19} traced the cloud in 
Planck 857-GHz emission \citep{pla14}, deriving a distance of $145^{+12}_{-16}$ pc, 
though they see evidence for two components.  Building on the work of \citet{sch14}, 
\citet{zuc19} used $Gaia$ DR2 results when available and allowed $R_V$ to vary.  
Their average distance was $141{\pm2}{\pm7}$ pc, where the first number gives the 
statistical uncertainty and the second the systematic uncertainty.  \citet{zuc19} also 
subdivided the clouds in their sample, finding a bimodel distribution for Taurus -- 
132 and 152 pc.  Using the interactive software given by 
\citet{zuc19}\footnote[10]{\url{https://dataverse.harvard.edu/dataverse/cloud_distances}}, 
we obtain distances to the clouds in front of 
our stars of ${\sim}162^{+5}_{-4}$ pc (HD~27778), $160^{+6}_{-8}$ pc 
(HD~28975), $123^{+11}_{-8}$ pc (HD~29647), and ${\sim}162^{+5}_{-4}$ pc 
(HD~30122).  The estimates for HD~27778 and HD~30122 lie somewhat beyond the region 
considered in this study.  Last, \citet{gal19} used $Gaia$ DR2 and VLBI techniques to 
obtain distances to Young Stellar Objects (YSOs) in Taurus.  They divided the TMC 
into clusters for different dark clouds and analyzed the results through Bayesian 
statistics.  For HD~29647 (represented by clusters called HCL-2), distances of 
140.2(139.9) pc, both with uncertainties of 1.3 pc were inferred.  The directions 
toward HD~27778 and HD~28975 (represented by a single cluster containing 
L1524 and L1529) had a distance of $129.0\pm0.8$ pc.  It is not clear that the same 
material was sampled in the three studies for the directions that are our focus, and 
there is the possibility small-scale structure is present in the chosen regions since our 
analysis is based on absorption along an infinitesmal pencil beam.  Still, we can infer 
two things of importance to our work: the TMC is about 150 pc away so that 
HD~29647 lies within the diffuse material on the far side of the cloud, and the other 
three stars are much beyond the cloud.

\section{Discussion}

Here we explore the implications for the transition from diffuse molecular gas to dark 
cloud arising from our results presented in the previous section.  We begin by 
providing an overall perspective gleaned from the results for the four directions 
toward HD~27778, HD~28975, HD~29647, and HD~30122, where the focus is on the presence 
of enhanced potassium depletion onto grains ($\S{4.1.1}$), the amount of gas phase carbon 
in CO ($\S{4.1.2}$), and the conversion of CH into other molecular species, including CO 
($\S{4.1.3}$).  This is followed by a discussion of a number of characteristics 
distinguishing the line of sight to HD~29647 in $\S{4.2}$ and a comparison of the 
material toward the four stars in $\S{4.3}$.  $\S{4.4}$ places our results into 
context with the findings from large-scale studies of the TMC ($\S{4.4.1}$) and from 
efforts with a focus on specific regions in this molecular cloud ($\S{4.4.2}$).

\subsection{Overall Perspective for the Four Directions in Taurus}
\subsubsection{Potassium Depletion}

The total number of protons, $N_{tot}$(H) = [$N$(H~{\small I}) $+$ 2$N$(H$_2$)], 
along a line of sight is directly obtained from UV observations of H~{\small I} and 
H$_2$ absorption.  Of the four sight lines in our study, only that toward HD~27778 has 
such a determination.  From the numerous measurements on diffuse molecular clouds, two 
other methods yield values of $N_{tot}$(H).  The first uses $E$($B-V$) from \citet{boh78}, 
$<N_{tot}$(H)/$E$($B-V$)$>$ $=$ $5.8\times10^{21}$ cm$^{-2}$ mag$^{-1}$.  
The second method is based on the column density of neutral potassium, 
$N$(K~{\small I}), from \citet{wel01}.  Welty \& Hobbs found linear relationships 
between log[$N$(K~{\small I})] and log[$N_{tot}$(H)]; we adopted the one that 
included all high-resolution measurements and the most reliable medium-resolution ones -- 
log[$N$(K~{\small I})] $=$ $-27.21\pm2.79$ $+$ ($1.84\pm0.13$)log[$N_{tot}$(H)].  
Toward HD~27778, the measured value for $N_{tot}$(H) is $2.3\times10^{21}$ 
cm$^{-2}$ \citep{rit18}; see also \citet{car01, car04}.  
Using $E$($B-V$) from Table 1 and the total $N$(K~{\small I}) 
from our analysis of the UVES spectrum, we obtain values for $N_{tot}$(H) 
of $2.1\times10^{21}$ and $1.9\times10^{21}$ cm$^{-2}$ respectively, within 20\% of 
the UV results.

Therefore, we apply the two methods to the data for HD~30122, HD~28975, and 
HD~29647.  For the gas toward HD~30122, we obtain 
$1.3\times10^{21}$/$1.6\times10^{21}$ cm$^{-2}$ from 
$E$($B-V$)/$N$(K~{\small I}), while for HD~28975 we find values of 
$3.5\times10^{21}$/$3.7\times10^{21}$ cm$^{-2}$ respectively.  The consistency found 
for the material toward HD~27778, HD~30122, and HD~28975 is not evident when applying 
the methods for the line of sight toward HD~29647, 
$6.3\times10^{21}$/$3.6\times10^{21}$ cm$^{-2}$.  Although there are numerous 
stellar features near the weak K~{\small I} doublet at 4044 and 4047 \AA\ in the UVES 
spectrum of HD 29647, we were able to measure $N$(K~{\small I}) from the optically thin 
line at 4047 \AA, obtaining a value of $(2.6\pm0.6)\times 10^{12}$ cm$^{-2}$.  This 
value agrees with the one obtained from our fit of the line at 7698 \AA, and so it 
appears the amount of depletion onto grains is enhanced by about a factor of two 
for potassium toward this star.

We can also estimate the extent of the molecular material along the four sight lines.  
For the very molecule-rich directions toward HD~28975 and HD~29647, we consider the 
total proton column densities discussed here, while for the diffuse molecular gas 
toward HD~27778 and HD~30122 we adopt the molecular hydrogen column densities 
from \citet{rac02} and \citet{she08}, respectively.  When dividing these column 
densities by the gas densities found from CO excitation and CN chemistry, we find 
that the molecular portions of the TMC in these directions is approximately 1 pc, 
varying by only 50\%.

\subsubsection{The Amount of Carbon in CO}

Using absorption from the C~{\small II} intercombination line 
at 2325 \AA, \citet{sof97} 
determined that the gas phase carbon abundance in diffuse clouds is 
$1.4\times10^{-4}$.  
Since C$^+$ is the main source for the electron fraction in diffuse material, 
our limits and values for $x$(e) from CN excitation ($\S{3.2.3}$) suggest higher 
precision CN measurements are needed to place constraints on C$^+$ for the carbon 
budget.  However, with the CO column density ranging from $10^{15}$ to 
$10^{18}$ cm$^{-2}$ for the four directions in the present study, we can compare 
the CO contribution to the carbon budget.  We adopt the values of $N_{tot}$(H) 
discussed in the previous section -- $2.3\times10^{21}$ cm$^{-2}$ 
(HD~27778) from observations as well as $1.3\times10^{21}$ (HD~30122), 
$3.5\times10^{21}$ (HD~28975), and $6.3\times10^{21}$ (HD~29647) 
cm$^{-2}$ from $E$($B-V$).  Besides singly-ionized carbon and CO, another 
significant contribution to the carbon budget comes from neutral carbon.  

For HD~27778, \citet{sof04} obtained an upper limit of $1.1\times10^{-4}$ for 
the fractional abundance of C$^+$, $x$(C$^+$) $=$ $n$(C$^+$)/$n_{tot}$(H).  
Though cruder, we can look at the 
estimate used in our chemical model to track the conversion of C$^+$ into CO 
\citep{fed94}, $\alpha$ $=$ [1 $+$ 14$\times$($\tau_{UV}-2$)/5], where 
$\alpha$ is the percentage of C$^+$ remaining.  For this direction, $\alpha$ is 0.27, 
or $x$(C$^+$) is $3.8\times10^{-5}$, comfortably below the upper limit of 
\citet{sof04}.  Moreover, \citet{bur10} quote log[$N$(C~{\small I})] of 15.06, 
or $x$(C) at least $5.0\times10^{-7}$, and \citet{she08} give 
log[$N$(CO)] $=$ 16.09, or $x$(CO) of $5.4\times10^{-6}$.  For this sight line 
only about 4\% of the gas phase carbon in diffuse gas is in CO; it is not clear where 
the missing carbon resides.

Because \citet{jen11} were not able to measure the amount of C~{\small I} toward 
HD~30122, a result of severe blending from stellar features, the other three 
directions in our sample only have information on CO (and the crude measure for 
C$^+$ from $\alpha$).  We start with the gas toward HD~30122, where 
$\tau_{UV}$  indicates no conversion of C$^+$.  Using the results of \citet{she08}, 
$N$(CO) of $7.04\times10^{14}$ cm$^{-2}$, we infer a CO abundance of 
$5.4\times10^{-7}$ relative to $N_{tot}$(H).  
Here, too, essentially all of the carbon is in C$^+$.  Turning to 
the sight line to HD~28975, $N$(CO) is $1.45\times10^{17}$ cm$^{-2}$ and 
$x$(CO) is $4.1\times10^{-5}$, or about 30\% of the available carbon.  With a 
value for $\alpha$ of 0.17, about 50\% of the carbon appears to be in neutral 
carbon.  For the gas toward HD~29647, $N$(CO) is $9.56\times10^{17}$ 
cm$^{-2}$, or $x$(CO) $=$ $1.5\times10^{-4}$; all of the interstellar carbon appears to 
be in the form of CO.  Because $\tau_{UV}$ is so large, $\alpha$ is only 0.07 and 
little of the carbon could be in C$^+$.  This is consistent with our conclusions reached 
through CO fractionation that C$^+$ is only present in the outer portion of the cloud, 
where the species CH$^+$, CH, and C$_2$ reside.  Moreover, \citet{whi89} found an upper 
limit for the CO column of $5\times10^{16}$ cm$^{-2}$ in solid form, or 5\% of the gas 
phase abundance.  Therefore, it seems that the line of sight toward HD~29647 probes 
the dark molecular TMC, with a gas temperature cold enough (10 K) to form solid 
CO but has not yet shown any evidence for it.  With a temperature of 30 K, the 
material toward HD~28975 is too warm for CO depletion onto grains.

\subsubsection{CO versus CH}

\citet{she08} sought relationships among observed quantities in their CO survey.  
Of particular interest for the present study is the correspondence between 
column densities of CO and CH for $N$(CO) $>$ $10^{13}$ cm$^{-2}$, which 
they represented as 

\begin{equation}
log[N({\rm CO})] = -(22.3\pm11.3) + (2.80\pm0.85)log[N({\rm CH})].
\end{equation}

\noindent Adopting our results for $N$(CH), the predicted values for $N$(CO) toward 
HD~27778, HD~30122, HD~28975, and HD~29647 are $2.6\times10^{15}$, 
$4.4\times10^{14}$, $1.14\times10^{16}$, and $1.79\times10^{16}$ cm$^{-2}$, 
respectively.  The observed CO column densities are $1.23\times10^{16}$ and 
$7.04\times10^{14}$ cm$^{-2}$ for the material toward HD~27778 \citep{she07} 
and HD~30122 \citep{she08}; Table 3 provides the column densities for the other 
two sight lines.  We first focus on the sight lines toward HD~27778 and HD~30122.  
The observed values for $N$(CO) are 4.7 and 1.6 times larger, respectively, than the 
values inferred from the relationship above.  Looking at Fig. 9 from \citet{she08}, a 
plot of log[$N$(CO)] vs. log[$N$(CH)], the data point for HD~27778 lies near the 
left upper boundary of the sample, while the data point for HD~30122 is in the 
middle of the sample.  This indicates that the material toward these two stars is 
representative of diffuse molecular gas and that the gas density is higher toward 
HD~27778 [as inferred from Fig. 17 in \citet{she08}].  This is borne out by our 
analyses of molecular excitation and CN chemistry.

The differences between observed and predicted $N$(CO) are greater for the 
gas toward HD~28975 and HD~29647, with ratios of 13 and 53.  The data points 
for both directions lie beyond the upper boundary shown in Fig. 9 of \citet{she08}.  
The data point for HD~28975 would appear approximately near the point for 
HD~200775, the illuminating source for the reflection nebula NGC~7023.  That 
for HD~29647 would appear about a dex higher, with a CO column density similar 
to that for a dark cloud [see Fig. 6b in \citet{she08}].  We note point 
[$N$(CO),$N$(H$_2$)] for HD~200775 in Fig. 6b of \citet{she08} is within the 
boundary represented by dark clouds.  However, the cause for these outliers differs.  
In NGC~7023, the enhanced flux of UV radiation from HD~200775 preferentially 
destroys CH relative to CO (and H$_2$), a consequence of the protection arising 
from CO (and H$_2$) self shielding.  Since there is no evidence for an 
enhancement in UV flux in its sight line, the extreme case of HD~29647 is 
likely the result of converting CH molecules into other molecules, including CO.  
In a study of CH emission from molecular clouds, \citet{mat86} found the CH 
column density leveled off in a plot $N$(CH) versus $A_B$, consistent with the 
chemical model of \citet{bol84} that suggested a value of about 
$2\times10^{13}$ cm$^{-2}$.  In their study of HCL-2 at 3.8 arcmin 
resolution, \citet{sak12} found CH column densities falling below the relationship 
between CH and H$_2$ from \citet{she08}.  The position of TMC-1(NH$_3$), 
which is closest to the sight line 
toward HD~29647, has a line width about 1.4 km s$^{-1}$ (or $b$-value 
of 0.8 km s$^{-1}$, similar to our observations) and $N$(CH) of 
$1.79\times10^{14}$ cm$^{-2}$.  This column density is three times larger than what 
we infer, but their measurements sample the full extent of the cloud.  Moreover, 
subsequent work by Magnani and colleagues \citep[e.g.,][]{mag95,mag98,mag03} 
presented evidence that CH emission was a reliable tracer of H$_2$ in the diffuse 
envelopes of dark clouds.  In light of our results for HD~29647, we believe that 
this star probes the outer predominantly molecular portion of the TMC, although 
it is embedded in the diffuse molecular material on the far side of the cloud 
\citep[][\S{3.5}]{whi04}.  This is described in more detail in the next section. 
 As for the portion of the TMC in front of HD~28975, we consider it a diffuse 
molecular cloud because only about 30\% of the available carbon is in the form 
of CO.

\begin{figure}
\begin{center}
\includegraphics[width=0.75\textwidth]{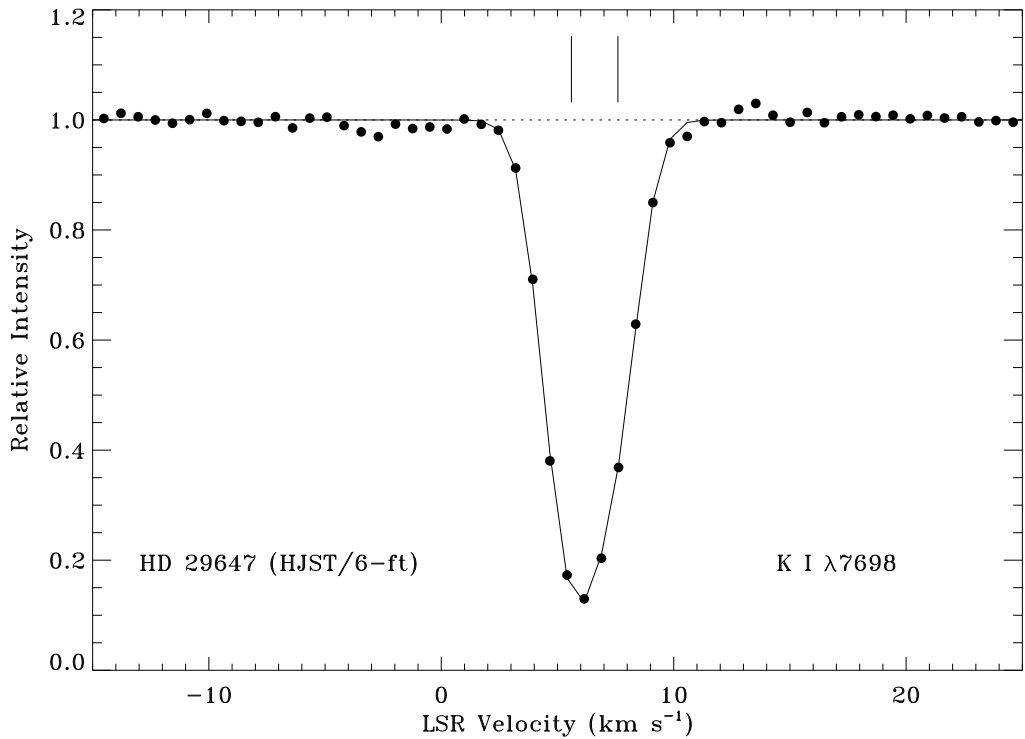}
\caption{High-resolution spectrum of K~{\sc I} absorption toward HD~29647 from previously unpublished data acquired with the coud\'{e} 6 foot camera on HJST. Our profile synthesis fit (solid line) indicates two velocity components at $+$5.6 and $+$7.6 km s$^{-1}$ (tick marks).}
\label{fig:HD29647_K1}
\end{center}
\end{figure}

Because \citet{wel01} concluded that there is a close correspondence between 
K~{\small I} and CH, we look at this for our lines of sight.  In particular, 
\citet{wel01} find a $N$(K~{\small I})/$N$(CH) ratio of 0.043 for their 
sample.  The ratios for gas toward HD~27778, HD~28975, HD~29647, and 
HD~30122 are 0.027, 0.052, 0.050, and 0.040.  Though the ratio toward 
HD~27778 is somewhat lower than the others, all four ratios are within about 
20\% of the mean value given by \cite{wel01}.  In light of our findings for 
K~{\small I} and CH toward HD~29647, we examine why the ratio does not differ, 
although evidence exists for smaller relative amounts of these two species, by 
considering the highest resolution spectra for K~{\small I} from our 
previously unpublished data acquired with the coud\'{e} 6 foot camera on HJST.  
There is a hint of an asymmetry in the profile (see Fig. 11).  
Fitting it with two velocity components yields velocities of $+$5.6 
and $+$7.6 km s$^{-1}$, $b$-values of 0.72 and 0.48 km s$^{-1}$, and 
relative fractions of 0.88 and 0.12 (yielding a total column of 
$2.58\times10^{12}$ cm$^{-2}$, much like the column from our most recent 
spectrum from HJST).  The component velocities are similar to those found for 
CO emission [$+$5.1 and $+$6.5 km s$^{-1}$ -- \citet{cru85}; $+$5.2 and 
$+$7.1 km s$^{-1}$ -- \citet{van91}; $+$5.4 and $+$6.7 km s$^{-1}$ -- our 
estimate from the $^{13}$CO spectrum in \citet{lis08}].  The Full 
Widths at Half Maximum of the emission lines are about 1 km s$^{-1}$ 
\citep{cru85}, equivalent to $b$-values of about 0.6 km s$^{-1}$.  
The redder component in the $^{13}$CO spectra of \citet{cru85} and 
\citet{lis08} is narrower, much like we infer from 
fitting the spectrum in Fig. 11.  Since 
Crutcher only detected HCO$^+$, HCN, and CN emission from the bluer component 
(with weaker wings in HCO$^+$ and CN), \citet{mes97} inferred that a 
dense clump with $v_{LSR}$ of $+$5.1 km s$^{-1}$ was present toward 
this star.  We suggest the presence of the dense clump results in lower 
column densities for both K~{\small I} and CH in this direction, and the 
relationship between these species in diffuse gas found by 
\citet{wel01} maintains their relative abundances.

\subsection{HD~29647 -- A Line of Sight Revealing the Transition from 
Diffuse Molecular to Dark Molecular Cloud}

Our results for the sight line toward HD~29647 differ significantly from those 
of molecule rich diffuse clouds like the material toward HD~27778.  The 
most obvious one is $N$(CO), where the value is about 100 times greater 
toward HD~29647.  There are a number of checks we can perform to 
verify our results.  The maximum optical depth at line center from the 
fitting of the IGRINS spectrum is a modest 1.7.  This optical depth 
arises from the required consistency when fitting the R(1)/P(1) and R(2)/P(2) 
pairs of lines.  The inferred $b$-value from the fit is 0.5 km s$^{-1}$.  
\citet{lac17} examined the amount of CO absorption with IGRINS for targets 
in the vicinity of HD~29647, obtaining column densities via a curve-of-growth 
analysis.  A comparison of the results with ours for HD~29647 appears in 
Table 6.  The focus of the comparison is on stars within 30 arcminutes 
of HD~29647, corresponding to a separation of 1.3 pc at 150 pc.  The results 
for $N$(CO), $T_{ex}$(CO), $v_{LSR}$, and $b$-value are very similar, 
indicating the material within about a 1 pc of the direction toward HD~29647 
is relatively homogeneous.  We note that HD~283809 and Tamura~8 lie beyond 
the IR nebula illuminated by HD~29647, and that the larger CO column toward 
Tamura~8 might arise from its probing one of the filaments mentioned above 
(see Fig. 10).

\begin{deluxetable}{lccccccc}[tbh]
\tablecolumns{8}
\tablewidth{0pt}
\tabletypesize{\small}
\tablecaption{CO Results for Embedded and Background Stars Near HD~29647}
\tablehead{\colhead{Star} & \colhead{R. A. (2000)\tablenotemark{a}} & 
\colhead{DEC (2000)\tablenotemark{a}} & 
\colhead{Distance\tablenotemark{b}} & \colhead{$N$(CO)} & 
\colhead{$T_{ex}$} & \colhead{$v_{LSR}$} & \colhead{$b$-value} \\
\colhead{} & \colhead{($^h$:$^m$:$^s$)} & 
\colhead{($^{\circ}$:$^{\prime}$:$^{\prime\prime}$)} & \colhead{(pc)} & 
\colhead{($10^{18}$ cm$^{-2}$)} & \colhead{(K)} & 
\colhead{(km s$^{-1}$)} & \colhead{(km s$^{-1}$)}
}
\startdata
Elias 3-16 & 04:39:39 & $+$26:11:27 & 275(58) & 2.6 & 9.8 & 6.8 & 
0.49 \\
Elias 3-18 & 04:39:56 & $+$25:45:02 & 149(5) & 0.86 & 10.0 & 6.4 & 
0.33: \\
Tamura 8 & 04:40:57 & $+$25:54:14 & $\ldots$ & 2.5 & 9.5 & 6.4 & 
0.61 \\
HD 283809 & 04:41:25 & $+$25:54:48 & 326(10) & 1.3 & 8.4 & 6.0 & 
0.41 \\
Kim 1-59 & 04:41:30 & $+$25:27:03 & 752(130)\tablenotemark{d} & 
1.1 & 10.0 & 6.4 & 1.5 \\
 & & & & & & & \\
HD~29647 & 04:41:08 & $+$25:59:34 & 155(2) & 1.0 & 9.5 & 5.4 & 
0.53 \\
\enddata
\tablenotetext{a}{SIMBAD -- Wenger et al. 2000.}
\tablenotetext{b}{{\it Gaia} DR2 -- Bailer-Jones et al. 2018.}
\tablenotetext{c}{Numbers in parentheses are uncertainties.}
\tablenotetext{d}{Source offset in {\it Gaia} DR2 catalog unusually large.}
\end{deluxetable}

Other differences involving the gas toward HD~29647 are the enhanced K 
depletion onto grains and the much lower predicted CO column density from our 
measured value for $N$(CH) from the relationship given by \citet{she08}.  
\citet{pan05} discussed the possibility of K depletion in terms of similarities 
derived from profile synthesis to the parameters extracted from Ca~{\small I} 
features.  While the same applies here (see Table 2), we focused on the 
lower $N_{tot}$(H) inferred from $N$(K~{\small I}) using the correspondence 
found by \citet{wel01}.  We deduce from our analyses on $N$(K~{\small I}) 
and $N$(CH) that the absorption from these species is restricted to the portion 
of the line of sight sampling typical diffuse molecular gas, while most of the CO 
(and CN) absorption arises from a denser clump of material in the picture 
developed by \citet{whi04}.  Further evidence for this scenario comes from the 
detection of solid H$_2$O in the form of an ice mantle toward HD~29647 
\citep{smi93,tei99}.  Clearly, HD~29647 probes material in a dark cloud, 
albeit over only about 4 mag of visual extinction.

We end the section with a comparison of results from analyses by others, mainly 
based on the material presented by \citet{cru85}.  In what follows, we have not 
tried to adopt a common set of input parameters, and so factors of two 
agreement are considered satisfactory.  We begin by summarizing the 
relevant findings of \citet{cru85}.  He provided estimates for gas density 
[$n$(H$_2$)], $T_k$, and $x$(e) in the gas toward HD~29647.  
From a Large Velocity Gradient (LVG) model of the emission 
from CO and its isotopologues, he obtained a combined value for $N$(CO) in 
the two components of $4.8\times10^{17}$ cm$^{-2}$, within a factor of two 
of ours.  The H$_2$ density and $T_k$ inferred from the model were 800 
cm$^{-3}$ and 10 K; these agree very well with our results taking into account 
that we adopted a correspondence, $n_{tot}$(H) $=$ 2$n$(H$_2$).  From 
HCO$^+$ emission and the ionization balance of K, he preferred the result for 
$x$(e) of $2\times10^{-7}\delta_{\rm K}^{-1}$, where 
$\delta_{\rm K}^{-1}$ is the enhancement in K depletion over the average 
interstellar value.  Since we find a factor of about 2 for this enhancement (see 
$\S{4.1.1}$), his estimate for $x$(e) becomes $10^{-7}$, supporting our 
conclusion that CO is the dominant constituent of the carbon budget.  Another 
effort based on analyzing excitation of CO emission lines is that of 
\citet{van91}.  The radiative transfer through a uniform spherical cloud was
determined via the mean escape probability, though other models were 
examined as well.  The focus was on reproducing observed line intensity 
ratios, $^{12}$CO(1--0)/(3--2), $^{12}$CO(2--1)/(3--2), and 
\linebreak $^{12}$CO(3--2)/$^{13}$CO(1--0) in light of 
results on C$_2$ excitation.  We converted their densities of collision partners 
into total gas densities by multiplying the C$_2$ results by a factor of 1.5 
and those for CO by a factor of 2 (H$_2$ excitation dominates) in what 
follows.  Of the three ratios, the one most consistent with the C$_2$ results 
for HD~29647 (as well as for most other directions in their survey) was the 
\linebreak $^{12}$CO(3--2)/$^{13}$CO(1--0) ratio: $525^{+450}_{-150}$ cm$^{-3}$ 
versus a preferred value of 1000 (with a range 400 to 2000) cm$^{-3}$.  These are 
also consistent with the present results.

Two modeling efforts \citep{nec88,van89} studied the chemistry of the gas 
in the $+$5.1 km s$^{-1}$ component toward HD~29647.  It is important 
to note that both efforts incorporated CO self shielding in their model clouds.  
\citet{nec88} adopted the gas density and temperatures from \citet{cru85} 
allowing elemental depletions and the cosmic ray ionization rate to vary, 
while \citet{van89} provided the best-fitting model described in their earlier 
work \citep{van88}.  A density of 1000 cm$^{-3}$ and temperature of 15 
K were adopted by \citet{van89}.  For $N$(C$_2$), \citet{van89} used the 
measurements of \citet{hob83}.  There is generally quite good agreement 
between available observations and predictions for the molecular species in 
our survey (CH, C$_2$, CN, and CO).  However, our column densities 
tend to be smaller than those quoted earlier, in large part because there is 
less confusion with stellar features in our spectra, except for CO where it 
is a factor of nearly 3 larger.  The prediction by \citet{van89} for 
$N$(CN) is much closer to ours.  For this set of species, the model 
predictions are within a factor of 2 to 3 of the column densities presented 
here.  Our measurements add NH to the mix for comparison; the NH 
column of $8.0\times10^{12}$ cm$^{-2}$ is not consistent with either 
model.  The electron abundance given in \citet{nec88} is consistent with our 
findings.  With the advancements in astrochemistry since these efforts and 
our more precise results, it might be worthwhile revisiting the cloud toward 
HD~29647.

\subsection{A Comparison of the Four Directions}

\begin{deluxetable}{lcccccccc}[tbh]
\tablecolumns{9}
\tablewidth{0pt}
\tabletypesize{\small}
\tablecaption{Summary of Results}
\tablehead{\colhead{Star} & \colhead{$N$(CO)} & 
\colhead{$^{12}$CO/$^{13}$CO} & \colhead{$T$(H$_2$)} & 
\colhead{$T_{01}$(CO)} & \colhead{$T$(C$_2$)} & 
\colhead{$n_{tot}^{ex}$(CO)} & \colhead{$n_{tot}$(Chem)\tablenotemark{a}} & 
\colhead{$n_{tot}^{ex}$(C$_2$)\tablenotemark{a}} \\
\colhead{} & \colhead{($10^{15}$ cm$^{-2}$)} & \colhead{} & 
\colhead{(K)} & \colhead{(K)} & \colhead{(K)} & 
\colhead{(cm$^{-3}$)} & \colhead{(cm$^{-3}$)} & 
\colhead{(cm$^{-3}$)}
}
\startdata
HD~29647 & 956. & $>$~9 & $\ldots$ & 9.5 & 10 & 1800 & 
$\sim$1000/$\sim$1000 & $\geq$~175/$\geq$~350 \\
HD~28975 & 145. & $\ldots$ & $\ldots$ & 11.5 & 30 & 1900 & 
475/1200 & 125/250 \\
 & & & & & & & \\
HD~27778 & 12.3 & 67(10) & 51 & 5.3 & 50 & 975 & 400/800 & 75/150 \\
HD~30122 & 0.70 & $\ldots$ & 61 & 3.8 & $\ldots$ & 450 & 275/525 & 
$\ldots$ \\
\enddata
\tablenotetext{a}{The first entry in the column for $n_{tot}$(Chem) applies 
to $I_{UV}$ $=$ 0.5, and the second 1.0.  For the next column, 
$n_{tot}^{ex}$(C$_2$), the two entries refer to $I_{IR}$ equal to 0.5 or 1.0.}
\end{deluxetable}

We describe our results for the other three directions in light of these findings for 
HD~29647.  They appear in Table 7, separated into two classes.  There are the 
more typical diffuse molecular clouds toward HD~27778 and HD~30122 
grouped at the bottom, and dark molecular cloud toward HD~29647 along with 
the intermediate case for the material toward HD~28975 at the top.  The main thing 
to note is the range in $N$(CO), over a factor of 1000.  In order to investigate 
the cause for this large variation, we consider the effects of density, $\tau_{UV}$, 
and the CO self-shielding factor 
\citep{van88,vis09}.  For gas density, we utilize $n_{tot}^{ex}$(CO) and 
$n_{tot}$(Chem), which vary by a factor of 3.  In diffuse molecular gas, the CO 
abundance varies roughly as the total gas density, but as the amount of self 
shielding increases, with $N$(CO), photodissociation becomes less important and 
then the CO abundance no longer depends on gas density (as in the case for CN 
toward HD~29647 discussed in $\S{3.4}$).  Because the chemistry in the dense 
clump toward HD~29647 is not sensitive to density, a factor of 2 
variation is more appropriate.  
We look at $\tau_{UV}$ next, where the range is 1.43 (to HD~30122) 
to 6.76 (HD~29647); the resulting variation in attenuation is a factor of $\sim$200.  
This is an upper limit, however, because CO photodissociation does not operate in 
the dense clump.  If we estimate a 25\% contribution of the clump to the reddening 
toward HD~29647, $\tau_{UV}$ is lowered to 5.07 and the variation in attenuation 
becomes of order 50.  We suggest the remaining factor of 10 comes from the 
self-shielding factor.  According to Fig. 13 in \citet{she08}, the self-shielding 
factor decreases from a value 0.30 at $N$(CO) of a few $10^{14}$ cm$^{-2}$ to 
0.01 at the edge of the region containing results from dark clouds with $N$(CO) 
about $10^{17}$ cm$^{-2}$.  A column density of $1.45\times10^{17}$ 
cm$^{-2}$ is found toward HD~28975, where $\tau_{UV}$ is 3.72 and the 
attenuation is now a factor of 10.  This column density is a factor of 
200 greater than that toward HD~30122.  Thus, the factor-of-5 decrease in grain 
attenuation toward HD~28975 accounts for the difference in results for HD~30122.  
This rough calculation indicates that grain attenuation and self shielding play 
comparable roles in CO photochemistry when $N$(CO) approaches the values 
associated with dark molecular clouds.

We conclude the comparison with a discussion of excitation and gas temperatures 
and $n_{tot}^{ex}$(C$_2$).  The CO excitation temperature seems to be higher toward 
the stars probing dark cloud material, while the gas temperature inferred from H$_2$ 
and C$_2$ excitation is higher along the other two directions.  Since dark molecular 
clouds have typical temperatures of 10 K, the $T_k$ should decrease as more of the 
dark cloud material is sampled.  Toward HD~29647, $T_k$ and $T_{01}$(CO) have similar 
values, and our analysis of CO excitation from the IR absorption spectrum provides 
confirmation of the near equality.  The CO molecule acts as a thermometer for dark 
clouds.  Our analysis using RADEX in $\S{3.2.2}$ 
suggests the need for optical depths approaching 100 in order to match 
our IGRINS results and Crutcher's (1985) for CO emission.  The direction toward 
HD~28975 appears as an intermediate case with $T_k$ somewhat higher than 
$T_{01}$(CO) and optical depths of about 10 for the predicted levels of emission.  
This indicates that $^{13}$CO cooling is comparable to that of the more common 
isotopologue.
  
An estimate for $I_{IR}$ toward HD~29647 is possible from the analysis 
in $\S{3.5.3}$.  In particular, adopting a stellar radius of 5.6
$R_{\sun}$ and the distance between the star and nebula yields a dilution factor 
of $\approx10^{-14}$, indicating the flux of radiation impinging on the nebula is 
about equal to the average interstellar flux or $I_{IR}$ is about 2.  However, 
C$_2$ excitation via optical pumping, which involves populating excited electronic 
levels followed by radiative cascades into the ground vibrational level, occurs by 
absorbing photons at wavelengths shorter than 1 $\mu$m.  As also noted in 
$\S{3.5.3}$, there is no evidence for a reflection nebula at visible wavelengths 
because the visual extinction is so large.  If the visual extinction between the nebula 
and the surface of the nearside of the TMC is appreciable, there may not be a 
significant amount of pumping.  Thus, an increase in the gas density derived from 
analysis of excited rotational levels in C$_2$ does not seem likely.  Instead, the 
similar results for $n_{tot}^{ex}$(C$_2$) among the four sight lines are possibly a 
consequence of the excitation taking place in diffuse molecular gas common to all 
our targets.

For the three directions mainly probing diffuse molecular gas (HD~27778, HD~30122, as 
well as HD~28975), an interesting dichotomy arises.  When adopting the average 
flux, $n_{tot}$(Chem) and $n_{tot}^{ex}$(CO) are 
more similar; however, $n_{tot}^{ex}$(C$_2$) is 
lower than the other densities.  This is probably a consequence of basing our 
analyses on homogeneous density clouds and that the molecular species are assumed 
to occupy similar volumes.  The responses to changes in the strength of the radiation 
field for $n_{tot}$(Chem) and $n_{tot}^{ex}$(C$_2$) arises from details of 
the processes involved.  An increase (decrease) in the strength requires an increase 
(decrease) in $n_{tot}$(Chem) to compensate for changes in the photodissociation 
rate when reproducing the observed column densities.  Similarly, an increase 
in $I_{IR}$ leads to a increase in $n_{tot}^{ex}$(C$_2$) as discussed in \citet{van82}.

The observational results from studies of atomic and 
molecular absorption are precise enough that more sophisticated approaches are 
required to achieve agreement between measurements and model predictions significantly 
better than a factor of 2.  As a first step, such models need to derive gas densities 
in a more consistent way by combining thermodynamics, chemistry, radiative transfer, 
and excitation.  Then for each species of interest, integrated results through the 
modeled cloud would reveal the regions where the various processes are mainly 
taking place.  Current PDR codes contain essentially all 
the necessary elements to perform such calculations.  
It is quite possible that we are seeing differences in total proton density because 
our diagnostics are located in different portions of a cloud.  Such a possibility 
was noted above for $n_{tot}^{ex}$(C$_2$) and schematically presented in 
\citet{pan05}.  They developed their picture of a diffuse cloud in part by comparing 
the $b$-values obtained for different species.  Since the $b$-values indicate that 
turbulence dominates over thermal motions, \citet{pan05} viewed $b$-values as 
revealing the number of clumps sampled by the lines from various species.  A smaller 
$b$-value arises because fewer clumps are intercepted, suggesting the observations 
probe a more restricted portion of the cloud.  The first step can be time 
independent, because many processes in diffuse atomic and molecular clouds (the 
envelopes of self-gravitating molecular clouds) take place under steady-state 
conditions.  Then time-dependent phenomena, such as turbulence, can be introduced 
for a more complete description of diffuse interstellar matter.

\subsection{A Comparison to Other Studies of the TMC}

\subsubsection{Observations of the Whole Cloud}

We now examine how our study fits into the global picture developed from 
measurements across the TMC \citep{fla09, gol08, nar08, pin10} as well as 
other observations of interstellar gas in Taurus \citep{par12, lee06}.  
While we already considered the results on dust emission from \citet{fla09} for 
our targets in $\S{3.5}$, some of their other findings are worth discussing.  
The average dust temperature across the TMC is $14.5\pm1.0$ K, with colder 
filaments seen in the far-IR emission maps in the middle of warmer material.  
One such filament lies in the vicinity of the sight line to HD~28975, 
and multiple filaments surround the sight line to HD 29647.  Emission from 
poly-cyclic aromatic hydrocarbons 
(or PAHs) and very small grains (VSGs) are correlated, but there is no 
relationship with emission from large grains.  Abundance variations on 
sub-pc scales of a factor of a few are seen in the PAH and VSG emission, 
including the more diffuse portions of the TMC with $A_V$ up to a few mag.  
The far-IR dust opacity was shown to correlate very well with the visual 
extinction derived from near-IR images and indicated an increase in dust emissivity 
relative to the diffuse ISM.  Their 
models of the large scale dust emission revealed an overall reduction in the 
interstellar radiation field, which was incorporated into our analyses above; 
the strength of the field also showed significant variations in different 
regions of the TMC.

\citet{gol08} created a 100 square degree map of the TMC in emission from the 
$J$ $=$ $1\rightarrow0$ lines of CO and $^{13}$CO, sampled on a 20$^{\prime\prime}$ 
grid.  The map was divided into three regions: one without molecular emission, 
one with only CO emission, and one showing emission from 
both isotopologues.  Intensities were 
integrated over velocity from $+2$ to $+9$ km s$^{-1}$.  As seen in our Fig. 1, 
the molecular emission is very filamentary.  Peak emission occurs between $+5$ 
and $+8$ km s$^{-1}$, the interval where we observe strong molecular absorption.  
\citet{gol08} note, however, that emission at about $+10$ km s$^{-1}$ (where we 
see absorption toward HD~26751 in Appendix C) might not be associated with 
the TMC.  The pixels with no emission suggested an upper limit of $N$(CO) of 
$7.5\times10^{15}$ cm$^{-2}$, indicating that the direction toward HD~30122 
probed this region of the map.  The region showing only CO emission contained 
about half the gas, with $N$(H$_2$) less than $2.1\times10^{21}$ cm$^{-2}$ 
and an average value for $N$(CO) of about $3.6\times10^{16}$ cm$^{-2}$.  The 
sight line toward HD~27778 may probe this type of material.  For the region 
containing both CO and $^{13}$CO emission, the average $N$(CO) spanned a range 
from $1.3\times10^{17}$ to approximately $1\times10^{18}$ cm$^{-2}$, 
consistent with our interpretation for the kind of material probed by HD~28975 
and HD~29647.  \citet{gol08} also used the models by \citet{van88} to place 
the results into context, finding (1) the region with no CO emission had H$_2$ 
columns less than $10^{21}$ cm$^{-2}$, (2) that with only CO emission had 
values between $10^{21}$ and about $3\times10^{21}$ cm$^{-2}$, and (3) that 
with emission from both CO and $^{13}$CO ranged over 1.5 to $10\times10^{21}$ 
cm$^{-2}$.  For the material analyzed in this paper, we have 
a measurement (toward HD~27778) and estimates (toward HD~29647 and HD~30122) 
for $N$(H$_2$) in units of $10^{20}$ cm$^{-2}$, 6.2 \citep{rac02}, $\sim$30 
(see $\S\S{4.1.1, 4.1.3}$), and 4.4 [from our $N$(CH) and a $N$(CH)/$N$(H$_2$) 
ratio of $3.5~\times~10^{-8}$ from \citet{she08}].  Clearly, the sight line 
toward HD~30122 is similar to regions without CO emission, and HD~29647 samples 
a region with both isotopologues.  It appears that HD~27778 probes gas that is 
somewhere between regions with and without CO emission based on 
the CO and H$_2$ results for this sight line.  \citet{nar08} 
described details of the observations used in creating the maps and showed 
the emission in 1 km s$^{-1}$ intervals.  For the four lines of sight 
discussed here, the emission peaked at velocities between 
5 and 7 km s$^{-1}$, where we see the strongest molecular absorption.

These CO and $^{13}$CO maps formed the basis for additional analyses.  In 
particular, \citet{pin10} used them for a comparison of H$_2$ columns derived 
from dust extinction with the Two Micron All Sky Survey (2MASS) on scales of 
200$^{\prime\prime}$.  Other improvements to the analyses included recent 
molecular data, adoption of RADEX for radiative transfer, and comparisons with 
the updated photochemical code of \citet{vis09}.  The TMC again was divided 
into three regions based on CO and $^{13}$CO emission.  Emphasis was placed on 
results for the CO-to-H$_2$ conversion factor, $x$(CO)~$=$~$N$(H$_2$)/$I_{CO}$, 
where $I_{CO}$ is the integrated intensity of line emission.   A typical 
value for the conversion factor throughout the cloud was $2.1\times10^{20}$ 
cm$^{-2}$(K km s$^{-1}$)$^{-1}$.  Using our estimate for $N$(H$_2$) from the 
previous paragraph and $I_{CO}$ from \citet{lis08}, we find a very similar 
value for $x$(CO) toward HD~29647, reinforcing our conclusion that this star 
probes dark cloud material.  For $A_V$ less than 3 mag, the conversion 
factor could be two orders of magnitude smaller.  
\citet{pin10} averaged nearly 10$^6$ spectra 
for the region showing no emission and obtained an average value for 
$N$(CO) of $7.8~\times~10^{14}$ cm$^{-2}$, much like the value toward 
HD~30122.  This reveals the effort needed to obtain CO column densities 
typically found in diffuse molecular clouds \citep[e.g.,][]{she08}.  \citet{pin10} 
found a linear correlation between visual extinction and $N$(CO) for $A_V$ from 3 
to 10 mag with a CO/H$_2$ ratio of $\sim10^{-4}$, as we found for the gas toward 
HD~29647.  Their modeling results based on the photochemistry described by 
\citet{vis09} reveal other similarities to our analyses.  For $A_V$ less than 
5 mag, their data are best described by a varying CO/H$_2$ ratio, and for 
$T_k$ of 15 K, the models yield a $n_{tot}$(H) of about 800 cm$^{-3}$; our 
analyses suggest such densities apply to the material toward HD~28975 and 
HD~29647 (and possibly toward HD 27778).  The models also indicate a reduced 
radiation field and that half the mass of the cloud is in the region without 
detectable amounts of $^{13}$CO.

There are two large-scale surveys including the TMC that provide useful 
comparisons.  First, \citet{par12} follow the approach taken by \citet{pla11}, 
except they use extinction instead of far IR emission to study the presence of 
CO-dark gas and the CO-to-H$_{2}$ conversion factor.  The visual extinction 
comes from \citet{dob11} and is based on 2MASS data.  The more diffuse regions 
give the optimal value for $A_V$/$N_{tot}$(H); for the solar neighborhood the 
value is $6.53\times10^{-22}$ mag cm$^2$, which agrees very well with the UV 
result of \citet{boh78} for $R_V$ of 3.1.  The global conversion factor is 
$1.67\times10^{20}$ cm$^{-2}$(K km s$^{-1}$)$^{-1}$, with significant 
variations, and it is 
$2.27\times10^{20}$ cm$^{-2}$(K km s$^{-1}$)$^{-1}$ for gas in Taurus, 
comparable to what \citet{pin10} found.  \citet{par12} obtained average 
values for the locations of the H-to-H$_2$ transition ($A_V$ of $\sim0.2$ mag) 
and the H$_2$-to-CO transition ($A_V$ of $\sim1.5$ mag).  The location of the 
H-to-H$_2$ transition agrees with the results from UV absorption 
\citep{sav77}, and that of the H$_2$-to-CO transition is consistent with our 
conclusions that the gas probed by HD~28975 and HD~29647 is predominantly 
molecular.  These authors also find a significant fraction of the molecular 
content in Taurus in the form of CO-dark gas.  Second, \citet{lee06} describe 
diffuse far UV observations of the Taurus region acquired with SPEAR/FIMS.  
Because the spatial resolution is coarse (pixel size 
of $0.2^{\circ}$ by $0.2^{\circ}$ 
smoothed by 3 pixels), detailed comparisons with our results are not possible.  
The measurements can distinguish cloud cores from halos surrounding the 
cores.  The map includes directions toward HD~27778, HD~28975, HD~29647, 
HD~30122, as well as HD~26571.  The sight lines toward HD~28975 and 
HD~29647 sample edges of cores, while the other sight lines lie within 
halos.  Cloud cores, with $A_V$ $>$ 1.5 mag, obscure background far UV 
radiation, while the flux seen from halos comes from scattered foreground 
light.  Molecular hydrogen fluorescence, which is only seen from halos, is 
examined with the model of \citet{bla87}.  The low values for gas density 
($\sim50$ cm$^{-3}$) and $N$(H$_2$) ($0.8\times10^{20}$) are likely caused 
by the large pixels.  Of note is that again a reduced radiation field is 
required to reproduce the observations.

\subsubsection{Observations of Specific Regions}

A number of efforts examined chemical species associated with diffuse molecular 
gas in specific dark clouds or regions of the TMC.  Since these species are 
seen along the lines of sight to our stellar background sources, we now describe 
how our results fit into the perspective of these findings for molecular cloud 
envelopes.  The works by \citet{sak12} 
and \citet{ebi15} studied CH emission from HCL-2 and OH emission from material 
to the east of the cloud, respectively.  We already noted that the CH 
column density lies below the extension of the relationship for diffuse clouds 
\citep{sak12}; here we focus on the kinematics.  For two of the four regions 
examined, both a narrow ($\sim0.3$ km s$^{-1}$) and broad component ($\sim1.5$ 
km s$^{-1}$) are seen.  Where a narrow component is observed, it contributes about 
20\% to the total CH column density.  Only the narrow component is present in the 
dense core, while the broad component is more extended and represents the diffuse 
envelope.  The transition from narrow to broad component is probably sharp, but 
their measurements with a $3.^{\prime}2$ beam could not resolve it.  The presence 
of the two components might be caused by dissipation of turbulence.  Only the 
broad component is observed in the region closest to HD~29647, TMC-1(NH$_3$).

All four hyperfine transitions in OH were measured by \citet{ebi15}.  The material 
to the east of HCL-2 is more diffuse and lies about 4.5 pc from the direction of 
HD~29647.  Intensity anomalies caused by non-LTE effects were found, allowing them 
to infer gas temperatures over a wide range in density ($10^2$ to $10^7$ cm$^{-3}$).  
Toward the center of their strip map acquired with a $8.^{\prime}2$ beam, $T_k$ is 
$53\pm1$ K and increases to about 60 K at the boundaries.  The averaged spectrum 
yields $60\pm3$ K and $N$(OH) of ($4.4\pm0.3$)$\times10^{14}$ cm$^{-2}$.  The lines 
only reveal the broader component ($\sim1.3$ km s$^{-1}$).  The emission appears to 
arise from CO-dark gas.  The line of sight toward HD~27778 probes similar material, 
and has a comparable amount of OH \citep{fel96}, ($1.02\pm0.04$)$\times10^{14}$ 
cm$^{-2}$.  The somewhat smaller value toward the star could be a consequence of 
the much larger beam used in the radio measurements.

The surveys by \citet{gol08}, \citet{nar08}, and \citet{pin10} were used in more 
focused efforts.  Of most relevance to our results, we discuss measurements of 
emission from H$_2$ \citep{gol10}, from C~{\small I}~and~{\small II} 
\citep{orr14}, and from CH and OH \citep{xu16a, xu16b}. \citet{gol10} obtained 
data on S(0) through S(3) transitions in H$_2$ for 
three boundary regions (see Fig. 1), a 
Linear Edge [RA(J2000) $=$ $4^h~38^m~00^s$ and DEC(J2000) $=$ 
$+27^{\circ}~00^{\prime}~00^{\prime\prime}$ at the center of the nominal slit 
position], the Filament [RA(J2000) $=$ $4^h~50^m~30^s$ and DEC(J2000) $=$ 
$+25^{\circ}~17^{\prime}~30^{\prime\prime}$], and the Globule [RA(J2000) $=$ 
$4^h~26^m~45^s$ and DEC(J2000) $=$ $+25^{\circ}~39^{\prime}~00^{\prime\prime}$].  
The Linear Edge and the Globule regions lie near the center of Fig. 1, while 
the Filament is near the eastern edge of the TMC.  
The study emphasized the Linear Edge because it was less 
complex.  The other studies acquired their data on the Linear Edge as well.  
The emission and relative populations were analyzed with the Meudon 
PDR code \citep{lep06}, but the high excitation temperature (210 K) suggested 
the presence of other processes such as the dissipation of turbulence.  The 
observed ortho-to-para ratio revealed the effects of turbulent diffusion, 
bringing colder interior gas to the surface.  The peak intensity occurred beyond 
the edge in $^{13}$CO emission for the Linear Edge, where the column densities 
for $J$ $=$ 2 through 5 were $1.3\times10^{18}$, $1.6\times10^{17}$,  
$\sim1.1\times10^{16}$, and $\sim8.5\times10^{15}$ cm$^{-2}$, respectively.  
These column densities can be compared with the results for the sight line 
toward HD~27778 \citep{jen10}, which samples similar material: 
$3.0\times10^{18}$, $3.5\times10^{17}$,  $4.4\times10^{15}$, and 
$2.1\times10^{14}$ cm$^{-2}$, respectively.  The UV data for HD~27778 show 
more excitation in the lower rotational levels and less in the highest ones.  
The line of sight toward HD~210839 ($\lambda$ Cep) in the survey by 
\citet{jen10} has column densities for these levels that are most similar to 
the $Spitzer$ measurements \citep{gol10}.  Somewhat lower column densities were derived 
for the Filament.

\citet{orr14} also utilized the Meudon PDR code in an attempt to reproduce their 
observations.  The most consistent model indicated the need for a weak radiation 
field, a low ambient $^{12}$C/$^{13}$C ratio of 43, a primary cosmic-ray 
ionization rate of about $5\times10^{-17}$ s$^{-1}$, and significant sulfur 
depletion.  The emission from neutral carbon and the CO isotopologues 
appears to come from dark cloud material.  While no C~{\small II} emission was 
seen, the modeling results reveal a rapid decline in C$^+$ abundance with depth 
into the cloud.  \citet{orr14} suggested ionized carbon was associated with the 
diffuse component of the cloud.  They provided integrated intensities for CO 
($\approx13$ K km s$^{-1}$) and $^{13}$CO (3 to 4 K km s$^{-1}$).  For two 
of the directions in our survey, HD~27778 and HD~29647, \citet{lis08} quoted 
respective values of 7.17(0.43) and 16.35(5.84) K km s$^{-1}$, where the numbers 
in parentheses are for $^{13}$CO.  Again these values imply the absorption 
toward HD~27778 samples diffuse material just beyond the edge, while that toward 
HD~29647 is more like a dark cloud.  The predictions for a rapid decline in 
C$^+$ abundance are consistent with the small amount of C$^+$ we infer for the 
gas toward HD~29647.

Observations of CH and OH emission from the Linear Edge were described by 
\citet{xu16a} and \citet{xu16b}.  Two velocity components are present, one 
changes from $+5.3$ to $+6.0$ km s$^{-1}$, while the other appears at $+6.8$ 
km s$^{-1}$.  The bluer emission comes from the region where $^{13}$CO 
emission is present; the redder component is mainly confined to the gas 
beyond the edge associated with only detectable amounts of CO.  The OH 
column density is about 50\% greater in the more diffuse portions of the 
strip map, with values of about $4\times10^{14}$ cm$^{-2}$.  In other words, 
the column densities along the strip intersecting the Linear Edge are about 
2 to 4 times greater than what is measured in the diffuse molecular gas toward 
HD~27778.  Column densities for CH were more uncertain because limited 
information on excitation temperature and optical depth was available.  The 
authors interpret the shift in velocity, the apparent excess in CH at $A_V$ 
less than about 1 mag, and line anomalies in the OH satellite line at 
1712 MHz as evidence for a continuous or C-shock caused by colliding streams 
or gas flow.  

Our CH$^+$ results can provide further insight into this phenomenon 
because C-shocks would be a site of significant production for 
this molecular ion \citep[e.g.,][]{flo85, dra86}.  Subsequent reactions can 
transform CH$^+$ into CH.  This route proceeds in low density gas 
($<100$ cm$^{-3}$) because CH$^+$ is destroyed rapidly through reactions with 
H, H$_2$, and electrons at higher densities.  Another route at higher 
densities synthesizes CH (as well as CN and most of the CO) and is initiated 
by a reaction involving radiative association,
C$^+$ $+$ H$_2$ $\rightarrow$ CH$_2^+$ $+$ photon.
According to \citet{pan05}, there is a linear relationship between 
CH associated with CH$^+$ and CH$^+$ in terms of column densities.  In other 
words when CH is made in the network involving CH$^+$, the molecules have 
comparable column densities.  According to Table 2, each of the velocity 
components containing absorption from both CH and CH$^+$ has a much smaller 
column of CH$^+$.  The same applies for the results on the sight line toward 
HD~26571 in Appendix C.  Thus, these five directions that sample diffuse and 
dark molecular clouds, the same type of material seen in CO and $^{13}$CO in 
the TMC \citep{gol08, pin10}, contain little CH$^+$.  While turbulence is 
clearly present in the form of non-thermal line widths, evidence for 
significant amounts of shocked gas appears less convincing.

\citet{ryb20} measured small-scale features traced by H~{\small I} absorption 
toward radio sources having component structure separated by less than a few 
arcminutes, or linear scales from 10$^3$ to 10$^5$ AU.  Resolved velocity 
components were treated separately in the analysis.  Both changes in peak 
optical depth and line width contribute to the observed variation.  These appear 
to be a transient feature of the cold neutral medium.  Differences 
in optical depth of about 0.05 are common, but reaching factors of 10 larger in 
some instances.  The median change in optical depth corresponds to changes in 
H~{\small I} column density of about $3.5\times10^{19}$ cm$^{-2}$, though for a 
sight line in Taurus (3C111) the variation in column density is $\sim2\times10^{20}$ 
cm$^{-2}$.  For gas in the TMC this corresponds to density variations of 1000 
cm$^{-3}$.  Only limits are available for another source in Taurus (3C123), which 
lies closer to the directions of the stars in our sample.  It is important to note 
that if the material is sheet-like, the density estimate could be a factor of a 
few smaller \citep{hei97}.  The inferred densities are similar to what we find for 
the molecular gas probed by HD~27778, HD~28975, HD~29647, and HD~30122.  
Considering absorption from the same set of species as in the present paper, 
\citet{pan01} detected spatial variations in molecular column densities across 
members of HD~206267 A/C/D and HD~217035 A/B with separations (10$^3$ to 10$^4$ 
AU) like those used in \citet{ryb20}.  Such a study is not possible for most of 
the material in the TMC because similar systems tend to be too faint or are in 
front of the TMC.  One interesting exception is the companion to HD~27778 
(BD$+23^{\circ}~683$), a star with $B$ and $V$ of 8.40 and 8.2 about 
30$^{\prime\prime}$ (4,600 AU) from the primary (see Fig. 7).

\section{Concluding Remarks}

We presented results on atomic and molecular absorption seen at UV, visible, and 
IR wavelengths that probe four portions of the TMC.  Gas temperatures and 
densities were inferred from analyses of molecular excitation and chemistry.  
The line of sight toward HD~29647 differed from the others in a number of ways.  
It was by far the most molecule rich, with a CO column density of $10^{18}$ 
cm$^{-2}$, and its gas temperature (derived from C$_2$ and CO excitation) was 
only 10 K.  It appears that essentially all the carbon is in CO.  The CH column 
density for the amount of CO observed revealed a deficit when comparison was 
made to typical diffuse molecular clouds.  The value of $N$(K~{\small I}) relative 
to the total proton column density to this star based on the amount of reddening
indicated an enhanced potassium depletion.  We surmised that this sight line 
mainly probes gas associated with the dark molecular cloud, HCL-2; the low value 
for $N$(CH) comes from the conversion of CH into other species, including CO.  
The directions toward HD~27778 and HD~30122 are like diffuse molecular clouds 
observed elsewhere, while the gas toward HD~28975 represents an intermediate 
case.  For instance, the fraction of carbon in CO for these three sight lines is 
about 4\%, 0.4\%, and 30\%, respectively.

We also placed our results for these directions within the context of the 
large-scale maps of CO and $^{13}$CO emission \citep{gol08} and dust extinction 
\citep{pin10}.  These authors distinguished three types of material -- no 
emission from CO, only emission from $^{12}$CO, 
and emission from both isotopologues.  
The four sight lines emphasized in the present paper can be associated with one 
of these types.  Clearly, HD~30122 probes gas without any CO emission, while 
HD~28975 and HD~29647 sample gas with emission from both isotopologues.  The 
situation for the gas toward HD~27778 is a little less clear; its properties 
place it at the interface between regions with and without CO 
emission.  Though only a statistical perspective, the comparison provides a 
useful means of connecting the two observational techniques.

The analyses conducted on excitation and chemistry yielded values for total proton 
density that spanned a range of a factor of 2.  We discussed the possibility in 
$\S{4.3}$ that this is due to the processes considered by us taking place in 
distinct regions (or portions) of the diffuse gas surrounding molecular clouds, 
and suggested that further progress in understanding the transition from diffuse 
material to molecular cloud requires more sophisticated treatments.  Available 
comprehensive PDR codes allow users to incorporate thermodynamics, chemistry, 
radiative transfer, and excitation into their models for interstellar material.  
Both generic clouds and focused efforts based on extensive observations for 
specific lines of sight are necessary.

\acknowledgments

This work used the Immersion Grating Infrared Spectrometer (IGRINS) that 
was developed under a collaboration between the University of Texas at 
Austin and the Korea Astronomy and Space Science Institute
(KASI) with the financial support of the US National Science Foundation 
under grants AST-1229522 and AST-1702267, of the University of Texas at 
Austin, and of the Korean GMT Project of KASI. This paper includes data 
taken at The McDonald Observatory of The University of Texas at Austin. 
These results made use of the Lowell Discovery Telescope. Lowell 
Observatory is a private, non-profit institution dedicated to 
astrophysical research and public appreciation of astronomy and operates the 
LDT in partnership with Boston University, the University of Maryland, the 
University of Toledo, Northern Arizona University, and Yale University.  This work 
was carried out in part at the Jet Propulsion Laboratory, which is operated for 
NASA by the California Institute of Technology.  This 
research made use of the SIMBAD database operated at CDS, France.  We 
thank Dan Welty for the many discussions, for comments on an earlier 
version of the paper, and for providing data for comparison.

\appendix

\section{Total Equivalent Widths}

Since previous results on interstellar absorption toward HD~28975 
and HD~29746 are available, we provide a comparison for total 
equivalent widths ($W_\lambda$) in Tables 8 and 9; Table 8 also 
lists our results for the sight line toward HD~30122.  The first 
comparison involves the species observed with TS and UVES, 
except C$_2$, where the results appear in the 
next table.  In both tables, additional previously 
unpublished results are presented.  These include our measurements 
on HJST with the coud\'{e} 6 foot camera ($R\sim200,000$) and the 
Sandiford Spectrograph ($R\sim60,000$) on the 2.1~m telescope at 
McDonald Observatory.  Also shown are C$_2$ measurements by 
D. Welty (2018, private communication) from his analysis of the 
UVES data as well as results acquired with the Astrophysical 
Research Consortium echelle spectrograph (ARCES) on the 3.5 m 
telescope at Apache Point Observatory \citep{tho03}, where 
total column densities were presented.

\begin{rotate}
\begin{deluxetable}{lccccccccccccc}[tbh]
\tablecaption{Compilation of Total Equivalent Widths at Visible 
Wavelengths}
\tabletypesize{\scriptsize}
\tablehead{ 
\\
\colhead{Line} & \multicolumn{2}{c}{HD~28975} & & 
\multicolumn{8}{c}{HD~29647} & & \multicolumn{1}{c}{HD~30122} \\ \cline{2-3} 
\cline{5-12} \cline{14-14}
  & \colhead{TS} & \colhead{FSLCVJ94\tablenotemark{a}} & & 
\colhead{TS} & \colhead{UVES} & \colhead{6 foot} & 
\colhead{Sandiford} & \colhead{CW82\tablenotemark{a}} & 
\colhead{C85\tablenotemark{a}} 
& \colhead{ARCES\tablenotemark{b}} & 
\colhead{SBGK08\tablenotemark{a}} & & \colhead{TS}} 
\startdata
K {\sc i} $\lambda$7698 & $151.6 \pm 1.1$\tablenotemark{c} & 
$\ldots$ & & $90 \pm 0.5$ & $80.8 \pm 0.2$ & $89.6 \pm 0.9$ & 
$\ldots$ & $93 \pm 9$ & $110 \pm 10$ & $\ldots$ & $\ldots$ & & $64.5 \pm 2.3$ \\
Ca {\sc ii} K & $132.6 \pm 4.6$ & $\ldots$ & & 
$(164.1 \pm 4.1)$\tablenotemark{d} & 
$(147.9 \pm 0.7)$\tablenotemark{d} & 
$\ldots$ & $(218)$/ $\le$ $7.2$\tablenotemark{d} & 
$\ldots$ & $(280 \pm 15)$\tablenotemark{d} & $\ldots$ & $\ldots$ & & $63.5 \pm 0.7$ \\
Ca {\sc ii} H & $\ldots$ & $\ldots$ & & $\ldots$ & 
$(117.1 \pm 1.0)$\tablenotemark{d} & $\ldots$ & 
$(150)$/ $\le$ $9.3$\tablenotemark{d} & 
$\ldots$ & $(150 \pm 10)$\tablenotemark{d} & $\ldots$ & $\ldots$ & & $\ldots$ \\
Ca {\sc i} $\lambda$4226 & $7.6 \pm 0.5$ & $\ldots$ & & $2.6 \pm 0.5$ & 
$4.0 \pm 0.3$ & $\ldots$ & $\ldots$ & $\ldots$ & $\le$ $8$ & $\ldots$ & 
$\ldots$ & & $\le$ $0.5$ \\
CH $\lambda$4300 & $36.9 \pm 0.8$ & $50.0 \pm 4.0$ & & $34.0 \pm 0.5$ & 
\tablenotemark{e} & $\ldots$ & $\ldots$ & $\ldots$ & $48 \pm 7$ & 
$65.0 \pm 5.0$ & $68.07 \pm 0.8$ & & $11.7 \pm 0.3$ \\
CH $\lambda$3886 & $\ldots$ & $31.0 \pm 6.0$ & & $\ldots$ & 
$12.5 \pm 0.4$ & $\ldots$ & $17.2 \pm 4.2$ & $\ldots$ & $19 \pm 3$ & 
$16.0 \pm 2.5$ & $14.60 \pm 0.6$ & & $\ldots$ \\
CH $\lambda$3890 & $\ldots$ & $\le~18.0$ & & $\ldots$ & $9.3 \pm 0.5$ & 
$\ldots$ & $10.8 \pm 4.4$ & $\ldots$ & $23 \pm 3$ & $\ldots$ & $\ldots$ & & $\ldots$ \\
CH $\lambda$3878 & $\ldots$ & $\le~18.0$ & & $\ldots$ & $5.3 \pm 0.5$ & 
$\ldots$ & $\le$ $14.9$ & $\ldots$ & $\ldots$ & $6.5 \pm 1.5$ & $\ldots$ & & $\ldots$ \\
CH$^+$ $\lambda$4232 & $11.8 \pm 1.1$ & $\ldots$ & & $4.4 \pm 0.8$ & 
$4.6 \pm 0.3$ & $\ldots$ & $\ldots$ & $\ldots$ & $\le$ $12$ & $\ldots$ & $\ldots$ & & 
$1.8 \pm 0.4$ \\
CH$^+$ $\lambda$3957 & $\ldots$ & $\ldots$ & & $\ldots$ & $3.2 \pm 0.3$ & 
$\ldots$ & $\le$ $5.2$ & $\ldots$ & $\le$ $11$ & $\ldots$ & $\ldots$ & & $\ldots$ \\
CN $B$--$X$ (0,~0) R(0) & $55.3 \pm 1.9$ & $30.0 \pm 6.0$ & & $45.0 \pm 1.4$ & 
$46.7 \pm 0.2$ & $\ldots$ & $54.4 \pm 3.6$ & $\ldots$ & $62 \pm 3$ & 
$44.0 \pm 1.5$ & $46.37 \pm 0.4$ & & $4.5 \pm 0.4$ \\
CN $B$--$X$ (0,~0) R(1) & $32.1 \pm 1.9$ & $11.0 \pm 3.0$ & & $34.4 \pm 1.4$ & 
$32.9 \pm 0.2$ & $\ldots$ & $31.5 \pm 2.6$ & $\ldots$ & $46 \pm 3$ & 
$30.0 \pm 1.0$ & $40.42 \pm 0.8$ & & $1.4 \pm 0.4$ \\
CN $B$--$X$ (0,~0) P(1) & $18.4 \pm 1.9$ & $\le~18.0$ & & $22.3 \pm 1.4$ & 
$23.9 \pm 0.2$ & $\ldots$ & $18.5 \pm 2.6$ & $\ldots$ & $31 \pm 3$ & 
$24.0 \pm 1.0$ & $30.34 \pm 0.5$ & & $0.8 \pm 0.4$ \\
CN $B$--$X$ (0,~0) R(2) & $\ldots$ & $\ldots$ & & $\ldots$ & $2.4 \pm 0.2$ & 
$\ldots$ & $\ldots$ & $\ldots$ & $\ldots$ & $\ldots$ & $6 \pm 2$ & & $\ldots$ \\
CN $B$--$X$ (0,~0) P(2) & $\ldots$ & $\ldots$ & & $\ldots$ & $1.6 \pm 0.2$ & 
$\ldots$ & $\ldots$ & $\ldots$ & $\ldots$ & $\ldots$ & $5 \pm 2$ & & $\ldots$ \\
CN $B$--$X$ (1,~0) R(0) & $\ldots$ & $\ldots$ & & $\ldots$ & $16.7 \pm 0.6$ & 
$\ldots$ & $\ldots$ & $\ldots$ & $\ldots$ & $\ldots$ & $21 \pm 4$ & & $\ldots$ \\
CN $A$--$X$ (2,~0) $^S$R$_{21}$(0) & $3.4 \pm 0.5$ & $\ldots$ & & 
$4.2 \pm 0.3$ & $\ldots$ & $\ldots$ & $\ldots$ & $\ldots$ & $\ldots$ & $\ldots$ & 
$\ldots$ & & $\ldots$ \\
NH $A$--$X$ (0,~0) R(1) & $\ldots$ & $\ldots$ & & $\ldots$ & 
$3.5 \pm 0.9$ & $\ldots$ & $\ldots$ & $\ldots$ & $\ldots$ & $\ldots$ & $\ldots$ & & $\ldots$ \\
\enddata
\tablenotetext{a}{References: FSLCVJ94, CW82, C85, and SBGK85 are 
Federman et al. 1994, Chaffee \& White 1982, Crutcher 1985, and S{\l}yk et al. 
2005, respectively.}
\tablenotetext{b}{ARCES data discussed in Thorburn et al. 2003 (Welty 
2019, private communication).}
\tablenotetext{c}{The units are m\AA.}
\tablenotetext{d}{Possible stellar contamination.  The upper limits for the Sandiford 
are our attempt to estimate the interstellar contribution.}
\tablenotetext{e}{Stellar contamination present.}
\end{deluxetable}
\end{rotate}

\begin{deluxetable}{lcccccccc}[tbh]
\tablecaption{Compilation of Total Equivalent Widths for the C$_2$ 
$A$--$X$ (2,~0) Band}
\tablehead{ 
\\
\colhead{Line} & HD~28975 & & 
\multicolumn{6}{c}{HD~29647} \\ \cline{2-2} \cline{4-9}
  & \colhead{TS} & & \colhead{TS} & \colhead{UVES\tablenotemark{a}} & 
\colhead{HBvD83\tablenotemark{b}} & \colhead{LC83\tablenotemark{b}} & 
\colhead{C85\tablenotemark{b}} & \colhead{ARCES\tablenotemark{c}}} 
\startdata
R(0) & $3.7 \pm 0.5$\tablenotemark{d} & & $17.1 \pm 0.4$ & 
$18.1 \pm 0.4$/$17.8 \pm 0.5$ & 
$21 \pm 2.5$ & $17.2 \pm 2.3$ & $17 \pm 5$ & $17.8 \pm 0.7$ \\
R(2) & $8.8 \pm 0.8$ & & $13.4 \pm 0.4$ & 
$14.5 \pm 0.4$/$14.5 \pm 0.4$ & 
$17 \pm 2.5$ & $13.9 \pm 1.8$ & $17 \pm 5$ & $13.0 \pm 0.4$ \\
Q(2) & $10.8 \pm 0.8$ & & $16.8 \pm 0.4$ & 
$17.6 \pm 0.4$/$17.2 \pm 0.5$ & 
$21 \pm 2.5$ & $16.7 \pm 2.0$ & $17 \pm 5$ & $14.8 \pm 0.5$ \\
P(2) & $\ldots$ & & $2.6 \pm 0.3$ & 
$3.6 \pm 0.3$/$3.1 \pm 0.3$ & $\ldots$ & 
$5.8 \pm 1.7$ & $\ldots$ & $3.0 \pm 0.4$ \\
R(4) & $6.2 \pm 0.9$ & & $5.0 \pm 0.4$ & 
$6.1 \pm 0.4$/$5.6 \pm 0.4$ & 
$\le~2.5$ & $5.7 \pm 1.8$ & $\le~10$ & 
$7.6 \pm 0.4$ \tablenotemark{e}\\
Q(4) & $9.5 \pm 0.9$ & & $7.8 \pm 0.4$ & 
$9.4 \pm 0.4$/$9.6 \pm 0.5$ & 
$12 \pm 2.5$ & $7.9 \pm 2.2$ & $\le~10$ & $8.1 \pm 0.5$ \\
P(4) & $\ldots$ & & $3.0 \pm 0.4$ & 
$2.8 \pm 0.3$/$2.6 \pm 0.3$ & $\ldots$ & 
\tablenotemark{f} & $\ldots$ & \tablenotemark{f} \\
R(6) & $\ldots$ & & $3.2 \pm 0.5$ & 
$3.0 \pm 0.3$/$3.3 \pm 0.4$ & $\le~2.5$ & 
$\ldots$ & $\ldots$ & $3.3 \pm 0.3$ \\
Q(6) & \tablenotemark{g} & & $4.4 \pm 0.4$ & 
$3.0 \pm 0.3$/$>~3.1 \pm 0.3$ & 
$7 \pm 2.5$ & $\ldots$ & $\ldots$ & $3.2 \pm 0.4$ \\
P(6) & $\ldots$ & & $\ldots$ & 
$1.9 \pm 0.4$/$<~2.0 \pm 0.3$ & $\ldots$ &
$\ldots$ & $\ldots$ & $\ldots$ \\
R(8) & $\ldots$ & & $1.6 \pm 0.4$ & 
$1.7 \pm 0.4$/$\ldots$ & $\le~2.5$ &
$\ldots$ & $\ldots$ & \tablenotemark{e} \\
Q(8) & $3.4 \pm 0.7$ & & $2.5 \pm 0.3$ & 
$2.8 \pm 0.4$/$2.6 \pm 0.3$ & 
$\le~2.5$ & $4.2 \pm 1.7$\tablenotemark{f} & $\ldots$ & 
$5.1 \pm 0.5$\tablenotemark{f} \\
P(8) & $\ldots$ & & $\ldots$ & 
$\ldots$/$\ldots$ & $\ldots$ & $\ldots$ & 
$\ldots$ & $2.1 \pm 0.2$ \\
R(10) & $\ldots$ & & $\le~1.2$ & 
$0.9 \pm 0.3$/$0.8 \pm 0.3$ & $\ldots$ & 
$\ldots$ & $\ldots$ & $\ldots$ \\
Q(10) & $\ldots$ & & $\le~1.2$ & 
$1.9 \pm 0.4$/$1.5 \pm 0.3$ & $\ldots$ & 
$\ldots$ & $\ldots$ & $\ldots$ \\
P(10) & $\ldots$ & & $\ldots$ & 
$\ldots$/$\ldots$ & $\ldots$ & $\ldots$ & 
$\ldots$ & $1.3 \pm 0.3$ \\
R(12) & $\ldots$ & & $\ldots$ & 
$\ldots$/$\ldots$ & $\ldots$ & $\ldots$ & 
$\ldots$ & $\ldots$ \\
Q(12) & $\ldots$ & & $\ldots$ & 
$\ldots$/$\ldots$ & $\ldots$ & $\ldots$ & 
$\ldots$ & $\ldots$ \\
P(12) & $\ldots$ & & $\ldots$ & 
$\ldots$/$\ldots$ & $\ldots$ & $\ldots$ & 
$\ldots$ & $1.1 \pm 0.1$ \\
\enddata
\tablenotetext{a}{Our results from UVES spectrum followed by those 
of D. Welty (Welty 2018, private communication).}
\tablenotetext{b}{References: HBvD83, LC83, and C85 are Hobbs et al. 
1983, Lutz \& Crutcher 1983, and Crutcher 1985, respectively.}
\tablenotetext{c}{ARCES data discussed in Thorburn et al. 2003 (Welty 
2018, private communication).}
\tablenotetext{d}{The units are m\AA.}
\tablenotetext{e}{Blended with R(8).}
\tablenotetext{f}{Blended with P(4).}
\tablenotetext{g}{Affected by cosmic ray.}
\end{deluxetable}

For the most part, the measurements in Tables 8 and 9 show good 
agreement, especially when considering the uncertainties.  The 
consistency involving the more recent and more precise results from TS and 
UVES toward HD~29647 should be noted.  The blends seen in C$_2$ 
with ARCES agree well with the results of summing the $W_{\lambda}$'s for 
the individual transitions when they are resolved; the same applies to the 
Q(8)/P(4) blend in the spectrum acquired by \citet{lut83}.  Where 
differences are seen, the numerous contaminating stellar features near the 
interstellar lines of CH and CN in the data of \citet{cru85} and \citet{sly08} 
toward HD~29647 or the quality of the CN data \citep{fed94} are the 
likely cause.  For completeness, the equivalent widths for the CO lines 
detected in IGRINS spectra of HD~28975 and HD~29647 are given in Table 10; 
we believe these to be the first detections of these lines for the two 
sight lines.

\begin{deluxetable}{lcc}[tbh]
\tablecaption{Total Equivalent Widths for the CO 
(2,~0) Rovibrational Band}
\setlength{\tabcolsep}{0.7in}
\tablehead{ 
\\
\colhead{Line} & \colhead{HD~28975} & \colhead{HD~29647}
}
\startdata
R(0) & $11.8 \pm 1.9$\tablenotemark{a} & $59.6 \pm 1.0$ \\
R(1) & $11.7 \pm 1.8$ & $69.4 \pm 1.1$ \\
P(1) & $\ldots$ & $37.8 \pm 1.1$ \\
R(2) & $5.5 \pm 1.8$ & $35.0 \pm 1.1$ \\
P(2) & $\ldots$ & $30.8 \pm 1.0$ \\
R(3) & $\ldots$ & $10.1 \pm 1.0$ \\
P(3) & $\ldots$ & $6.5 \pm 1.0$ \\
\enddata
\tablenotetext{a}{The units are m\AA.}
\end{deluxetable}

\begin{deluxetable}{lccccc}[tbh]
\tablecaption{Same as Table 8 for HD~27778: Previously Unpublished Data}
\tablehead{
\\
\colhead{Line} & \colhead{UVES} & \colhead{6 foot} & \colhead{Sandiford} & 
\colhead{Coude Feed\tablenotemark{a}} & \colhead{ARCES\tablenotemark{b}} 
} 
\startdata
K {\sc i} $\lambda$7698 & $84.2 \pm 0.2$ & $87.3 \pm 1.1$ & 
$\ldots$ & $84.0 \pm 1.2$/$83.8 \pm 1.0$ & $88.1 \pm 0.2$/$88 \pm 2$ \\
Ca {\sc ii} K & $60.3 \pm 0.3$ & $\ldots$ & $61.9 \pm 1.2$ & 
$61$/$63.1 \pm 0.9$ & $62.4 \pm 0.5$/$62 \pm 1$ \\
Ca {\sc ii} H & $35.7 \pm 0.3$ & $\ldots$ & $40.1 \pm 2.1$ & 
$\ldots$/$\ldots$ & $\ldots$/$\ldots$ \\
Ca {\sc i} $\lambda$4226 & $1.2 \pm 0.2$ & $\le~3.3$ & $\le~5.1$ & 
$<$2.4/$\ldots$ & $1.3 \pm 0.1$/$1.5 \pm 0.3$ \\
CH $\lambda$4300 & $22.5 \pm 0.3$ & $\ldots$ & $20.2 \pm 0.9$ & 
$21.0 \pm 0.9$/$21.6 \pm 0.6$ & $22.3 \pm 0.3$/$22.5 \pm 1.0$ \\
CH $\lambda$3886 & $6.1 \pm 0.2$ & $\ldots$ & $7.6 \pm 1.3$ & 
$7.4 \pm 0.8$/$\ldots$ & $6.1 \pm 0.3$/$5.5 \pm 1.0$ \\
CH $\lambda$3890 & $3.8 \pm 0.2$ & $\ldots$ & $4.3 \pm 1.0$ & 
$\ldots$/$\ldots$ & $4.3 \pm 0.4$/$4.5 \pm 1.5$ \\
CH $\lambda$3878 & $2.1 \pm 0.2$ & $\ldots$ & $\le~5.1$ & 
$\ldots$/$\ldots$ & $2.1 \pm 0.3$/$2.1 \pm 0.7$ \\
CH$^+$ $\lambda$4232 & $6.2 \pm 0.2$ & $3.7 \pm 0.7$ & 
$5.4 \pm 1.0$ & $9.5 \pm 0.8$/$\ldots$ & $6.6 \pm 0.2$/$6.7 \pm 0.5$ \\
CH$^+$ $\lambda$3957 & $3.7 \pm 0.2$ & $\ldots$ & $\ldots$ & 
$4.4 \pm 0.7$/$\ldots$ & $4.4 \pm 0.3$/$4.5 \pm 1.5$ \\
CN $B$--$X$ (0,~0) R(0) & $30.6 \pm 0.3$ & $\ldots$ & $32.5 \pm 1.3$ & 
$30.9 \pm 0.6$/$30.3 \pm 1.2$ & $30.6 \pm 0.5$/$31.0 \pm 1.0$ \\
CN $B$--$X$ (0,~0) R(1) & $11.2 \pm 0.2$ & $\ldots$ & $9.9 \pm 1.4$ & 
$12.5 \pm 0.6$/$11.9 \pm 1.4$ & $11.1 \pm 0.5$/$11.0 \pm 1.0$ \\
CN $B$--$X$ (0,~0) P(1) & $6.0 \pm 0.2$ & $\ldots$ & $5.2 \pm 1.2$ & 
$7.2 \pm 0.6$/$6.1 \pm 1.1$ & $6.1 \pm 0.5$/$5.5 \pm 0.7$ \\
CN $B$--$X$ (1,~0) R(0) & $3.3 \pm 0.1$ & $\ldots$ & $\ldots$ & 
$\ldots$/$\ldots$ & $\ldots$/$\ldots$ \\
CN $A$--$X$ (1,~0) R(1) & $0.9 \pm 0.1$ & $\ldots$ & $\ldots$ & 
$\ldots$/$\ldots$ & $\ldots$/$\ldots$ \\
NH $A$--$X$ (0,~0) R(1) & $1.2 \pm 0.1$ & $\ldots$ & $\ldots$ & 
$\ldots$/$\ldots$ & $\ldots$/$\ldots$ \\
\enddata
\tablenotetext{a}{The first entry under Coude Feed involves results with 
Camera No. 5 and the second with Camera No. 6 (Welty 2019, private 
communication).}
\tablenotetext{b}{ARCES data analyzed by D. Welty (first entry) and 
J. Thornburn (second entry); the molecular results were discussed in 
Thorburn et al. 2003 and Fan et al. 2017.  Fan et al. 2017 also analyzed 
the atomic data.  (Welty 2019, private communication).}
\end{deluxetable}

\begin{deluxetable}{lcccccccc}[tbh]
\tablecaption{Same as Table 8 for HD~27778: Published Data}
\tablehead{ 
\\
\colhead{Line} &  \colhead{UVES} & 
\colhead{FSLCVJ94\tablenotemark{a}} & 
\colhead{RM95\tablenotemark{a}} & 
\colhead{MSBMHKG05\tablenotemark{a}} & 
\colhead{WGMK08\tablenotemark{a}} & 
\colhead{SBGK08\tablenotemark{a}} & 
\colhead{WGBK09a,b\tablenotemark{a}} &
\colhead{WGGK14\tablenotemark{a}}} 
\startdata
K {\sc i} $\lambda$7698 & $84.2 \pm 0.2$ & 
$\ldots$ & $\ldots$ & $87.7 \pm 4.1$ & $\ldots$ & $\ldots$ & 
$\ldots$ & $\ldots$ \\
Ca {\sc ii} K & $60.3 \pm 0.3$ & $\ldots$ & 
$\ldots$ & $51.9 \pm 6.1$ & $\ldots$ & $\ldots$ & $\ldots$ & 
$\ldots$ \\
Ca {\sc ii} H & $35.7 \pm 0.3$ & $\ldots$ & $\ldots$ & 
$27.7 \pm 4.7$ & $\ldots$ & $\ldots$ & $\ldots$ & $\ldots$ \\
Ca {\sc i} $\lambda$4226 & $1.2 \pm 0.2$ & $\ldots$ & $\ldots$ & 
$\ldots$ & $\ldots$ & $\ldots$ & $\ldots$ & $\ldots$ \\
CH $\lambda$4300 & $22.5 \pm 0.3$ & 
$21.7 \pm 0.6$ & $\ldots$ & $22.1 \pm 0.3$ & $23.3 \pm 0.3$ & 
$23.57 \pm 0.8$ & $23.82 \pm 2.67$\tablenotemark{b} & 
$22.42 \pm 0.28$ \\
CH $\lambda$3886 & $6.1 \pm 0.2$ & 
$6.7 \pm 0.8$ & $\ldots$ & $\ldots$ & $7.7 \pm 0.9$ & 
$7.59 \pm 0.6$ & $9.75 \pm 0.78$ & $7.04 \pm 0.86$ \\
CH $\lambda$3890 & $3.8 \pm 0.2$ & $\le~6.3$ & $\ldots$ & 
$\ldots$ & $5.1 \pm 0.6$ & $\ldots$ & $4.82 \pm 0.64$ & 
$4.47 \pm 0.78$ \\
CH $\lambda$3878 & $2.1 \pm 0.2$ & $\le~8.1$ & $\ldots$ & 
$\ldots$ & $\ldots$ & $\ldots$ & $\ldots$ & $2.83 \pm 0.69$ \\
CH$^+$ $\lambda$4232 & $6.2 \pm 0.2$ & 
$\ldots$ & $\ldots$ & $\ldots$ & $\ldots$ & $\ldots$ & $9.3 \pm 0.3$ & 
$6.55 \pm 0.27$ \\
CH$^+$ $\lambda$3957 & $3.7 \pm 0.2$ & $\ldots$ & 
$\ldots$ & $\ldots$ & $\ldots$ & $\ldots$ & $4.9 \pm 0.3$ & 
$3.39 \pm 0.28$ \\
CN $B$--$X$ (0,~0) R(0) & $30.6 \pm 0.3$ & 
$27.3 \pm 0.5$ & $30.64 \pm 0.13$ & $\ldots$ & $32.6 \pm 0.7$ &
$31.14 \pm 0.4$ & $\ldots$ & $\ldots$ \\
CN $B$--$X$ (0,~0) R(1) & $11.2 \pm 0.2$ & 
$10.5 \pm 0.5$ & $11.73 \pm 0.13$ & $\ldots$ & $12.4 \pm 0.3$ & 
$11.89 \pm 0.6$ & $\ldots$ & $\ldots$ \\
CN $B$--$X$ (0,~0) P(1) & $6.0 \pm 0.2$ & 
$4.9 \pm 0.4$ & $6.44 \pm 0.23$ & $\ldots$ & $7.1 \pm 0.5$ & 
$6.65 \pm 0.5$ & $\ldots$ & $\ldots$ \\
CN $B$--$X$ (1,~0) R(0) & $3.3 \pm 0.1$ & $\ldots$ & 
$3.44 \pm 0.23$ & 
$\ldots$ & $\ldots$ & $\ldots$ & $10.48 \pm 2.44$ & $\ldots$ \\
CN $A$--$X$ (1,~0) R(1) & $0.9 \pm 0.1$ & $\ldots$ & 
$1.03 \pm 0.22$ & 
$\ldots$ & $\ldots$ & $\ldots$ & $3.38 \pm 1.97$ & $\ldots$ \\
NH $A$--$X$ (0,~0) R(1) & $1.2 \pm 0.1$ & $\ldots$ & 
$\ldots$ & $\ldots$ & 
$\ldots$ & $\ldots$ & $\ldots$ & $\ldots$ \\
\enddata
\tablenotetext{a}{References: FSLCVJ94, RM95, MSBMHKG05, WGMK08, 
SBGK85,WGBK09a,b, and WGGK14 are Federman et al. 1994, Roth \& 
Meyer 1995, Megier et al. 2005, Weselak et al. 2008, S{\l}yk et al. 2008, 
Weselak et al. 2009a,b, and Weselak et al. 2014, respectively.}
\tablenotetext{b}{Since the other entries are based on data from the same 
spectroscopic setup with the same measures of $W_{\lambda}$, we listed 
the results for CH $\lambda$4300 from Weselak 2019 in this column.}
\end{deluxetable}

The diffuse gas toward HD~27778 has been the subject of many efforts.  We compare 
our analysis of UVES spectra with our previously unpublished results with the 6 foot 
camera and the Sandiford Spectrograph, along with those of D. Welty and colleagues 
with the Coude Feed Telescope at Kitt Peak National Observatory and ARCES (Welty 
2019, private communication), in Table 11.  The comparison with published 
results appears in Table 12; the results from more focused studies, those 
involving only one or two lines, are given here.  \citet{jos86} reported a 
value of 29 m\AA\ for the $W_{\lambda}$ of the CN (0,~0) R(0) line, \citet{mey91} 
reported the first detection of NH absorption from diffuse gas, with 
$W_{\lambda}$ of $1.1~\pm~0.3$ m\AA, and \citet{kre99} obtained measures of 
$W_{\lambda}$ of 24.0 and 23.8 m\AA\ for CH $\lambda$4300 and of 7.8 m\AA\ for 
CH$^+$ $\lambda$4232.  For atomic species, we note two additional 
studies: \citet{wel01} gave $W_{\lambda}$ of $83.6~\pm~1.3$ m\AA\ for the 
K~{\small I} line and \citet{meg09} provided values of $52~\pm~6.6$ and 
$28~\pm~4.9$ m\AA\ for the Ca~{\small II} K and H lines.  This extensive 
set of results shows very good agreement, with variations within 2$\sigma$ 
levels.

\section{Calcium Ions and the Electron Density}

The emphasis for the main text was molecules and related species like 
K~{\small I}.  Here we briefly interpret results associated with absorption from 
calcium atoms and ions in our spectra.  In the picture of a diffuse cloud presented 
by \citet{pan05}, Ca~{\small I} and Ca~{\small II} absorption mainly arises 
from the outer regions of the cloud where the gas densities are lower 
[$n_{tot}$(H)~$<$~100 cm$^{-3}$, e.g., \citet{ric18a}] and the gas is 
predominantly atomic.  The ionization balance for calcium allows us to estimate 
the electron density for this material.

The component structures in terms of column densities for Ca~{\small I} and 
Ca~{\small II} appear in Table 2.  For the lines of sight toward HD~27778 and 
HD~28975, the components with the largest Ca~{\small II} columns and the 
only ones indicating absorption from Ca~{\small I} are the molecular ones, a 
consequence that most of the material is associated with these components.  For 
the clouds toward HD~30122, the sight line with the least amount of molecular 
material in our sample, the individual components have comparable amounts of 
singly-ionized Ca and only a limit on the amount 
of neutral calcium.  The two directions 
resembling a typical diffuse molecular cloud have total column densities of 
Ca~{\small II} similar to the well-studied examples toward $o$ Per, $\zeta$ Per, 
and $\zeta$ Oph \citep[e.g.,][]{wel96}; the Ca~{\small I} column density 
toward HD~27778 is also similar \citep[see][]{wel03}.  The material toward 
HD~28975 has factor of a few larger total $N$(Ca~{\small II}) and is more 
closely associated with the molecular components; a similar enhancement in 
$N$(Ca~{\small I}) is found.  According to \citet{ric18a}, these Ca~{\small II} 
components are atomic gas associated with molecular cloud envelopes, in this 
case TMC.  The other components seen in Ca~{\small II} toward HD~28975 and 
HD~30122 likely represent absorption from the cold (100 K) neutral medium or 
the warm (10,000 K) medium detected at radio wavelengths in H~{\small I}.

For the diffuse molecular clouds toward HD~27778 and HD~28975,  ionization balance 
provides us with an estimate of $n$(e).  Since the recombination rates are 
temperature dependent \citep[e.g.,][]{shu82}, we adopt a kinetic temperature of 50 K 
for both directions. The outcome also depends on the radiation field enhancement 
factor $I/I_0$ and we perform 
calculations for both $I/I_0=1.0$ and 0.5 (just as we did for the chemical 
analysis discussed in Section 3.4). For HD~27778, these calculations 
yield $n$(e) of 0.18 and 0.09 cm$^{-3}$ for $I/I_0=1.0$ and 0.5, respectively. 
Similar calculations for HD~28975 yield $n$(e) of 0.24 and 0.12 cm$^{-3}$. For HD~27778, 
the estimates for $n$(e) obtained here are somewhat lower than those derived from the 
analysis of CN rotational excitation, which indicated $n$(e) in the range 0.21 to 0.34 
cm$^{-3}$ (Section 3.2.3). For HD~28975, 
the above estimates for $n$(e) are consistent 
with the upper limit provided by the CN analysis. The electron densities could be larger 
if the temperature of the Ca~{\small I}- and Ca~{\small II}-bearing gas is higher than 
50 K. For example, the estimates for $n$(e) would increase by $\sim$50\% if a 
kinetic temperature of 80 K were adopted instead.  Because 
severe stellar contamination prevents us from measuring interstellar 
Ca~{\small II} toward HD~29647, we are not able to extract an electron 
density for this sight line.

While the values for $n$(e) derived from CN excitation and ionization balance 
between neutral and singly-ionized calcium are comparable, analyses of other ion 
pairs tend to suggest lower electron densities \citep{wel03}.  Welty et al. found 
that the results for $N$(Ca~{\small I})/$N$(Ca~{\small II}) were typically a 
factor of ten larger than those for other pairs, such as 
$N$(C~{\small I})/$N$(C~{\small II}), $N$(Mg~{\small I})/$N$(Mg~{\small II}), 
$N$(S~{\small I})/$N$(S~{\small II}), and $N$(Fe~{\small I})/$N$(Fe~{\small II}).  
These relative results did not depend on whether or not charge exchange between 
ions and small grains \citep[e.g.,][]{wei01} was included.  It is also important 
to remember that our analyses refer to diffuse molecular gas (CN) and atomic gas 
(ionization balance), but the gas densities are a factor of several or more 
greater in the molecular material.  Thus, the ionization fraction, $x$(e), appears 
to be much larger than that obtained from C$^+$ observations \citep{sof97} although 
C$^+$ is often assumed to be the main contributor of electrons in neutral diffuse gas.  
\citet{wel03} noted that inclusion of charge exchange with small grains leads to 
more consistent results for the amount of ionization, but not for all ion pairs.

\section{The Line of Sight toward HD~26571}

A number of observational studies included measurements on atomic and molecular 
species for the gas in front of HD~26571, a B8III star \citep{moo13} with 
$E$($B-V$) of 0.27 and an $R_V$ comparable to the average interstellar 
value \citep{weg03}.  While the line of sight to this star 
is located in the lower right-hand corner of Fig. 1, where at higher contrast 
CO emission is seen \citep{gol08, nar08, pin10}, there are no high-resolution 
measurements for the type of analyses conducted above.  Of particular relevance 
to the work presented here are the data on absorption from \citet{dic83}, 
\citet{jos86}, \citet{cra90}, and \citet{tho03} and on CO emission by 
\citet{dic83} and \citet{van91}. Tables 13 and 14 provide compilations of 
published and previously unpublished total equivalent widths.  The molecular 
absorption occurs at $+10$ km s$^{-1}$.  The maps of CO and $^{13}$CO emission 
\citep{nar08} peak at $+9$ to $+11$ km s$^{-1}$ along the line of sight.  In 
what follows, we use data from ARCES on CH, C$_2$, and CN from the analysis 
by D. Welty (2020, private communication).  The higher resolution spectra of 
\citet{cra90} (3 km s$^{-1}$) reveal only one molecular component, and ours 
from the HJST coud\'{e} 6 foot camera (1.5 km s$^{-1}$) show an additional very 
weak component in K~{\small I} absorption with $W_{\lambda}$ of 2.5(0.4) m\AA.  
For K~{\small I}, our $W_{\lambda}$ associated with molecular absorption is 
adopted.  Table 14 also includes the column densities for each rotational level.

\begin{rotate}
\begin{deluxetable}{lccccccc}[tbh]
\tablecaption{Same as Table 8 for HD~26571}
\tablehead{
\\
\colhead{Line} & \colhead{6 foot} & \colhead{Sandiford} & 
\colhead{JSSC86\tablenotemark{a}} & \colhead{c90\tablenotemark{a}} & 
\colhead{FSLCVJ94\tablenotemark{a}} & \colhead{Coude Feed\tablenotemark{b}} & 
\colhead{ARCES\tablenotemark{c}} 
} 
\startdata
K {\sc i} $\lambda$7698 & $94.9 \pm 0.8$ & $\ldots$ & $\ldots$ & $\ldots$ &	
$\ldots$ & $94.1 \pm 1.1$ & $96.9 \pm 0.5$/$99.3 \pm 0.4$ \\
Ca {\sc ii} K & $\ldots$ & $186.7 \pm 3.3$\tablenotemark{d} & $\ldots$ & 
$90 \pm 23$ & $\ldots$ & $\ldots$ & 
$174 \pm 1$\tablenotemark{d}/$175 \pm 6$\tablenotemark{d} \\
Ca {\sc ii} H & $\ldots$ & $104.4 \pm 4.6$\tablenotemark{d} & $\ldots$ & 
$\ldots$ & $\ldots$ & $\ldots$ & $\ldots$/$108 \pm 9$\tablenotemark{d} \\
Ca {\sc i} $\lambda$4226 & $\le~1.5$ & $\le~1.7$ & $\ldots$ & 
$\ldots$ & $\ldots$ & $\ldots$ & $\ldots$/$\le~2.0$ \\
CH $\lambda$4300 & $\ldots$ & $16.1 \pm 0.6$ & $\ldots$ & 
$13 \pm 2$ & $10.3 \pm 1.3$ & $\ldots$ & $13.8 \pm 0.3$/$13.6 \pm 0.8$ \\
CH $\lambda$3886 & $\ldots$ & $5.2 \pm 0.9$ & $\ldots$ & 
$\ldots$ & $\ldots$ & $\ldots$ & $4.7 \pm 0.4$/$3.7 \pm 1.0$ \\
CH $\lambda$3890 & $\ldots$ & $1.3 \pm 0.5$ & $\ldots$ & 
$\ldots$ & $\ldots$ & $\ldots$ & $\ldots$/$\le~4.0$ \\
CH $\lambda$3878 & $\ldots$ & $\le~3.3$ & $\ldots$ & 
$\ldots$ & $\ldots$ & $\ldots$ & $\ldots$/$1.7 \pm 0.5$ \\
CH$^+$ $\lambda$4232 & $3.9 \pm 0.6$ & \tablenotemark{d} & $\ldots$ & 
$\le~6.0$ & $\ldots$ & $\ldots$ & $3.5 \pm 0.3$/$\le~6.9$ \\
CH$^+$ $\lambda$3957 & $\ldots$ & $\le~2.2$\tablenotemark{d} & $\ldots$ & 
$\ldots$ & $\ldots$ & $\ldots$ & $\ldots$/$\le~4.0$\tablenotemark{d} \\
CN $B$--$X$ (0,~0) R(0) & $\ldots$ & $21.7 \pm 0.7$ & $22$ & 
$21 \pm 3$ & $\ldots$ & $\ldots$ & $22.7 \pm 0.6$/$21.5 \pm 1.0$ \\
CN $B$--$X$ (0,~0) R(1) & $\ldots$ & $5.2 \pm 0.9$ & $\ldots$ & 
$11 \pm 4$ & $\ldots$ & $\ldots$ & $9.3 \pm 0.5$/$9.5 \pm 0.7$ \\
CN $B$--$X$ (0,~0) P(1) & $\ldots$ & $7.0 \pm 0.8$ & $\ldots$ & 
$8 \pm 3$ & $\ldots$ & $\ldots$ & $3.7 \pm 0.4$/$4.1 \pm 0.7$ \\
CN $A$--$X$ (2,~0) R$_1$(0) & $\ldots$ & $\ldots$ & $\ldots$ & 
$\ldots$ & $\ldots$ & $\le~4.5$ & $2.4 \pm 0.4$/$3.0 \pm 0.6$ \\
\enddata
\tablenotetext{a}{References: JSSC86, C90, and FSLCVJ94 are Joseph et al. 1986, 
Crawford 1990, and Federman et al. 1994, respectively.}
\tablenotetext{b}{The entry under Coude Feed involves results with 
Camera No. 5 (Welty 2020, private communication).}
\tablenotetext{c}{ARCES data analyzed by D. Welty (first entry) and 
J. Thornburn (second entry); the molecular results were discussed in 
Thorburn et al. 2003.  (Welty 2020, private communication).}
\tablenotetext{d}{Likely affected by stellar contamination.}
\end{deluxetable}
\end{rotate}

\begin{deluxetable}{lccccc}[tbh]
\tablecaption{Compilation of Results for the C$_2$ $A$--$X$ (2,~0) Band Toward 
HD~26571}
\tablehead{ 
\\
Line  & \colhead{FSLCVJ94\tablenotemark{a}} & \colhead{ARCES\tablenotemark{b}} & 
 & \colhead{Column Density} \\
 & & & & \colhead{$10^{12}$ cm$^{-2}$}
} 
\startdata
R(0) & $4.2 \pm 1.0$\tablenotemark{c} & 
$5.0 \pm 0.4$/$4.5 \pm 1.0$\tablenotemark{c} & 
 & $5.4 \pm 0.4$\tablenotemark{d} \\
R(2) & $7.3 \pm 1.1$ & $5.9 \pm 0.7$/$6.9 \pm 1.3$ & & $15.3 \pm 0.8$ \\
Q(2) & $8.2 \pm 1.1$ & $6.9 \pm 0.4$/$6.9 \pm 1.0$ & & $\ldots$ \\
P(2) & $\ldots$ & $2.7 \pm 1.0$/$5.5 \pm 1.3$ & & $\ldots$ \\
R(4) & $7.3 \pm 1.2$ & $5.0 \pm 0.3$\tablenotemark{e}/$4.0 \pm 1.3$ & 
 & $11.7 \pm 1.4$ \\
Q(4) & $9.5 \pm 1.2$ & $5.4 \pm 0.6$/$4.8 \pm 1.2$ & & $\ldots$ \\
P(4) & $\ldots$ & $3.3 \pm 0.3$\tablenotemark{f}/$6.0 \pm 1.5$ & 
 & $\ldots$ \\
R(6) & $\ldots$ & $1.0 \pm 0.3$/$\le~3.0$ & 
 & $3.4 \pm 1.0$ \\
Q(6) & $\ldots$ & $\le~2.4$/$\le~4.0$ & & $\ldots$ \\
P(6) & $\ldots$ & $\le~1.2$/$\le~3.0$ & & $\ldots$ \\
R(8) & $\ldots$ & \tablenotemark{e}/$\ldots$ & & $3.7 \pm 1.1$ \\
Q(8) & $\ldots$ & \tablenotemark{f}/$\le~3.0$ & & $\ldots$ \\
P(8) & $\ldots$ & $\ldots$/$\le~3.0$ & & $\ldots$ \\
R(10) & $\ldots$ & $\le~1.2$/$\ldots$ & & $\le~4.4$ \\
Q(10) & $\ldots$ & $\ldots$/$\le~3.0$ & & $\ldots$ \\
\enddata
\tablenotetext{a}{Reference: FSLCVJ94 is Federman et al. 1994.}
\tablenotetext{b}{ARCES data analyzed by D. Welty (first entry) and J. 
Thornburn (second entry); the molecular results were discussed in Thorburn 
et al. 2003.  (Welty 2020, private communication).}
\tablenotetext{c}{The units for $W_{\lambda}$ are m\AA.}
\tablenotetext{d}{Column density appears in first row for that rotational 
level.}
\tablenotetext{e}{Blended with R(8).}
\tablenotetext{f}{Blended with Q(8).}
\end{deluxetable}

The results displayed in Tables 13 and 14 are generally consistent, with a 
few exceptions.  Like HD~29647, HD~26571 is a late-type peculiar B star, causing 
severe stellar contamination especially in the vicinity of the interstellar 
Ca~{\small II}, CH$^+$, and Na~{\small I} D lines \citep[see][]{cra90}.  The data 
for the Na~{\small I} transitions are not included in the tables.  We used curves 
of growth to convert equivalent widths into column densities.  For CH and CN, 
consistency was sought between weak and strong lines, thereby providing an 
estimate for the $b$-value.  The CH transitions at 3886 and 4300 \AA\ were 
used, as were CN $\lambda\lambda$7906,3874 from $N^{\prime\prime}$ $=$ 0.  
Because CH $\lambda$3886 only represents one of the $\Lambda$-doubling 
components in the ground state and because the members of the doublet should 
have equal populations, the value for $N$(CH) was multiplied by 2.  The 
lines associated with the other component, $\lambda\lambda$3878,3890, were not 
seen in the ARCES spectrum.  A $b$-value of 1.2 km s$^{-1}$ best describes the 
data and is consistent with the analysis performed by \citet{cra90} and with 
the line widths seen in CO emission \citep{dic83, van91}.  This $b$-value was 
also adopted for absorption from C$_2$ and K~{\small I}, while a slightly 
larger value for CH$^+$ of 1.5 km s$^{-1}$ was chosen in light of the findings by 
\citet{pan05}.  Since the optical depth at line center for CH$^+$ $\lambda$4232 
is only 0.10, the curve-of-growth results do not differ much from considering an 
optically thin line in the calculation.  When more than one 
line yielded a column density for a rotational level, we quote the weighted 
mean.  Moreover, the estimates for CO column density from UV absorption 
\citep{dic83, jos86} and millimeter-wave  emission \citep{dic83, van91} are 
similar, yielding values of $10^{16}$ cm$^{-2}$ to within 30\%.

The column densities for CH, C$_2$, CN, CH$^+$, and K~{\small I}, along with 
the column densities for rotational levels in C$_2$ and CN, are used in our 
analyses on excitation and chemistry.  We obtain respective values for $N$(CH), 
$N$(C$_2$), $N$(CN), $N$(CH$^+$), and $N$(K~{\small I}) of 
$20.8\times10^{12}$, $39.5\times10^{12}$, $10.7\times10^{12}$, 
$4.2\times10^{12}$, and $1.91\times10^{12}$ cm$^{-2}$.  The column 
densities for the C$_2$ rotational levels appear in Table 14.  Those for CN 
are $7.52\times10^{12}$ cm$^{-2}$ for $N$~$=$~0 and $3.22\times10^{12}$ 
cm$^{-2}$ for $N$~$=$~1, and $T_{01}$(CN) is $2.8 \pm 0.1$ K.  The analysis 
for C$_2$ excitation reveals a gas temperature of 20 (10 to 30) K and total 
proton density of $\sim$100(225) cm$^{-3}$ when adopting $I_{IR}$ of 0.5(1.0).  
A temperature of 20 K and $\tau_{UV}$ of 1.65 were used in our simple chemical 
model.  The model yields total proton densities of 275(550) cm$^{-3}$ when 
considering similar strengths for the UV field permeating the gas.  For these 
physical conditions, the predictions for $N$(C$_2$) and $N$(CN) are 
$45.2\times10^{12}$ and $8.7\times10^{12}$ cm$^{-2}$, respectively, independent 
of the strength of the radiation field.  The value for $N_{tot}$(H) ranges from 
$1.6\times10^{21}$ to $2.9\times10^{21}$ cm$^{-2}$, depending on whether 
it is based on $E$($B-V$) or $N$(K~{\small I}).  This indicates that the 
fractional abundance of CO is between $3.4\times10^{-6}$ and 
$6.2\times10^{-6}$, only a few percent of the carbon budget.  Finally, using 
$N$(CH) to predict the CO column density \citep{she08} leads to $N$(CO) of 
about $10^{15}$ cm$^{-2}$, a factor of 10 less than observed.  It appears that 
the line of sight toward HD~26571 is very similar to the one toward HD~27778, 
with a somewhat lower gas temperature, as shown in Table 7.

Earlier studies of the interstellar material toward HD~26571 gave estimates of 
gas densities that we can compare with ours.  As in our earlier comparisons, 
we do not adjust the input parameters.  When the results are given in 
terms of density of collision partners, we multiply these by a factor of 1.5. 
\citet{jos86} used C~{\small I} excitation to infer a pressure and assumed a 
gas temperature of 40 K to infer a total proton density between 180 and 
2000 cm$^{-3}$.  They also obtained a density from C~{\small I} excitation 
toward HD~27778 of 175 to 800 cm$^{-3}$; the range is similar to what we find 
from our analyses (see Table 7).  \citet{cra90} used an earlier version 
of our chemical model to extract a value for $n_{tot}$(Chem) from his CN 
measurements, obtaining 1300 cm$^{-3}$ which is similar to the value quoted 
by \citet{fed94}.  From their analysis of CO excitation, \citet{van91} 
obtained $n_{tot}^{ex}$(CO) between 750 and 2250 cm$^{-3}$ with a preferred 
density of 1200 cm$^{-3}$.  Besides modeling the CN chemistry, \citet{fed94} 
found a gas density from C$_2$ excitation of 800 cm$^{-3}$ and a gas 
temperature of about 50 K, higher than we obtain.  These results and ours are 
in reasonable agreement, especially when one considers the precision of the 
earlier absorption-line measurements.  In light of the current results for 
$T_{01}$(CN) showing no excess over the CMB, we suggest treating the electron 
and molecular hydrogen densities, 0.59 and 5900 cm$^{-3}$, determined by 
\citet{bla91} as upper limits. 

\facilities{McDonald:HJST(TS, IGRINS), VLT(UVES), Lowell:LDT(IGRINS)}

\software{IRAF\citep{tod86, tod93}, ISMOD\citep{she08}, RADEX\citep{van07}}

\end{document}